\documentclass[11pt,epsf]{article}
 \usepackage{amsmath}
 \usepackage{graphicx}
 \usepackage[merge,numbers,compress]{natbib}
 \usepackage[T1]{fontenc}
 \usepackage{booktabs}
 \usepackage{xcolor} 
 \usepackage{xspace}
 \usepackage{dcolumn}
 \usepackage{hyperref}
 \usepackage{caption}
 \usepackage
 [subrefformat=parens,position=top,skip=-15pt,margin=15pt,justification=justified,singlelinecheck=false]
 {subcaption}

\def\refeq#1{\mbox{(\ref{#1})}}

\def\refta#1{\mbox{Table~\ref{#1}}}

\def\citere#1{\mbox{Ref.~\cite{#1}}}
\def\citeres#1{\mbox{Refs.~\cite{#1}}}

\newcommand{\ie}{\emph{i.e.}\ }
\newcommand{\eg}{\emph{e.g.}\ }

\def\be{\begin{equation}}
\def\ee{\end{equation}}

\newcommand{\PH}{\ensuremath{\text{H}}\xspace}
\newcommand{\Pj}{\ensuremath{\text{j}}\xspace}
\newcommand{\Pp}{\ensuremath{\text{p}}\xspace}
\newcommand{\Pe}{\ensuremath{\text{e}}\xspace}
\newcommand{\Pb}{\ensuremath{\text{b}}\xspace}
\newcommand{\Pq}{\ensuremath{\text{q}}\xspace}
\newcommand{\Pt}{\ensuremath{\text{t}}\xspace}
\newcommand{\Pu}{\ensuremath{\text{u}}\xspace}
\newcommand{\Pd}{\ensuremath{\text{d}}\xspace}
\newcommand{\Ps}{\ensuremath{\text{s}}\xspace}
\newcommand{\Pc}{\ensuremath{\text{c}}\xspace}
\newcommand{\Pg}{\ensuremath{\text{g}}\xspace}

\newcommand{\PW}{\ensuremath{\text{W}}\xspace}
\newcommand{\PZ}{\ensuremath{\text{Z}}\xspace}

\newcommand{\Mt}{\ensuremath{m_\Pt}\xspace}

\newcommand{\MWOS}{\ensuremath{M_\PW^\text{OS}}\xspace}
\newcommand{\MW}{\ensuremath{M_\PW}\xspace}
\newcommand{\MZOS}{\ensuremath{M_\PZ^\text{OS}}\xspace}
\newcommand{\MZ}{\ensuremath{M_\PZ}\xspace}

\newcommand{\GZOS}{\ensuremath{\Gamma_\PZ^\text{OS}}\xspace}

\newcommand{\GWOS}{\ensuremath{\Gamma_\PW^\text{OS}}\xspace}

\newcommand{\MeV}{\ensuremath{\,\text{MeV}}\xspace}
\newcommand{\GeV}{\ensuremath{\,\text{GeV}}\xspace}
\newcommand{\TeV}{\ensuremath{\,\text{TeV}}\xspace}

\newcommand{\alphas}{\ensuremath{\alpha_\text{s}}\xspace}
\newcommand{\order}[1]{\ensuremath{\mathcal{O}{\left(#1\right)}}\xspace}

\newcommand{\GF}{\ensuremath{G_\mu}}

\newcommand{\MVOS}{\ensuremath{M_{V}^\text{OS}}\xspace}%
\newcommand{\GVOS}{\ensuremath{\Gamma_{V}^\text{OS}}\xspace}%

\newcommand{\newc}{\newcommand}
\newc{\bi}{\begin{itemize}}
\newc{\ei}{\end{itemize}}
\newc{\benu}{\begin{enumerate}}
\newc{\eenu}{\end{enumerate}}
\newc{\bc}{\begin{center}}
\newc{\ec}{\end{center}}
\newc{\bfig}{\begin{figure}}
\newc{\efig}{\end{figure}}
\newc{\qbar}{\bar{q}}
\newc{\go}{\tilde{g}}
\newc{\PB}{\textsc{Powheg-Box}}

\newcommand{\Recola}{{\sc Recola}\xspace}
\newcommand{\Sherpa}{{\sc Sherpa}\xspace}
\newcommand{\Rivet}{{\sc Rivet}\xspace}
\newcommand{\Amegic}{A\protect\scalebox{0.8}{MEGIC}\xspace}
\newcommand{\Comix}{C\protect\scalebox{0.8}{OMIX}\xspace}
\newcommand{\OpenLoops}{O\protect\scalebox{0.8}{PEN}L\protect\scalebox{0.8}{OOPS}\xspace}
\newcommand{\Njet}{N\protect\scalebox{0.8}{JET}\xspace}
\newcommand{\BlackHat}{B\protect\scalebox{0.8}{LACK}H\protect\scalebox{0.8}{AT}\xspace}
\newcommand{\Gosam}{G\protect\scalebox{0.8}{O}S\protect\scalebox{0.8}{AM}\xspace}

\newcommand{\collier}{{\sc Collier}\xspace}
\newcommand{\CutTools}{{\sc CutTools}\xspace}
\newcommand{\OneLOop}{{\sc OneLOop}\xspace}
\newcommand{\madgraph}{{\sc\small MadGraph5\_aMC@NLO}\xspace}
\newcommand{\madgraphbis}{{\sc\small MG5\_aMC@NLO}\xspace}

\newcolumntype{.}{D{.}{.}{-1}}
\newcolumntype{d}[1]{D{.}{.}{#1}}

\colorlet{tableoverheadcolor}{gray!37.5}
\colorlet{tableheadcolor}{gray!25}
\colorlet{tablerowcolor}{gray!12.5}

\newlength{\width}
\newlength{\height}

\def\draftdate{\relax}
\def\mda{\relax}
\def\mua{\relax}
\def\mla{\relax}
\def\draft{
\def\thtystars{******************************}
\def\sixtystars{\thtystars\thtystars}
\typeout{}
\typeout{\sixtystars**}
\typeout{* Draft mode!
         For final version remove \protect\draft\space in source file *}
\typeout{\sixtystars**}
\typeout{}
\def\draftdate{\today}
\def\mua{\marginpar[\boldmath\hfil$\uparrow$]%
                   {\boldmath$\uparrow$\hfil}\color{black}%
                    \typeout{marginpar: $\uparrow$}\ignorespaces}
\def\mda{\color{red}\marginpar[\boldmath\hfil$\downarrow$]%
                   {\boldmath$\downarrow$\hfil}%
                    \typeout{marginpar: $\downarrow$}\ignorespaces}
\def\mla{\marginpar[\boldmath\hfil$\rightarrow$]%
                   {\boldmath$\leftarrow $\hfil}%
                    \typeout{marginpar: $\leftrightarrow$}\ignorespaces}
\def\Mua{\marginpar[\boldmath\hfil$\Uparrow$]%
                   {\boldmath$\Uparrow$\hfil}\color{black}%
                    \typeout{marginpar: $\uparrow$}\ignorespaces}
\def\Mda{\color{red}\marginpar[\boldmath\hfil$\Downarrow$]%
                   {\boldmath$\Downarrow$\hfil}%
                    \typeout{marginpar: $\downarrow$}\ignorespaces}
\def\Mla{\marginpar[\boldmath\hfil\textcolor{red}{$\Rightarrow$}]%
                   {\boldmath\textcolor{red}{$\Leftarrow $}\hfil}%
                    \typeout{marginpar: $\leftrightarrow$}\ignorespaces}
\overfullrule 5pt
\oddsidemargin 15mm
\marginparwidth 29mm
}

\begin{document}

\title{\hfill ~\\[-30mm]
\phantom{h} \hfill\mbox{\small MCNET-17-05}
\\[1cm]
\vspace{13mm}   \textbf{Automation of NLO QCD and EW corrections \\ with \Sherpa and \Recola}}

\date{}
\author{
Benedikt Biedermann$^{1\,}$\footnote{E-mail: \texttt{benedikt.biedermann@physik.uni-wuerzburg.de}},
Stephan Br\"auer$^{2\,}$\footnote{E-mail: \texttt{stephan.braeuer@phys.uni-goettingen.de}},
Ansgar Denner$^{1\,}$\footnote{E-mail:
  \texttt{ansgar.denner@physik.uni-wuerzburg.de}},\\
Mathieu Pellen$^{1\,}$\footnote{E-mail:
  \texttt{mathieu.pellen@physik.uni-wuerzburg.de}},
Steffen Schumann$^{2\,}$\footnote{E-mail: \texttt{steffen.schumann@phys.uni-goettingen.de}},
Jennifer M. Thompson$^{3\,}$\footnote{E-mail: \texttt{thompson@thphys.uni-heidelberg.de}}
\\[9mm]
{\small\it
$^1$Universit\"at W\"urzburg, %
        Institut f\"ur Theoretische Physik und Astrophysik,} \\ %
{\small\it Emil-Hilb-Weg 22, \linebreak %
        97074 W\"urzburg, %
        Germany}\\[3mm]
{\small\it
$^2$Georg-August Universit\"at G\"ottingen, %
II. Physikalisches Institut,} \\ %
{\small\it Friedrich-Hund-Platz 1, \linebreak %
        37077 G\"ottingen, %
        Germany}\\[3mm]
{\small\it
$^3$Universit\"at Heidelberg, %
Institut f\"ur Theoretische Physik,} \\ %
{\small\it Philosophenweg 16, \linebreak %
        69120 Heidelberg, %
        Germany}\\[3mm]
}

\maketitle

\begin{abstract}
\noindent
This publication presents the combination of the one-loop matrix-element generator \Recola with the 
multipurpose Monte Carlo program \Sherpa.
Since both programs are highly automated, the resulting
\Sherpa\!\!+\Recola framework allows for the computation of---in principle---any Standard Model process at both NLO QCD and EW accuracy.
To illustrate this, three representative LHC processes have been computed at NLO QCD and EW: vector-boson production in association with jets, off-shell ${\rm Z}$-boson pair production, and
the production of a top-quark pair in association with a Higgs boson.
In addition to fixed-order computations, when considering QCD corrections, 
all functionalities of \Sherpa, \emph{i.e.} particle decays, QCD parton showers, hadronisation, underlying events, etc.\ can be used in combination with \Recola.
This is demonstrated by the 
merging and matching of one-loop QCD matrix elements for Drell--Yan
production in association with jets to the parton shower. The 
implementation is fully automatised, thus
making it a perfect tool for both experimentalists and theorists 
who want to use state-of-the-art predictions at NLO accuracy.
\end{abstract}
\thispagestyle{empty}
\vfill
\newpage
\setcounter{page}{1}

\tableofcontents
\newpage

\section{Introduction}

With Run~II, the Large Hadron Collider (LHC) has fully entered the 
precision era. An unprecedented amount of data is being collected, 
which allows the experimental collaborations to perform very 
precise measurements. In order to match this precision on the theory side, 
appropriate and accurate predictions need to be available. To that 
end, Monte Carlo programs are the perfect tool to 
provide a bridge between theoretical predictions and experimental 
measurements. The implementation of state-of-the-art theoretical 
calculations in public Monte Carlo programs is, therefore, of prime 
importance for the analysis and interpretation of the wealth of 
LHC data.

In recent years, there has been enormous progresses towards the 
automated computation of next-to-leading-order (NLO) QCD and 
electroweak (EW) corrections for Standard Model  
processes. In particular, several one-loop matrix-element generators 
have been developed~\cite{Hahn:1998yk,Berger:2008sj,Cascioli:2011va,Bevilacqua:2011xh,Hirschi:2011pa,Badger:2012pg,Cullen:2014yla,Actis:2016mpe}
and used in various multipurpose Monte Carlo programs~\cite{Cafarella:2007pc,Kilian:2007gr,Bahr:2008pv,Gleisberg:2008ta,Alwall:2014hca}.

This article describes and validates the interface between the one-loop 
matrix-element generator \Recola~\cite{Actis:2016mpe,Actis:2012qn} and
the Monte Carlo generator \Sherpa~\cite{Gleisberg:2008ta,Gleisberg:2003xi},
which allows NLO QCD and EW corrections to arbitrary Standard Model 
processes to be computed. The calculational scheme relies on
  the use of the Catani--Seymour dipole subtraction method for dealing with
  QCD \cite{Catani:1996vz,Catani:2002hc} and QED \cite{Dittmaier:1999mb,Dittmaier:2008md}
infrared singularities. Furthermore, this framework can be employed
to calculate loop-induced processes.  The \Sherpa program is a
state-of-the-art multipurpose event generator that allows for a complete 
description of LHC processes, from calculating the hard matrix element 
to modelling the hadronisation. On the other side, \Recola has been 
demonstrated to be a reliable and efficient one-loop matrix-element generator 
for several non-trivial
processes~\cite{Denner:2014wka,Denner:2015yca,Denner:2014ina,Biedermann:2016guo,Biedermann:2016yvs,Denner:2016jyo,Biedermann:2016yds,Biedermann:2016lvg,Denner:2016wet} at both NLO QCD and EW.

To illustrate the possibilities offered by the combination of these tools, 
three phenomenologically significant LHC processes have been computed at both 
NLO QCD and NLO EW accuracy in a fully automatic way.
These comprise the off-shell production of a vector boson in association with 
jets, the production of two off-shell ${\rm Z}$ bosons, and the on-shell production of 
a top-quark pair in association with a Higgs boson. Because these processes 
have been evaluated recently by other groups at both NLO QCD and EW, they 
provide crucial benchmarks for the assessment of the current implementation.
In addition to fixed-order computations, all functionalities of \Sherpa 
(parton shower, hadronisation etc.) can be used in association with 
\Recola. This is demonstrated by the merging and matching of NLO QCD
matrix elements for Drell--Yan production in association with multiple jets 
to the QCD parton shower. These 
predictions are then compared to experimental data.

Both the \Sherpa\footnote{\Sherpa is publicly available at \url{https://sherpa.hepforge.org}.} and \Recola\footnote{\Recola is publicly available at \url{https://recola.hepforge.org}.} 
programs can be downloaded and readily used to compute NLO QCD corrections.
The \Sherpa implementation of NLO EW corrections used in this
  work and as well as in Refs.~\cite{Kallweit:2014xda,Kallweit:2015dum,Kallweit:2017khh,Chiesa:2017gqx}
  will soon be publically released.
As the implementation is fully automatised and publicly available, 
\Sherpa\!\!+\Recola is an ideal tool for both experimentalists 
and theorists who want to have state-of-the-art predictions at NLO accuracy.

This article is organised as follows: in Section~\ref{sec:implementation}, a 
brief review of the methods used in \Sherpa and \Recola is provided, as well as a 
description of the interface. Section~\ref{sec:QCDvalidation} 
describes the tests performed to fully validate the interface for NLO QCD 
calculations based on several challenging physics cases involving massive coloured particles and
high final-state multiplicity across a large phase space. These
tests comprise comparisons of squared matrix elements for individual 
phase-space points, fixed-order cross-section computations, and the 
matching and merging to the parton shower.
Section~\ref{sec:processes} then begins with the validation of the NLO EW 
computations of these benchmark processes before presenting combined 
predictions for NLO QCD and EW corrections to all of the
process classes considered. After the conclusions in
Section~\ref{sec:conclusion}, details on the installation procedures and
run-card commands are given in the Appendix. 

\section{Details of the implementation}
\label{sec:implementation}

This section presents some details of the techniques and algorithms used in both
\Sherpa\!\! and \Recola, as well as information on the interface between them.
For a detailed description of the methods employed by these programs,
the reader is invited to explore the references given below.

\subsection{The \Sherpa framework}

\Sherpa~\cite{Gleisberg:2008ta,Gleisberg:2003xi} is a multipurpose event generator, which
aims to simulate the entirety of exclusive scattering events in high-energy particle 
collisions. In order to achieve this, \Sherpa contains several modules and
algorithms to cope with the many different physics challenges of collider physics.
These include methods for the generation and integration of hard-scattering 
matrix elements, QCD parton-shower simulations, and the modelling of the parton-to-hadron 
fragmentation process and the underlying-event activity. 

Leading-order matrix elements are provided by the built-in generators \Amegic~\cite{Krauss:2001iv} 
and \Comix~\cite{Gleisberg:2008fv}. For virtual matrix elements contributing at one-loop order, \Sherpa
relies on interfaces to dedicated tools. In particular, for the evaluation of QCD corrections, 
\Sherpa has interfaces to \BlackHat~\cite{Berger:2008sj}, \Gosam~\cite{Cullen:2014yla}, 
\Njet~\cite{Badger:2012pg}, \OpenLoops~\cite{Cascioli:2011va} as well
as the BLHA interface~\cite{Binoth:2010xt}.
Infrared divergences appearing in the QCD virtual and real-emission amplitudes are treated by the 
Catani--Seymour dipole-subtraction method~\cite{Catani:1996vz,Catani:2002hc} that has been automated 
in the \Sherpa framework~\cite{Gleisberg:2007md}. The default parton-shower algorithm in \Sherpa is 
based on Catani--Seymour factorisation \cite{Schumann:2007mg,Hoeche:2009xc}. 

\Sherpa can also correctly combine the matrix elements for multijet-production processes at both 
LO and NLO QCD with the parton shower. There are several techniques available on the market for
this task. The method employed in \Sherpa
for the merging of varying parton-multiplicity tree-level processes, called MEPS@LO, is based on shower 
truncation and is presented in Ref.~\cite{Hoeche:2009rj}. Its generalisation to NLO QCD matrix elements, 
dubbed MEPS@NLO, is discussed in Ref.~\cite{Hoeche:2012yf}. The latter is based on the MC@NLO-style 
matching of NLO QCD matrix elements to the Catani--Seymour dipole shower~\cite{Hoeche:2011fd}. In
Ref.~\cite{Krauss:2016orf} the inclusion of finite-mass effects in particular for bottom-quark-initiated processes has been discussed. To efficiently provide 
matrix-element and parton-shower computations with
theoretical uncertainties, usually estimated by renormalisation and factorisation
scale variations, or modified input parameters such as the strong coupling $\alphas$ or the
parton-density functions (PDFs), an event-wise reweighting approach has recently been implemented~\cite{Bothmann:2016nao}.

The present article contributes to the recent efforts to extend the \Sherpa framework to the automated 
computation of NLO EW corrections, thus further improving the perturbative accuracy of the predictions. To 
this end, based on the results presented in Refs.~\cite{Dittmaier:1999mb,Dittmaier:2008md}, the implementation of the
Catani--Seymour dipole-subtraction formalism has been extended to account for QED
corrections~\cite{Kallweit:2014xda,Schonherr:2017}. 
Furthermore, the interfaces to the available one-loop amplitude generators had to be generalised to account
for the variable orders in the strong and EW coupling parameters. In Refs.~\cite{Kallweit:2014xda,Kallweit:2015dum}
this has been presented for the \OpenLoops generator and applied to the production of on- and off-shell
gauge-boson production in association with jets. Here, the corresponding developments for 
the \Recola program are presented, along with the validation of the NLO QCD and EW corrections for a variety of
phenomenologically important processes.  

\subsection{The matrix-element generator \Recola}

\Recola \cite{Actis:2016mpe,Actis:2012qn} is a public matrix-element generator that 
is able to compute all tree and one-loop contributions to matrix elements squared in the Standard Model.
For a detailed description of the code, the interested reader is referred to the manual of the program 
\cite{Actis:2016mpe} as only the features relevant for the interface are discussed here.
All amplitude computations are performed in the 't Hooft--Feynman gauge, and
ultraviolet (UV) and infrared (IR) divergences are treated by dimensional regularisation.
All tree and one-loop matrix elements squared are summed over spin and colour.
In \Recola everything is generated 
dynamically, and no process-specific libraries are needed to compute arbitrary processes at NLO QCD and EW accuracy in the Standard 
Model.
Several schemes for the renormalisation of the electromagnetic coupling are available and are briefly discussed below.
Furthermore, \Recola allows for a flexible and generic treatment of
the flavour scheme used for the running of the strong coupling 
constant as detailed in the following.
An important feature of the code is its general implementation of the complex-mass scheme \cite{Denner:1999gp,Denner:2005fg,Denner:2006ic}, 
which allows for the consistent computation of processes featuring resonant massive particles.

The computation of both the tree and one-loop amplitudes is performed in a recursive fashion~\cite{Actis:2012qn}.
For the tree-level amplitudes, the algorithm employed is inspired by the Dyson--Schwinger equation.
At one-loop level, the amplitude can be written as the sum of tensor integrals $T^{\hat{\mu}_1\cdots\hat{\mu}_{r_t}}_{(t)}$ multiplied with tensor coefficients $c^{(t)}_{\hat{\mu}_1\cdots\hat{\mu}_{r_t}}$,
\begin{equation}
 \mathcal{A}_1 = \sum_t c^{(t)}_{\hat{\mu}_1\cdots\hat{\mu}_{r_t}} T^{\hat{\mu}_1\cdots\hat{\mu}_{r_t}}_{(t)} + \mathcal{A}_{\mathrm{CT}} + \mathcal{A}_{\mathrm{R2}},
\end{equation}
where $\mathcal{A}_{\mathrm{CT}}$ and $\mathcal{A}_{\mathrm{R2}}$ 
denote the counter terms and the rational terms \cite{Ossola:2008xq}, respectively.
The former cancel the UV divergences present in the tensor integrals,
the latter provide additional finite terms originating from
dimensional regularisation.  
The core algorithm \cite{Actis:2012qn} used for the recursive computation of the tensor coefficients numerically is based on an idea of van Hameren~\cite{vanHameren:2009vq}.
To numerically evaluate the one-loop scalar \cite{'tHooft:1978xw,Beenakker:1988jr,Dittmaier:2003bc,Denner:2010tr} and tensor integrals \cite{Passarino:1978jh,Denner:2002ii,Denner:2005nn}, \Recola relies on the \collier library \cite{Denner:2014gla,Denner:2016kdg}.
Finally, we note that the implementation of the interface is valid for \Recola-1.2 and subsequent releases.

\subsection{The interface}
\label{sec:interface}

The purpose of the interface is to provide arbitrary Standard Model 
one-loop matrix elements, generated with \Recola, to \Sherpa.
This section outlines the choices made for the implementation of \Recola in 
\Sherpa in order to compute the virtual piece of NLO QCD and EW corrections.
The other tools and ingredients needed for NLO computations, such as 
tree-level matrix elements or the subtraction method, are not addressed here.
Indeed, the tree-level, integrated subtraction term and the real-subtracted parts are all computed by \Sherpa without the help of \Recola.
The details of these implementations can be found in the related publications cited above.

In \Recola, the generation of the matrix elements is done on the fly, and once \Recola is installed and linked to \Sherpa, any processes can be computed 
readily.\footnote{The installation procedures are described in Appendix~\ref{sec:installation}.}
In the initialisation phase of a \Sherpa integration run, all necessary partonic one-loop processes are registered in \Recola and automatically generated as soon as the actual integration starts. No extra process-specific libraries are needed.
The combination \Sherpa\!\!+\Recola allows thus to compute---in
principle---any tree-level-induced Standard Model process at NLO QCD
and EW accuracy and any one-loop-induced process at LO.

As for any computation in \Sherpa, the processes, event selection, parameters,
etc.\ are defined in a run card.
The masses and widths issued in the run card are by default the on-shell 
masses, apart from the ${\rm Z}$ and ${\rm W}$ masses which are assumed to be the pole masses.
Switching to on-shell masses also for the latter is possible through specific 
keywords (see Appendix \ref{sec:commands}).

For the renormalisation of the electromagnetic coupling constant, three schemes are 
available \cite{Actis:2016mpe}.
In the $G_\mu$ scheme, the electromagnetic coupling $\alpha$ 
is derived from the Fermi constant $G_\mu$. 
In the $\alpha(0)$ scheme, $\alpha$ is fixed from the value measured 
in Thomson scattering at $p^2 = 0$, while in the $\alpha(M_{\rm Z})$ 
scheme, $\alpha$ is renormalised at the ${\rm Z}$ pole and thus implicitly takes into 
account the running from $p^2 = 0$ to $p^2 = M^2_{\rm Z}$.
Due to the ambiguity in defining a real value of $\alpha$ in the complex-mass
 scheme, the numerical value of $\alpha$ computed by \Sherpa is taken as an 
input by \Recola to ensure compatibility.

Concerning the strong coupling, $\alphas$, the computation of the 
corresponding counter term is done consistently according to the PDF
set used.
The strong coupling constant is extracted from the PDF set as well as the 
flavour scheme and the quark masses used for the PDF evolution.
These parameters are then used for the computation of the strong-coupling 
counter term $\delta Z_{g_\mathrm{s}}$ which reads \cite{Actis:2016mpe}
\begin{equation}
\label{eq:gsct}
    \delta Z_{g_\mathrm{s}} = - \frac{\alphas \left( Q^2 \right)}{4
      \pi} \left[ \left( \frac{11}{2} - \frac{N_{\rm l}}{3} \right) \left( \Delta_{\rm UV} + \ln \frac{\mu^2_{\rm UV}}{Q^2}   \right) - \frac13 \sum_q \left(\Delta_{\rm UV} + \ln \frac{\mu^2_{\rm UV}}{m_q^2} \right) \right] ,
\end{equation}
where $N_l$ is the number of light quark flavours and the index $q$ runs over the heavy flavours.
The parameter $\Delta_{\rm UV}$ contains the poles in $D-4$ as described in Ref.~\cite{Actis:2016mpe}.
In variable-flavour schemes, all quarks lighter than the scale $Q$ are considered light while the remaining ones are treated as heavy.
In fixed $N_f$-flavour schemes, the $N_f$ lightest quarks are considered light while the others are treated as heavy.
As emphasised above, the flavour schemes and quarks masses used to compute $\delta Z_{g_s}$ are set consistently in \Recola according to the PDF set.
Nevertheless, specific commands described in Appendix \ref{sec:commands} allow the user to choose all possible flavour schemes in combination with arbitrary quark masses.
Finally, note that the quark masses used to compute $\delta Z_{g_s}$ can, in principle, differ from the ones used in the rest of the matrix element.
Ensuring consistent mass values between the run card and the PDF set used is left to the user.

Finally, the IR and UV renormalisation scales are fixed to $100\GeV$
by default in the \Sherpa\!\!+\Recola interface.
Thus in general, the virtual part calculated with 
\Sherpa\!\!+\Recola cannot be directly compared to the virtual part from a
\Sherpa\!\!+\OpenLoops \cite{Kallweit:2014xda} calculation.
However, the sum of the virtual and integrated subtraction-term contributions 
is independent of the regulators. It is also possible to set the IR scale in \Recola equal to a fixed renormalisation scale, 
in which case a direct comparison of the virtual corrections of \OpenLoops and \Recola is possible (see Appendix \ref{sec:commands}).

\section{NLO QCD validation}
\label{sec:QCDvalidation}

This section is devoted to the validation of the \Sherpa\!\!+\Recola interface for NLO QCD computations.
This is accomplished by direct comparisons to results obtained from public\footnote{Even if not explicitly reported in this article, several NLO QCD checks have been performed against the code \madgraph~\cite{Alwall:2014hca}.}
and private codes as well as published work at three different levels.
First, for a broad range of processes, squared matrix elements for individual phase-space points are compared with \Sherpa\!\!+\OpenLoops. 
Next, a comparison of cross sections and differential distributions with NLO QCD accuracy is presented where
the results are compared with those obtained from public codes, private codes, and the literature.
Finally, to illustrate the applicability of the \Sherpa\!\!+\Recola framework for NLO QCD calculations matched to parton-shower simulations,
results for Drell--Yan lepton-pair production in association with jets at MEPS@NLO QCD accuracy are presented.
Furthermore, information on the memory
consumption and run times are provided.

\subsection{Phase-space point comparison}
\label{sec:phase_space_points}

As a first validation, the sum of the virtual and integrated dipole part of the NLO QCD corrections to a 
wide variety of processes, calculated with both \Sherpa\!\!+\OpenLoops and \Sherpa\!\!+\Recola, are compared at the level of individual phase-space points.
Using identical set-ups, this provides a stringent test of the
implementation of the two one-loop generators in \Sherpa.
 \Sherpa's Python interface~\cite{Hoche:2014kca} has been extended for this purpose.
For each process, \ie individual partonic channel, $1000$ randomly chosen phase-space points are considered.
These correspond to the parton momenta in proton--proton collisions with a hadronic centre-of-mass energy of $13\TeV$.
Both for the factorisation and renormalisation scales, we choose
$\mu_{\rm R}=\mu_{\rm F}=\sqrt{\hat s}$. We employ the NNPDF-3.0 NNLO set~\cite{Ball:2014uwa, Ball:2013hta}, featuring $\alphas(M_{\rm Z})=0.118$
with a variable number of active flavours up to $N_\text{F}=5$ and two-loop running of $\alphas$. Electroweak input parameters
are defined in the $G_\mu$ scheme. For all considered
phase-space points all final-state particles have to pass the following set of cuts:
\begin{align}
&p_{\rm T} > 25 \GeV, \qquad M_{\ell^+ \ell^-} > 60 \GeV, \notag\\
&\Delta R_{ij} > 0.4, \qquad 
\Delta R_{i\gamma} > 0.2, \qquad
\Delta R_{\gamma\gamma} > 0.2, 
\end{align}
where $i,j$ are any particles apart from a photon.
These cuts regulate any potential soft or collinear singularities at the Born level.
For on-shell massive particles, \ie ${\rm Z}$, ${\rm W}$, Higgs bosons and top quarks, vanishing widths are assumed.
However, for intermediate resonances finite-width effects are included, using the complex-mass scheme.
We have used \OpenLoops version 1.3.1 in the default configuration with \CutTools~\cite{Ossola:2007ax} (version 1.9.5) and the \OneLOop library~\cite{vanHameren:2010cp} (version 3.6.1) in this comparison.

In \refta{tab:pspt_validation} the logarithmically averaged relative deviation, $\Delta_{\text{VI}}$, of the sum of the virtual corrections and integrated dipoles
between \Sherpa\!\!+\Recola and \Sherpa\!\!+\OpenLoops is presented for 62 partonic processes.
The logarithmically averaged relative deviation $\Delta_{\text{VI}}$ is defined as
\begin{equation}
\label{eq:logavr}
 \log_{10}\Delta_{\text{VI}} = \frac{1}{N_p} \sum_{i=1}^{N_p} \log_{10}{\left| \frac{{\rm d}\sigma^i_{\text{VI, \OpenLoops}} - {\rm d}\sigma^i_{\text{VI, \Recola}}}{{\rm d}\sigma^i_{\text{VI, \OpenLoops}}} \right|},
\end{equation}
where ${\rm d}\sigma^i_{\text{VI}}$ is the sum of virtual and
integrated-dipole contributions at the phase-space point $i$, and $N_p$ is the number of phase-space points.
In \refta{tab:loop-induced_pspt_validation} the squared
one-loop amplitudes for 13 loop-induced processes are compared.
The logarithmically averaged relative deviation $\Delta_{\text{LI}}$ is defined similarly as in Eq.~\refeq{eq:logavr} with ${\rm d}\sigma^i_{\text{VI}}$ replaced by the loop-induced cross section ${\rm d}\sigma^i_{\text{LI}}$.

\begin{table}[p] 
 \begin{minipage}{.5\linewidth}
 \centering
 \begin{tabular}{|l|c|}
 \hline
  process &  $\Delta_{\text{VI}}$ \\
 \hline
$ \Pd \bar \Pd \rightarrow \Pe^+ \Pe^-      $    \rule{0ex}{2.4ex}      & $1.338\cdot 10^{-12}$ \\          
$ \Pd \Pg \rightarrow \Pe^+ \Pe^- \Pd       $     & $5.664\cdot 10^{-12}$ \\            
$ \bar \Pu \Pg \rightarrow \Pe^+ \Pe^- \bar \Pu   $       & $5.676\cdot 10^{-12}$ \\        
$ \Pd \bar \Pd \rightarrow \Pe^+ \Pe^- \Pg    $         & $3.260\cdot 10^{-12}$ \\         
$ \Pd \bar \Pd \rightarrow \Pe^+ \Pe^- \Pd \bar \Pd $       & $2.861\cdot 10^{-12}$ \\        
$ \Pg \Pg \rightarrow \Pe^+ \Pe^- \Pu \bar \Pu $           & $1.735\cdot 10^{-09}$ \\        
$ \Pd \Pg \rightarrow \Pe^+ \Pe^- \Pg \Pd   $       & $1.719\cdot 10^{-09}$ \\               
$ \Pd \bar \Pd \rightarrow \Pe^+ \Pe^- \Pg \Pg  $         & $4.518\cdot 10^{-09}$ \\        
$ \Pu \bar \Pd \rightarrow \Pe^+ \nu_{\Pe}        $       & $1.158\cdot 10^{-12}$ \\       
$ \Ps \bar \Pc \rightarrow \mu^- \bar \nu_{\mu}    $      & $3.723\cdot 10^{-12}$ \\ 
$ \Pu \bar \Pd \rightarrow \Pe^+ \nu_{\Pe} \Pg        $          & $7.207\cdot 10^{-12}$ \\ 
$ \Ps \bar \Pc \rightarrow \mu^- \bar \nu_{\mu} \Pg  $      & $7.327\cdot 10^{-12}$ \\ 
$ \Pu \bar \Pd \rightarrow \Pe^+ \nu_{\Pe} \Ps \bar \Ps  $        & $5.971\cdot 10^{-12}$ \\  
$ \Ps \bar \Pc \rightarrow \mu^- \bar \nu_{\mu} \Pg \Pg $      & $3.068\cdot 10^{-11}$ \\ 
\hline                                                                 
$ \Pd \bar \Pd \rightarrow \Pe^+ \Pe^- \mu^+ \mu^-$ \rule{0ex}{2.4ex} & $7.652\cdot 10^{-11}$ \\     
\hline                                                                 
$ \Pd \bar \Pd \rightarrow \Pd \bar \Pd     $ \rule{0ex}{2.4ex} & $9.650\cdot 10^{-13}$ \\   
$ \Pu \bar \Pu \rightarrow \Pu \bar \Pu     $           & $9.413\cdot 10^{-13}$ \\       
$ \Pu \bar \Pu \rightarrow \Ps \bar \Ps     $     & $2.115\cdot 10^{-12}$ \\             
$ \Pu \bar \Pu \rightarrow \Pg \Pg          $     & $2.410\cdot 10^{-12}$ \\             
$ \Pd \bar \Pd \rightarrow \Pg \Pd \bar \Pd   $     & $6.223\cdot 10^{-12}$ \\             
$ \Pd \bar \Pd \rightarrow \Pg \Pg \Pd \bar \Pd $     & $2.857\cdot 10^{-09}$ \\             
$ \Pu \bar \Pu \rightarrow \Pg \Pg \Pu \bar \Pu $     & $3.751\cdot 10^{-09}$ \\             
\hline                                                                 
$ \Pd \bar \Pd \rightarrow \Pt \bar \Pt   $ \rule{0ex}{2.4ex}    & $1.861\cdot 10^{-12}$ \\             
$ \Pu \bar \Pu \rightarrow \Pt \bar \Pt \Pg     $   & $1.854\cdot 10^{-12}$ \\             
$ \Pg \Pg \rightarrow \Pt \bar \Pt      $    & $1.682\cdot 10^{-12}$ \\             
$ \Pg \Pg \rightarrow \Pt \bar \Pt \Pg   $    & $5.363\cdot 10^{-09}$ \\             
\hline                                                                 
$ \Pd \bar \Pd \rightarrow \gamma \Pg     $   \rule{0ex}{2.4ex}    & $1.784\cdot 10^{-12}$ \\             
$ \Pd \bar \Pd \rightarrow \gamma \Pd \bar \Pd $         & $3.455\cdot 10^{-12}$ \\      
$ \Pu \bar \Pu \rightarrow \gamma \gamma $            & $2.577\cdot 10^{-12}$ \\     
$ \Pu \bar \Pu \rightarrow \gamma \gamma \Pg $         & $1.845\cdot 10^{-07}$ \\     
$ \Pd \bar \Pd \rightarrow \gamma \gamma \gamma $         & $4.193\cdot 10^{-11}$ \\
\hline
 \end{tabular}
 \end{minipage}%
 \ 
 \begin{minipage}{.5\linewidth}
  \centering
\begin{tabular}{|l|c|}
 \hline
 process &  $\Delta_{\text{VI}}$ \\
 \hline
 $ \Pd \bar \Pd \rightarrow \Pe^+ \Pe^- \PH \Pd \bar \Pd $ \rule{0ex}{2.4ex}  & $3.745\cdot 10^{-11}$ \\          
 $ \Pu \bar \Pd \rightarrow \Pe^+ \nu_{\Pe} \PH \Pg  $ & $9.927\cdot 10^{-12}$ \\              
 $ \Pu \bar \Pu \rightarrow \PH \PZ        $    & $1.415\cdot 10^{-12}$ \\               
 $ \Pu \bar \Pu \rightarrow \PH \PZ \Pg      $   & $1.147\cdot 10^{-10}$ \\              
 $ \Pd \bar \Pd \rightarrow \PH \PZ \Pg \Pg    $         & $7.334\cdot 10^{-07}$ \\              
 $ \Pc \bar \Ps \rightarrow \PH \PW^+      $       & $1.173\cdot 10^{-12}$ \\            
 $ \Pu \bar \Pd \rightarrow \PH \PW^+ \Pd \bar \Pd    $ & $3.577\cdot 10^{-11}$ \\           
 \hline                                                                
 $ \Pd \bar \Pd \rightarrow \PZ \Pg $  \rule{0ex}{2.4ex}   & $1.787\cdot 10^{-12}$ \\           
 $ \Pu \bar \Pu \rightarrow \PZ \Pg \Pg        $           & $7.399\cdot 10^{-09}$ \\           
 $ \Pu \bar \Pu \rightarrow \PZ \Pd \bar \Pd      $   & $3.285\cdot 10^{-12}$ \\           
 $ \Pu \bar \Pu \rightarrow \PZ \gamma     $      & $2.795\cdot 10^{-12}$ \\           
 $ \Pu \bar \Pu \rightarrow \PZ \gamma \gamma  $  & $2.231\cdot 10^{-11}$ \\          
 $ \Pu \bar \Pd \rightarrow \PW^+ \Pg        $             & $1.903\cdot 10^{-12}$ \\    
 $ \Pu \bar \Pd \rightarrow \PW^+ \Pg \Pg       $           & $3.203\cdot 10^{-11}$ \\           
 $ \Pu \bar \Pd \rightarrow \PW^+ \gamma \Pg $      & $1.287\cdot 10^{-11}$ \\           
 $ \Ps \bar \Pc \rightarrow \PW^- \gamma \gamma $  & $4.306\cdot 10^{-11}$ \\           
 \hline                                                                
 $ \Pu \bar \Pu \rightarrow \PZ \PZ          $ \rule{0ex}{2.4ex}  & $3.127\cdot 10^{-12}$ \\                
 $ \Pu \bar \Pu \rightarrow \PZ \PZ \Pg        $         & $2.244\cdot 10^{-08}$ \\             
 $ \Pd \bar \Pd \rightarrow \PZ \PZ \Pg \Pg      $       & $3.256\cdot 10^{-07}$ \\              
 $ \Pd \bar \Pd \rightarrow \PZ \PZ \gamma   $ & $2.423\cdot 10^{-11}$ \\               
 $ \Pu \bar \Pu \rightarrow \PW^+ \PW^-      $ & $2.499\cdot 10^{-11}$ \\                
 $ \Pu \bar \Pu \rightarrow \PW^+ \PW^- \Pg    $ & $2.473\cdot 10^{-08}$ \\                
 $ \Pd \bar \Pd \rightarrow \PW^+ \PW^- \Pg \Pg  $ & $7.163\cdot 10^{-07}$ \\                
 $ \Ps \bar \Ps \rightarrow \PW^+ \PW^- \gamma   $  & $7.727\cdot 10^{-11}$ \\           
 $ \Pu \bar \Pd \rightarrow \PZ \PW^+ \Pg      $    & $6.724\cdot 10^{-11}$ \\            
 $ \Pu \bar \Pd \rightarrow \PZ \PW^+ \gamma   $    & $1.431\cdot 10^{-10}$ \\          
 $ \Pd \bar \Pu \rightarrow \PZ \PW^- \gamma     $  & $1.570\cdot 10^{-10}$ \\           
 \hline                                                                
 $ \Pd \bar \Pd \rightarrow \PZ \PZ \PZ      $  \rule{0ex}{2.4ex}  & $1.689\cdot 10^{-10}$ \\                
 $ \Pu \bar \Pd \rightarrow \PZ \PZ \PW^+      $ & $2.622\cdot 10^{-10}$ \\               
 $ \Pu \bar \Pu \rightarrow \PZ \PW^+ \PW^-    $ & $1.458\cdot 10^{-10}$ \\                
 $ \Pu \bar \Pd \rightarrow \PW^+ \PW^+ \PW^- $   & $1.150\cdot 10^{-10}$ \\                
 \hline
 \end{tabular}
\end{minipage}
 \caption{Average relative deviations  $\Delta_{\text{VI}}$ between \Sherpa\!\!+\Recola and \Sherpa\!\!+\OpenLoops
   for the sum of the NLO QCD virtual and integrated-dipole contributions evaluated for 1000 phase-space points. 
   The average relative deviations are calculated by averaging the
   logarithms of the relative deviations between the two results for the considered phase-space points.}
 \label{tab:pspt_validation}
\end{table}

\begin{table}[hptb]
 \centering
 \begin{tabular}{|l|c|}
 \hline
  process & $\Delta_{\text{LI}}$ \\
 \hline
$ \Pg \Pg \rightarrow \Pe^+ \Pe^- \gamma     $ \rule{0ex}{2.4ex}  & $1.825\cdot 10^{-07}$ \\ 
$ \Pg \Pg \rightarrow \Pe^+ \Pe^- \mu^+ \mu^- \Pg $ & $4.570\cdot 10^{-06}$ \\
$ \Pg \Pg \rightarrow \mu^+ \mu^- \mu^+ \mu^- $ & $4.853\cdot 10^{-07}$ \\
$ \Pg \Pg \rightarrow \nu_{\Pe} \bar \nu_{\Pe} \gamma \Pg  $ & $4.960\cdot 10^{-06}$ \\
$ \Pg \Pg \rightarrow \gamma \Pg    $ & $2.169\cdot 10^{-07}$ \\
$ \Pg \Pg \rightarrow \gamma \Pg \Pg $ & $1.145\cdot 10^{-06}$ \\ 
$ \Pg \Pg \rightarrow \gamma \gamma   $  & $1.522\cdot 10^{-07}$ \\
$ \Pg \Pg \rightarrow \PH \PZ $  & $3.541\cdot 10^{-10}$ \\
$ \Pg \Pg \rightarrow \PH \PZ \Pg $  & $3.013\cdot 10^{-07}$ \\
$ \Pg \Pg \rightarrow \PH \PH      $  & $4.023\cdot 10^{-10}$ \\
$ \Pg \Pg \rightarrow \PH \PH \Pg     $ & $1.420\cdot 10^{-09}$ \\
$ \Pg \Pg \rightarrow \PZ \PZ $  & $5.576\cdot 10^{-08}$ \\
$ \Pg \Pg \rightarrow \PW^+ \PW^-          $ & $6.350\cdot 10^{-08}$ \\
\hline
 \end{tabular}
 \caption{Average relative deviations $\Delta_{\text{LI}}$ between \Sherpa\!\!+\Recola and \Sherpa\!\!+\OpenLoops
   for the matrix element squared evaluated for 1000 phase-space points for QCD loop-induced processes. 
   The average relative deviations are calculated by averaging the
   logarithms of the relative deviations between the two results for the considered phase-space points.}
 \label{tab:loop-induced_pspt_validation}
\end{table}

For all processes considered, good agreement is found between the results of \Sherpa\!\!+\Recola 
and those of the public \Sherpa\!\!+\OpenLoops.
For most processes the average relative
deviation lies between $\mathcal{O}(10^{-9})$ and
$\mathcal{O}(10^{-12})$, corresponding to an agreement to 9--12 digits
on average.

As one can expect, the agreement decreases for processes with higher final-state particle
multiplicity as well as for the loop-induced processes. This
originates from the increase in complexity with the number of
external particles and the fact that one-loop amplitudes appear
squared in loop-induced processes.
In addition, the presence of external gauge bosons, in particular gluons, worsens the agreement due to the additional spin and colour degrees of freedom.

In addition to the one-loop results presented here, the squared tree-level matrix
elements of \Recola have further been compared against the ones provided by \Sherpa, through the matrix-element generators
\Amegic~\cite{Krauss:2001iv} and \Comix~\cite{Gleisberg:2008fv}. 
For all the processes listed in \refta{tab:pspt_validation} the
logarithmically averaged relative differences per phase-space point are well below $\mathcal{O}(10^{-11})$. 
The number of phase-space points with differences above $\mathcal{O}(10^{-8})$ amounts to at most a few per cent in the worst channels and for the vast majority of processes all 1000 phase-space points have differences below $\mathcal{O}(10^{-8})$.

Regarding the time spend on the evaluation of the one-loop matrix elements, the performances of the
generators are overall very similar. 
We have compared \Recola against \OpenLoops in the default configuration with \CutTools, as well as against \OpenLoops in combination with the tensor reduction of the \collier library (albeit using version 1.0 instead of version 1.1 which is used by \Recola).
While \OpenLoops with \CutTools is slightly slower, especially for the loop-induced processes, the performance of both generators with the \collier library is comparable.
\Recola features a longer initialisation time as it does not rely on pre-compiled process libraries
as \OpenLoops does. However, this initialisation phase becomes negligible compared to the overall run time in
realistic applications. Overall, no significant differences in the run times have been observed.
In Section~\ref{sec:mepsnlo}, the performance and memory usage for both one-loop providers for a full-fledged matrix-element plus parton-shower simulation
is presented.

\subsection{NLO QCD fixed-order calculations}
\label{sec:NLOQCDfixedorder}

Next, we present results
for cross sections and differential distributions at NLO QCD accuracy. 
Using the \Sherpa\!\!+\Recola framework, four process classes are
considered, namely off-shell
${\rm Z}$-boson production in association with up to two additional jets
(a full MEPS@NLO set-up of this process is presented in
Section~\ref{sec:mepsnlo}), off-shell ${\rm Z}$-boson pair production,
on-shell top-quark pair production in association with a Higgs boson,
and Higgs-boson production
in association with an on-shell ${\rm Z}$ boson.

\subsubsection{${\rm Z}$-boson production in association with jets}
\label{DYJets:sec}

\paragraph{Input parameters:} 
Proton--proton collisions at a centre-of-mass energy of $13\TeV$ are considered, and the \Rivet
analysis package~\cite{Buckley:2010ar} is used to analyse events. As before, the NNPDF-3.0 NNLO PDF set
with up to five active flavours $\alphas(M_{\PZ})=0.118$ and two-loop
running is employed. EW input parameters are defined in the $G_\mu$ scheme.

In the Drell--Yan plus jets processes, the QCD jets are reconstructed by means of the
anti-$k_\text{T}$ jet algorithm~\cite{Cacciari:2008gp,Cacciari:2011ma} with radius parameter
$R = 0.4$, $p_\text{T,j} > 25\GeV$, and $\left| \eta_\text{j} \right| < 3.5$. The ${\rm Z}$~boson 
is assumed to decay into an electron--positron pair with a dilepton
invariant mass in the range
$66\GeV < M_{\Pe^- \Pe^+} < 116\GeV$. The renormalisation and
factorisation scales are event-wise set to $M_{\Pe^- \Pe^+}$.

\paragraph{NLO QCD validation:} 
The Drell--Yan-plus-jets processes are considered at the orders $\order{\alpha^2}$, $\order{\alphas \alpha^2}$, and $\order{\alphas^2 \alpha^2}$ for the LO cross sections for 0, 1, and 2 jets, respectively. 
The corresponding NLO QCD cross sections contribute at the orders $\order{\alphas \alpha^2}, \order{\alphas^2 \alpha^2}$, and $\order{\alphas^3 \alpha^2}$.
The total cross sections calculated with \Sherpa\!\!+\Recola and
\Sherpa\!\!+\OpenLoops presented in \refta{tab:DYJets_crosssections}
turn out to be identical within the given accuracy as they have been
obtained using the very same phase-space points and \OpenLoops and
\Recola agree to more than 9 digits.
\begin{table}
\begin{center}
\begin{tabular}{|l|c|c|}
\hline\rule[-1.4ex]{0ex}{3.8ex}%
$\sigma^{\mathrm{NLO}}_{\mathrm{QCD}}$ & \Sherpa\!\!+\Recola [pb] & \Sherpa\!\!+\OpenLoops [pb] \\
\hline \rule{0ex}{2.4ex}%
$ \Pp \Pp \to \Pe^+ \Pe^- $ & $1976.0(3)$ & $1976.0(3)$ \\
$ \Pp \Pp \to \Pe^+ \Pe^- \Pj $ & $414.3(4)$ & $414.3(4)$ \\
$ \Pp \Pp \to \Pe^+ \Pe^- \Pj \Pj $ & $123.5(5)$ & $123.5(5)$ \\
\hline
\end{tabular}
\end{center}
\caption{\label{tab:DYJets_crosssections} Integrated cross sections calculated for ${\rm pp}\to {\rm e}^+{\rm e}^- + \text{jets}$ at NLO QCD with the \Sherpa\!\!+\Recola interface compared against \Sherpa\!\!+\OpenLoops.}
\end{table}

In Fig.~\ref{fig:ZPT}, the resulting transverse-momentum distribution for the Drell--Yan pair
is shown, considering the production processes with zero, one and two additional jets
evaluated at NLO QCD accuracy. Evidently, the low-$p_{\rm T}$ region is dominated by the pure
Drell--Yan process, \ie without additional final-state jets at the
Born level, where the finite
recoil originates solely from the real radiative corrections. 
On the other hand, the tail of the transverse-momentum distribution is dominated by Drell--Yan-plus-one-jet processes, where the ${\rm Z}$ transverse momentum results
from the recoil against the Born-level jet and the real radiation.
\bfig
  \center
  \includegraphics[width=0.5\textwidth]{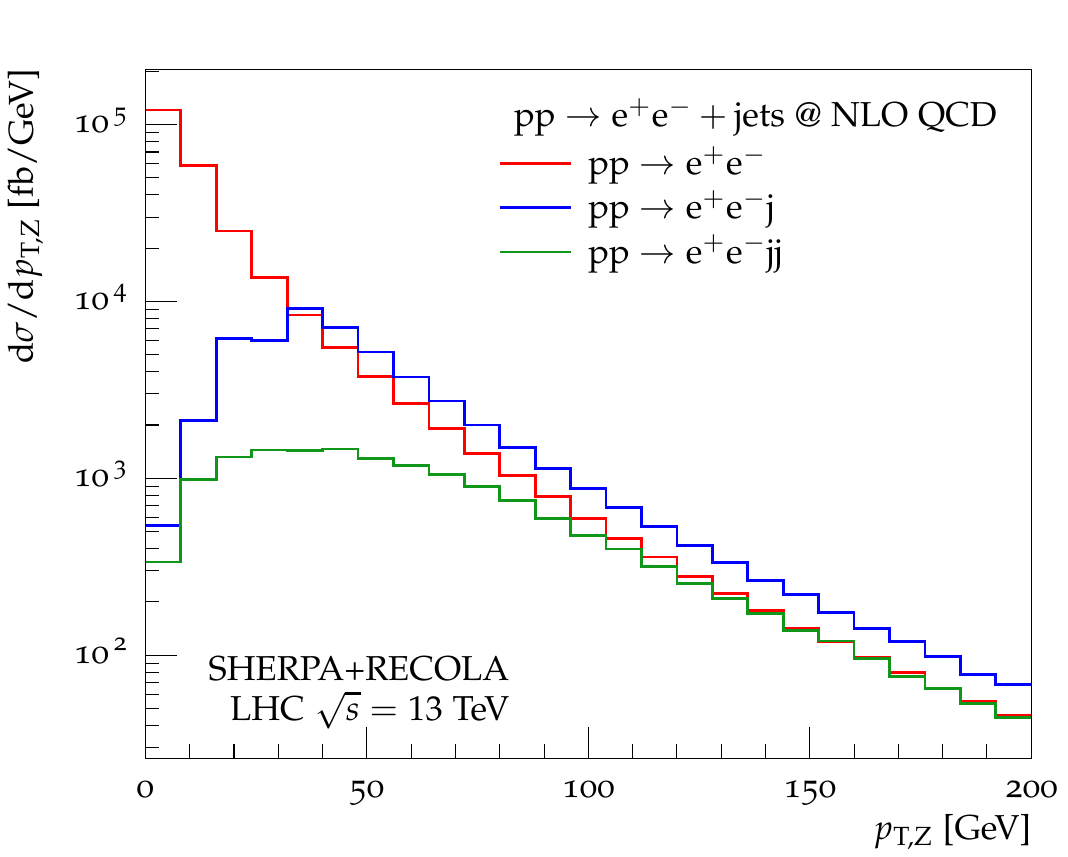}
  \caption{\label{fig:ZPT} The transverse-momentum distribution of the reconstructed
  ${\rm Z}$ boson in Drell--Yan plus zero, one and two-jet production at NLO QCD.}
\efig

Here only  the results obtained with
\Sherpa\!\!+\Recola are displayed. All results have been cross-checked against \Sherpa\!\!+\OpenLoops, using
the identical phase-space points, and no significant deviations have been observed. In fact,
for each observable bin the relative difference of the two predictions is well below
$10^{-5}$.

\subsubsection{${\rm Z}$-boson pair production}
\label{sec:ZZQCDvalidation}

\paragraph{Input parameters:}
The set-up employed is the one of Ref.~\cite{Biedermann:2016lvg} which is, for completeness, repeated in the following. 
The predictions are for the LHC operating at a centre-of-mass energy of $\sqrt{s}=13\TeV$. 
The NNPDF-2.3QED NLO PDF set \cite{Ball:2013hta} with a variable flavour-number scheme and $\alphas(M_{\PZ})=0.118$ has been used for all computations both
at LO and NLO. Furthermore, a fixed factorisation and renormalisation scale at the ${\rm Z}$-boson pole mass $\mu_{\rm R}= \mu_{\rm F} = M_{\rm Z}$ is employed.
The strong coupling $\alphas$ is extracted from the PDF set at the renormalisation scale $\mu_{\rm R}$, and
the electromagnetic coupling $\alpha$ is calculated in the $G_\mu$ scheme according to

\begin{equation}\label{eqn:FermiConstant}
  \alpha = \frac{\sqrt{2}}{\pi} G_\mu \MW^2 \left( 1 - \frac{\MW^2}{\MZ^2} \right),
  \qquad \text{with}  \qquad   \GF    = 1.16637\times 10^{-5}\GeV^{-2}
\end{equation}
denoting the Fermi constant.
The on-shell (OS) values for the masses and widths of the massive vector bosons read
\begin{alignat}{2} 
                \MZOS &=  91.1876\GeV,  \qquad    & \GZOS &= 2.4952\GeV,  \nonumber \\
                \MWOS &=  80.385\GeV,   \qquad    & \GWOS &= 2.085\GeV. 
\end{alignat}
They are converted into pole masses and pole widths according to \citere{Bardin:1988xt},
\begin{equation}
M_V = \MVOS/\sqrt{1+(\GVOS/\MVOS)^2}, \qquad  \Gamma_V = \GVOS/\sqrt{1+(\GVOS/\MVOS)^2} \qquad \text{with}  \quad V=\PW,\PZ.
\end{equation}
Since the top quark and the Higgs boson do not appear as internal resonances, their widths are set equal to zero. The corresponding masses read
\begin{alignat}{2}
  \Mt   &=  173\GeV, \qquad      & M_{\rm H} &=  125 \GeV.
\end{alignat}
The masses and widths of all other quarks and leptons have been neglected.
As in Ref.~\cite{Biedermann:2016lvg}, the following acceptance cuts are imposed on the 
charged leptons $\ell^\pm$:
 \begin{equation}
               p_{{\rm T}, \ell}>  15 \GeV,  \qquad |y_{\ell}| < 2.5,\qquad\Delta R_{\ell\ell} >  0.2.
 \end{equation}
The jets from real QCD radiation are treated fully inclusively.

\paragraph{NLO QCD validation:}
The \Sherpa\!\!+\Recola interface has been cross-checked for this process against the independent private Monte Carlo program that had been used for the computations in Refs.~\cite{Biedermann:2016guo,Biedermann:2016yvs,Biedermann:2016yds,Biedermann:2016lvg}.

\begin{table}
\begin{center}
\begin{tabular}{|l|c|c|c|}
\hline  \rule[-1.2ex]{0ex}{1.8ex}%
$\Pp\Pp\to\mu^+\mu^-\Pe^+\Pe^-$ & \Sherpa\!\!+\Recola [fb] & private MC+\Recola [fb] & std. dev. [$\sigma$] \\
\hline \rule{0ex}{2.4ex}%
$\sigma^{\rm LO}$ & $11.498(1)$ & $11.4964(1)$ & $1.6$ \\
 \rule[-1.4ex]{0ex}{1.8ex}%
$\sigma^{\rm NLO}_{\rm QCD}$ & $15.79(1)$ & $15.801(2)$ & $1.0$ \\
\hline
\end{tabular}
\end{center}
\caption{\label{tab:ZZQCDvalidation} Integrated cross sections calculated at $\sqrt{s} = 13\TeV$ for ${\rm pp}\to\mu^+\mu^-{\rm e}^+{\rm e}^-$ at LO and NLO QCD with the \Sherpa\!\!+\Recola interface, compared against the independent private multi-channel Monte Carlo program that was employed for the computations in Ref.~\cite{Biedermann:2016lvg}.
The difference is expressed in standard deviations.}
\end{table}
\bfig
  \center
  \includegraphics[width=0.45\textwidth]{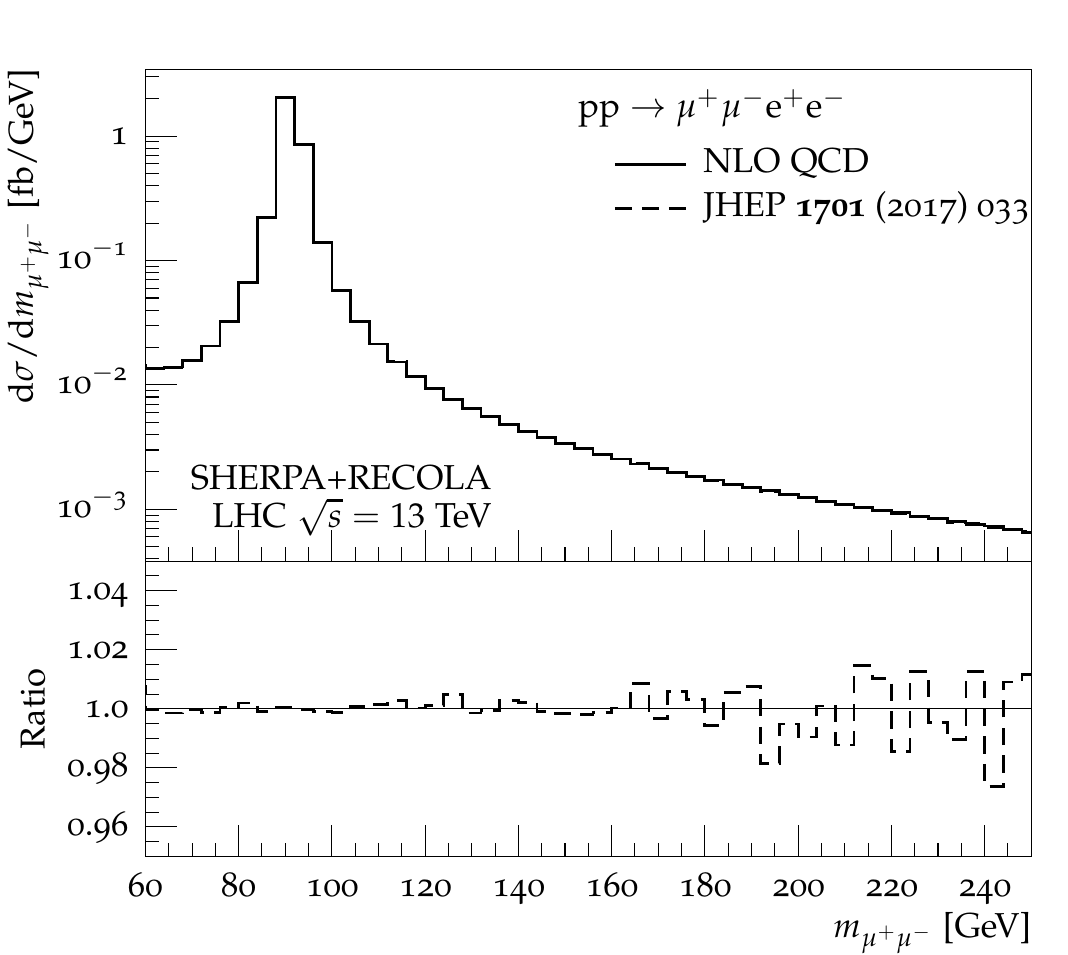}
  \includegraphics[width=0.45\textwidth]{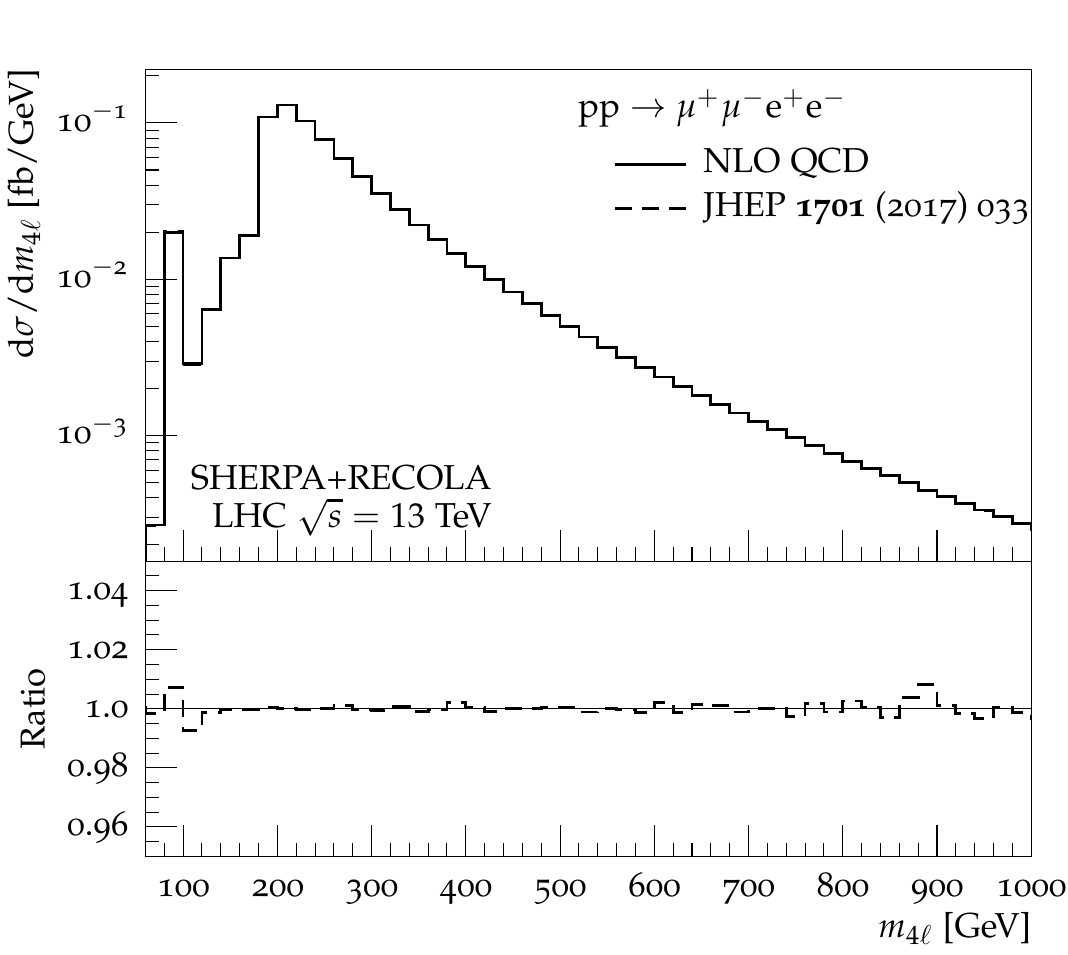}
  \caption{\label{fig:ZZQCDvalidation} NLO QCD predictions of the
    di-muon (left) and 4-lepton invariant-mass (right) distributions in
    $\Pp\Pp \to \mu^+\mu^-\Pe^+\Pe^-$ at $\sqrt{s}=13\TeV$. The plots
    show a
    direct comparison of \Sherpa\!\!+\Recola to the Monte Carlo program employed in Ref.~\cite{Biedermann:2016lvg}.
 }
\efig
The comparison for the total cross section is shown in
\refta{tab:ZZQCDvalidation}. Agreement to four digits is found between the two 
independent calculations within the statistical uncertainty. Figure~\ref{fig:ZZQCDvalidation}, shows a comparison at the level of 
differential distributions between the two Monte Carlo integrations,
once for the distribution in the di-muon mass $m_{\mu^+\mu^-}$ (left plot) and once for 
the four-lepton invariant mass $m_{4\ell}$ (right plot). The upper panels plot the absolute predictions of the two integrations on top 
of each other, the differences being almost invisible. The lower panels show the bin-wise ratio between the two calculations. 
The fluctuations in the four-lepton invariant mass are almost everywhere below 0.5\%, from the far off-shell region over the on-shell 
production threshold at $m_{4\ell}=2M_\PZ$ up to $1\TeV$. A similar pattern is observed for the di-muon mass in the left plot with slightly 
larger fluctuations.

\subsubsection{Higgs production in association with a top-quark pair}
\label{sec:tthQCD}

\paragraph{Input parameters:} In Ref.~\cite{Badger:2016bpw}, a comparison
between \madgraph and \Sherpa\!\!+\OpenLoops for various cross sections at LO, NLO QCD, and NLO EW has been
presented for the process ${\rm p} {\rm p} \to {\rm t} {\rm \bar t} {\rm H}$.
In particular, five different cross sections have been reported. The first one is fully inclusive, and
no event selections are applied. Two of them are computed when applying a cut on the transverse momentum of the three
massive final-state particles at $200\GeV$ and $400\GeV$, respectively.
Furthermore, a cross section with a higher transverse-momentum cut
($500\GeV$) on the Higgs boson only is presented and
finally the cross section obtained by excluding events with a top quark with absolute rapidity lower than $2.5$.
The computations are done for a centre-of-mass energy of $\sqrt{s}=13\TeV$ at the LHC.
The masses of the involved particles read  
\begin{equation}
 M_{\rm Z}=91.188\GeV,\qquad M_{\rm W}=80.385\GeV,\qquad M_{\rm
   H}=125\GeV,\qquad m_{\rm t}=173.3\GeV ,
\end{equation}
and the corresponding widths are all set to zero.
The bottom quark is considered massless and its PDF contribution is included.
The NNPDF-2.3QED PDF set with a variable flavour-number scheme and $\alphas(M_{\rm Z})=0.118$ \cite{Ball:2013hta,Carrazza:2013bra,Carrazza:2013wua}
has been used. The renormalisation as well as the factorisation scales are set to a common scale $\hat{H}_{\rm T} / 2$, defined as
\begin{equation}
\label{eq:tthscale}
\hat{H}_{\rm T} = \sum_i \sqrt{p^2_{{\rm T},i} +m^2_i} ,
\end{equation}
where the index $i$ runs over all the final-state particles.
The considered LO production cross section is at the order $\order{\alphas^2\alpha}$.

\paragraph{NLO QCD validation:}
The obtained LO cross sections listed in \refta{table:ttHLO} show a generally good agreement.
\begin{table}
\begin{center}
\begin{tabular}{|c|c|c|c|}
\hline \rule{0ex}{2.4ex}%
 $\sigma^\mathrm{LO}$~[pb]  \rule{0ex}{2.4ex} & \Sherpa\!\!+\Recola & \madgraphbis & \Sherpa\!\!+\OpenLoops \\
\hline \rule{0ex}{2.4ex}%
inclusive & $3.612(4) \cdot 10^{-1}$ & $3.617 \cdot 10^{-1}$ & $3.617 \cdot 10^{-1}$ \\
$p_{{\rm T}, {\rm t/\bar t/H}}>200\GeV$ & $1.338(2) \cdot 10^{-2}$ & $1.338 \cdot 10^{-2}$ & $1.338 \cdot 10^{-2}$ \\
$p_{{\rm T}, {\rm t/\bar t/H}}>400\GeV$ & $4.001(4) \cdot 10^{-4}$ & $3.977 \cdot 10^{-4}$ & $3.995 \cdot 10^{-4}$ \\
$p_{{\rm T}, {\rm H}}>500\GeV$ & $2.015(3) \cdot 10^{-3}$ & $2.013 \cdot 10^{-3}$ & $2.014 \cdot 10^{-3}$ \\
$\left| y_{\rm t} \right| > 2.5$ & $5.017(5) \cdot 10^{-3}$ & $4.961 \cdot 10^{-3}$ & $5.006 \cdot 10^{-3}$ \\
\hline
\end{tabular}
\end{center}
\caption{\label{table:ttHLO}
Integrated cross sections for ${\rm p} {\rm p} \to {\rm t} {\rm \bar
  t} {\rm H}$ at LO for a centre-of-mass energy of $\sqrt{s} = 13\TeV$ calculated with \Sherpa\!\!+\Recola, \madgraph, and \Sherpa\!\!+\OpenLoops for the set-up of Ref.~\cite{Badger:2016bpw}.
The cross sections are expressed in pb.
The integration errors of the last digits are given in parentheses for the \Sherpa\!\!+\Recola predictions.}
\end{table}
The corresponding NLO QCD predictions of order
$\order{\alphas^3\alpha}$ are reproduced in \refta{table:ttHNLOQCD}.
\begin{table}
\begin{center}
\begin{tabular}{|c|c|c|c|c|}
\hline \rule{0ex}{2.4ex}%
& \multicolumn{2}{c|}{\Sherpa\!\!+\Recola} & \madgraphbis & \Sherpa\!\!+\OpenLoops \\
\cline{2-5} \rule[-1.4ex]{0ex}{3.9ex}%
 & $\sigma^\mathrm{NLO}_{\rm QCD}$~[pb] & $\delta_{\rm QCD} [\%]$ & {$\delta_{\rm QCD} [\%]$} & $\delta_{\rm QCD} [\%]$ \\
\hline \rule{0ex}{2.4ex}%
inclusive & $4.6407(9)\cdot 10^{-1}$ & $28.5(2)$ & $28.9$ & $28.3$ \\
$p_{{\rm T}, {\rm t/\bar t/H}}>200\GeV$ & $1.630(2)\cdot 10^{-2}$ & $21.9(2)$ & $23.4$ & $22.5$ \\
$p_{{\rm T}, {\rm t/\bar t/H}}>400\GeV$ & $4.289(5)\cdot 10^{-4}$ & $7.2(2)$ & $9.6$ & $10.4$ \\
$p_{{\rm T}, {\rm H}}>500\GeV$ & $2.747(3)\cdot 10^{-3}$ & $36.3(2)$ & $37.8$ & 3$7.3$ \\
$\left| y_{\rm t} \right| > 2.5$ & $6.840(7)\cdot 10^{-3}$ & $36.3(2)$ & $37.5$ & $36.9$ \rule[-1.2ex]{0ex}{1.8ex}\\
\hline
\end{tabular}
\end{center}
\caption{\label{table:ttHNLOQCD}
Integrated cross sections for ${\rm p} {\rm p} \to {\rm t} {\rm \bar
  t} {\rm H}$ at NLO QCD for a centre-of-mass energy $\sqrt{s} = 13\TeV$ calculated with \Sherpa\!\!+\Recola, \madgraph, and \Sherpa\!\!+\OpenLoops for the set-up of Ref.~\cite{Badger:2016bpw}.
The cross sections are expressed in pb while the relative corrections are expressed in per cent.
The integration errors of the last digits are given in parentheses for the \Sherpa\!\!+\Recola predictions.}
\end{table}
The agreement is reasonable but a definite statement is not possible as the predictions made by \madgraph and \Sherpa\!\!+\OpenLoops do not have statistical errors.
In addition we have cross-checked the fully inclusive set-up at the level of distributions
against the public \Sherpa\!\!+\OpenLoops implementation, finding perfect agreement in each bin.
The distributions for \Sherpa\!\!+\Recola are shown in the plots of Fig.~\ref{fig:ttHQCD}.

\bfig
  \center
   \includegraphics[width=0.45\textwidth]{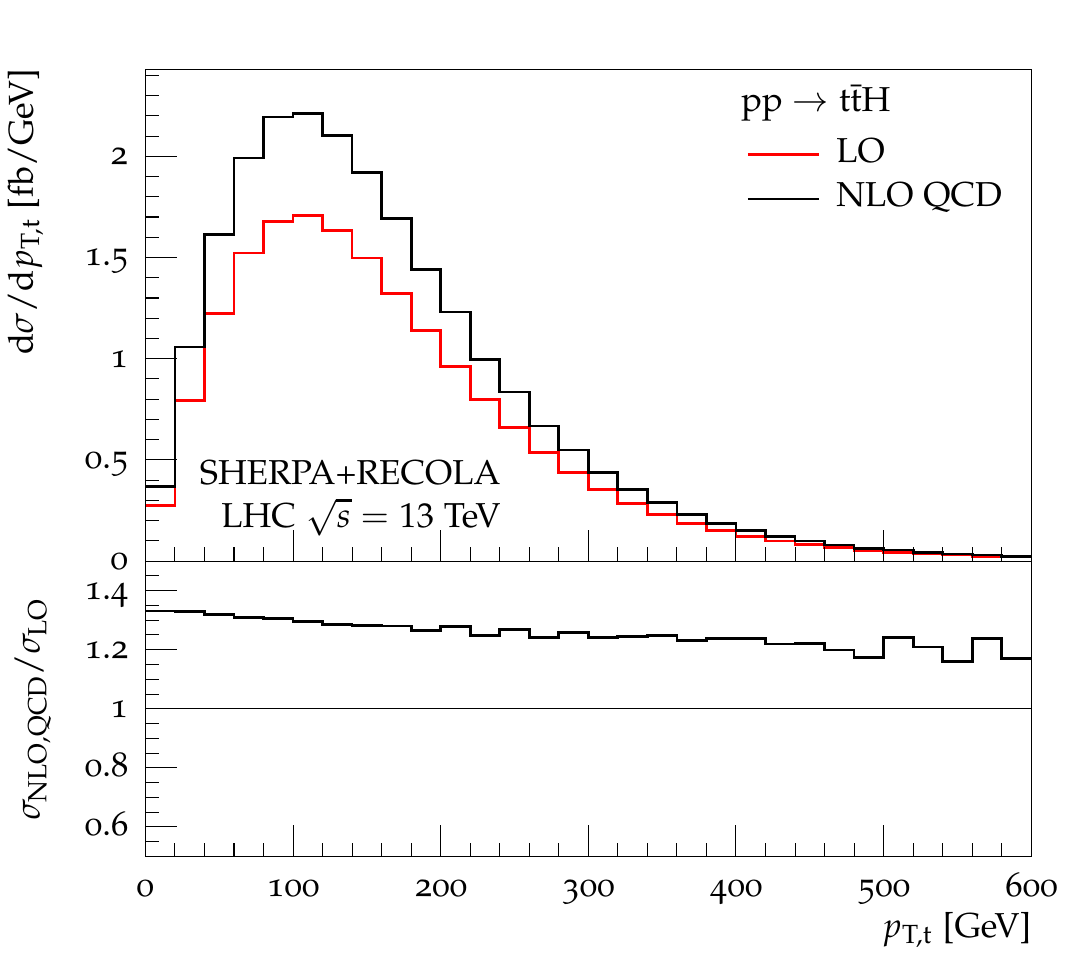}
   \includegraphics[width=0.45\textwidth]{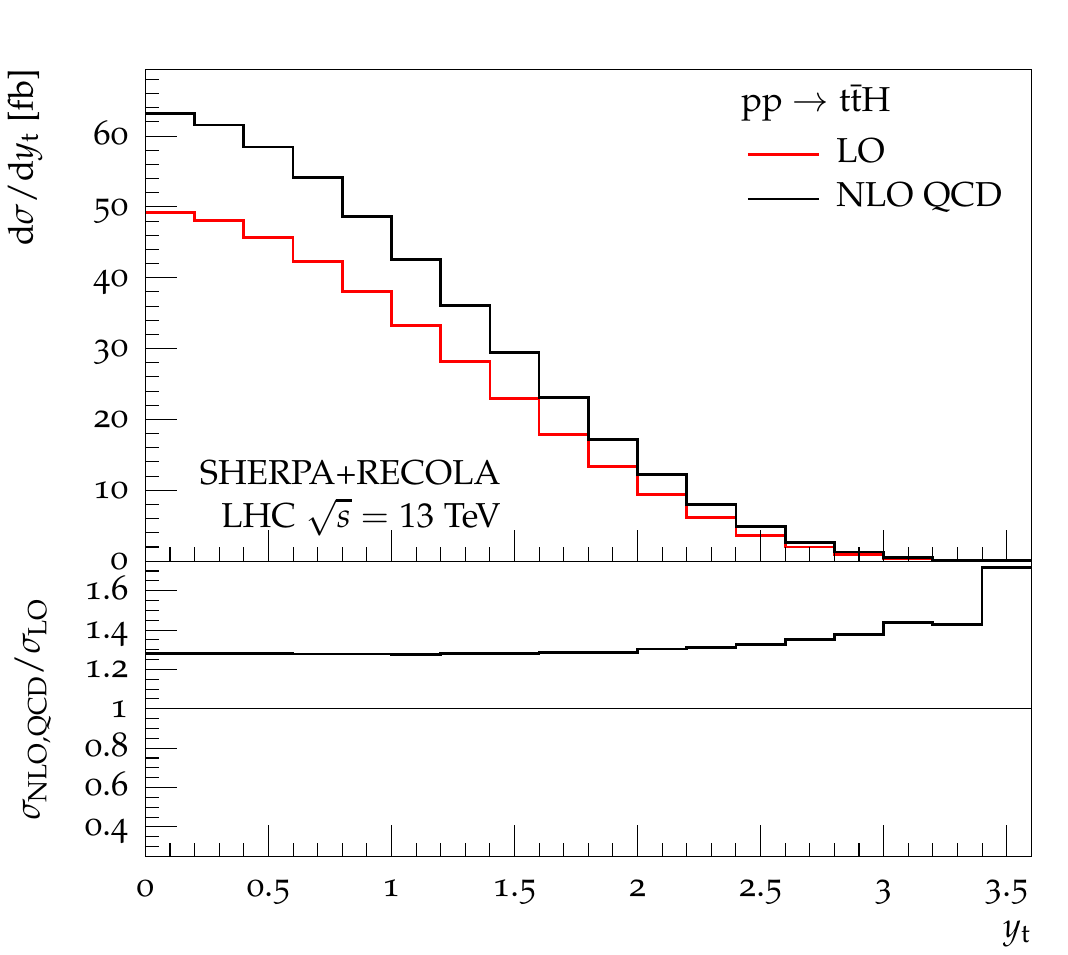}
  \caption{\label{fig:ttHQCD}
  Differential distributions at a centre-of-mass energy $\sqrt{s} = 13\TeV$ for ${\rm pp} \to {\rm t \bar t H}$ at the LHC in a fully inclusive set-up:
  transverse momentum distribution of the top quark (left) and rapidity distribution of the top quark (right).}
\efig

\subsubsection{Higgs-boson production in association with a ${\rm Z}$ boson}

As last example, Higgs-boson production in association with an on-shell ${\rm Z}$ boson is considered.
This channel is particularly interesting, as it receives sizeable contributions from
the loop-induced $\Pg\Pg\to\PH\PZ$ channel. 

\paragraph{Input parameters:} The set-up used is similar to the one of the Drell--Yan example
described in Section~\ref{DYJets:sec}. However, here we use the partonic centre-of-mass energy
$\sqrt{\hat s}$ as the renormalisation and factorisation scale. For the Higgs boson we assume
the decay into a pair of bottom quarks, that is, however, fully factorised from the production
process. The mass of the bottom quark is thereby taken to be $m_{\Pb}=4.8\GeV$.

\paragraph{NLO QCD validation:}
The LO contribution appears at order $\order{\alpha^2}$ for the quark-initiated channel,
hence the NLO QCD cross section is of order $\order{\alphas
  \alpha^2}$. The loop-induced $\Pg\Pg\to\PH\PZ$ process contributes at order $\order{\alphas^2
  \alpha^2}$ but is enhanced by the gluon PDF.
The comparison for the total cross sections calculated with \Sherpa\!\!+\Recola and
\Sherpa\!\!+\OpenLoops can be found in \refta{tab:HZ_crosssections}.
Again, perfect agreement is found, originating from the fact that both cross sections
have been evaluated using the same phase-space points such that there is no statistical
difference, and the deviations per point are below $10^{-9}$ (cf.\ Section~\ref{sec:phase_space_points}).
\begin{table}
\begin{center}
\begin{tabular}{|l|c|c|}
\hline \rule{0ex}{2.4ex}%
$\sigma^{\mathrm{NLO}}_{\mathrm{QCD}}$ & \Sherpa\!\!+\Recola [pb] & \Sherpa\!\!+\OpenLoops [pb] \\
\hline \rule{0ex}{2.4ex}%
$ \Pq \bar \Pq \to \PH \PZ $ & $0.41012(9)$ & $0.41012(9)$ \\
$ \Pg \Pg \to \PH \PZ $ & $0.029482(2)$ & $0.029482(2)$ \\
\hline
\end{tabular}
\end{center}
\caption{\label{tab:HZ_crosssections} Total cross sections calculated for ${\rm pp}\to {\rm H}{\rm Z}$ at NLO QCD with the \Sherpa\!\!+\Recola interface, compared against \Sherpa\!\!+\OpenLoops.}
\end{table}

In Fig.~\ref{fig:HPTPLOT} the transverse-momentum distribution of the reconstructed Higgs
boson is presented, where the quark- and the gluon-initiated channels are shown separately.
Though much smaller than the quark-induced contribution, the loop-induced
gluon-initiated channel still contributes around $10\%$ to the total cross section for Higgs transverse
momenta in the range $100$--$200\GeV$. This nicely illustrates the importance of precise
predictions of this phenomenologically important Higgs-production process. Considering
even larger values of $p_\text{T,H}$ the gluon contribution falls off more steeply, due
to the different slope of the quark and gluon parton-density functions. Let us note that
the results for the differential distributions have explicitly been checked against
\Sherpa\!\!+\OpenLoops, and using identical phase-space points perfect agreement has
been found.

\bfig
  \center
  \includegraphics[width=0.5\textwidth]{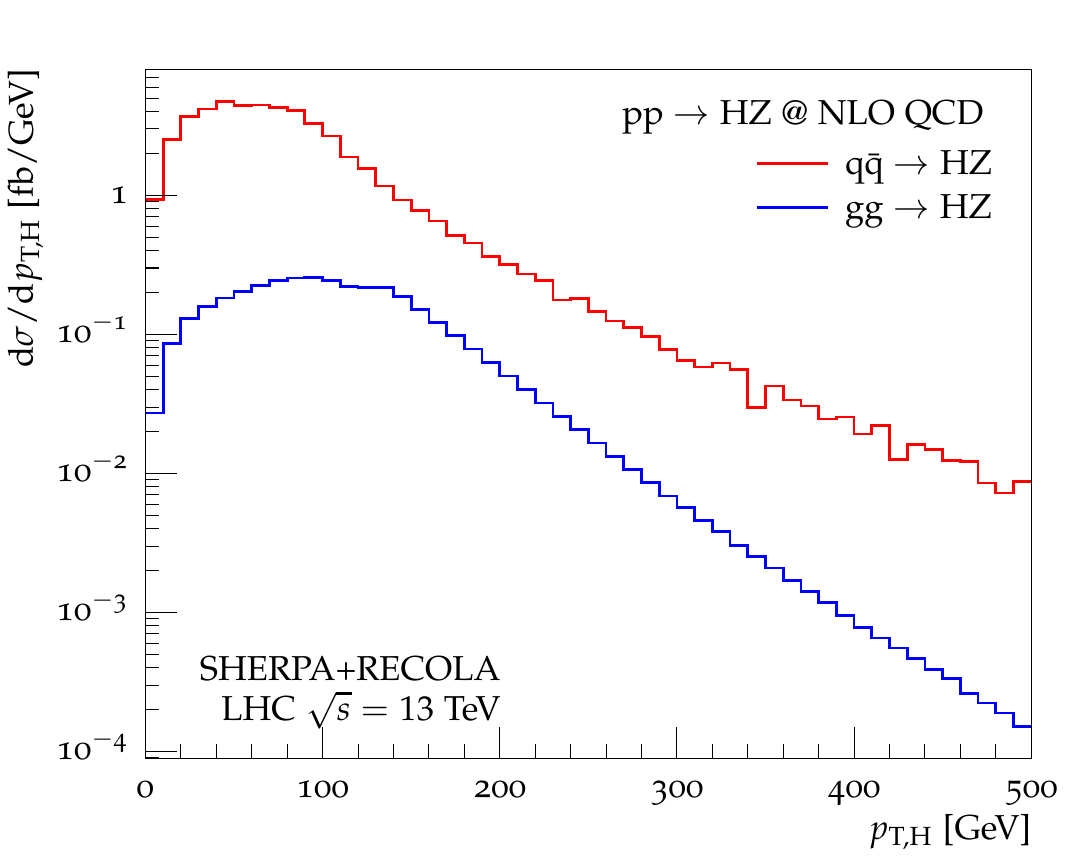}
  \caption{\label{fig:HPTPLOT} The transverse-momentum distribution of the reconstructed $\Pb \bar \Pb$-decayed $\PH$ boson in the $\Pp\Pp \to {\rm HZ}$ process at NLO QCD.}
\efig

\subsection{Matching to parton shower}
\label{sec:mepsnlo}

To illustrate and validate the use of virtual QCD matrix elements from \Recola for full
particle-level simulations, a state-of-the-art QCD calculation is presented where NLO QCD matrix
elements of varying final-state parton multiplicity get matched to the QCD parton shower of
\Sherpa \cite{Schumann:2007mg} and subsequently hadronised. In particular, the
Drell--Yan process is investigated, \ie$\Pp\Pp \to \gamma^*/{\rm Z}^* \to \Pe^+\Pe^-/\mu^+\mu^-$, in association with jets in the
MEPS@NLO scheme \cite{Hoeche:2012yf}. To this end, NLO QCD matrix elements are considered for
up to two additional jets and the tree-level contribution for three final-state partons.
The phase-space slicing parameter of the merging scheme is set to $Q_{\rm{cut}}=20\GeV$. The
NNPDF-3.0 NNLO PDF set is employed with $\alphas(M_{\rm Z})=0.118$ and five active flavours.
For the electromagnetic coupling constant $\alpha$, the $\alpha(\MZ)$ scheme is used with a numerical value of $\alpha(\MZ) = 0.007764$.
In the MEPS@NLO approach the renormalisation and factorisation
scales are chosen dynamically, based on the determination of an event-wise $2\to 2$ core
process and an associated sequence of nodal splitting scales, obtained through a clustering
of the matrix-element partons that effectively inverts the \Sherpa parton
shower~\cite{Hoeche:2009xc, Hoeche:2009rj, Carli:2010cg}. To estimate the scale uncertainties,
a 7-point scale variation is considered for $\mu_{\rm{R}}$ and
$\mu_{\rm{F}}$, computed by reweighting
the central prediction ~\cite{Bothmann:2016nao}. Both scales are varied independently by
factors of $1/2$ and $2$, thereby omitting the variations with ratios of 4 between the two
scales. The corresponding uncertainty is then taken as the envelope of all variations considered.

The \Sherpa\!\!+\Recola MEPS@NLO results are compared against corresponding data from the
ATLAS and CMS experiments taken at $\sqrt{s}=7\TeV$. The CMS analysis of jet-associated
Drell--Yan production presented in \citere{Khachatryan:2014zya} assumes pairs of electrons
or muons with an invariant mass between $71$ and $111\GeV$. Furthermore,
the leptons have to exhibit a transverse momentum of $p_{{\rm T},{\ell}}>20\GeV$. Jets
are reconstructed according to the anti-$k_{\rm T}$ algorithm with a radius parameter
of $R=0.5$. A transverse-momentum threshold for jets of $p_{{\rm T},{\rm j}}>30\GeV$ is
assumed, and only jets with $|\eta_{\rm j}|<2.4$ are considered. In addition, all
jets are required to be separated from the leptons by $\Delta R_{{\rm j}{\ell}}\geq 0.5$.
Similarly, in the ATLAS analysis presented in \citere{Aad:2013ysa} electron and muon
pairs within a mass range of $66\GeV \leq m_{\ell\ell} \leq 116\GeV$
are selected and
the leptons must have $p_{{\rm T},{\ell}}>20\GeV$. Anti-$k_{\rm T}$ jets with a radius
parameter of $R=0.4$, $p_{{\rm T},{\rm j}}>30\GeV$ and $|y_{\rm j}|<4.4$ are considered.
Each jet candidate needs to be separated from the reconstructed leptons by
$\Delta R_{{\rm j}{\ell}}\geq 0.5$. For the study presented here, the public
\Rivet implementations of the two analyses were used, allowing for a direct comparison to
particle-level results. Further details on the selections can be found in the
respective publications. 

Figures~\ref{fig:MATCHMERGEMULTIZ}~and~\ref{fig:MATCHMERGEJETPT} provide examples
for the comparison of particle-level MEPS@NLO simulations based on
\Sherpa\!\!+\Recola against data. In the left panel of Fig.~\ref{fig:MATCHMERGEMULTIZ}
the inclusive transverse-momentum distribution of the reconstructed ${\rm Z}$
bosons is compared to the ATLAS data from~\citere{Aad:2013ysa}. The inclusion
of the ${\rm Z}+{\Pj}$ and ${\rm Z}+{\Pj\Pj}$ matrix elements at NLO QCD accuracy provides a good description of
events with sizeable recoil of the Drell--Yan pair. The parton-shower component, on
the other hand, dominates the low-$p_{{\rm T},{\rm Z}}$ region. Notably, the theoretical
scale uncertainties are of order $\pm 20\%$ and well overlap with the experimental
uncertainty band, indicated in yellow. The right panel of
Fig.~\ref{fig:MATCHMERGEMULTIZ} presents the comparison of the jet-multiplicity
distribution against the data from CMS~\cite{Khachatryan:2014zya}. By construction of
the MEPS@NLO algorithm, in the given set-up, the first two bins have
NLO QCD accuracy, while the third has LO QCD precision. All higher jet multiplicities
solely originate from the QCD parton-shower component. The central theoretical
predictions agree well with the data though there is a tendency to overestimate higher
jet counts. However, taking into account theoretical and experimental uncertainties,
simulation and data agree well. 

Finally, Fig.~\ref{fig:MATCHMERGEJETPT} presents the transverse-momentum
distributions of the two leading jets. The set-up provides NLO QCD
accuracy for both observables. The predictions are compared to the respective LHC
data from CMS~\cite{Khachatryan:2014zya}. Very good agreement of data and
theoretical prediction is found. For the latter the theoretical uncertainty estimates from
variations around the central scale choice increase for larger jet
transverse momenta, exceeding the $\pm 20\%$ range seen in the more inclusive boson
transverse momentum distribution. 

\bfig
  \center
  \includegraphics[width=0.45\textwidth]{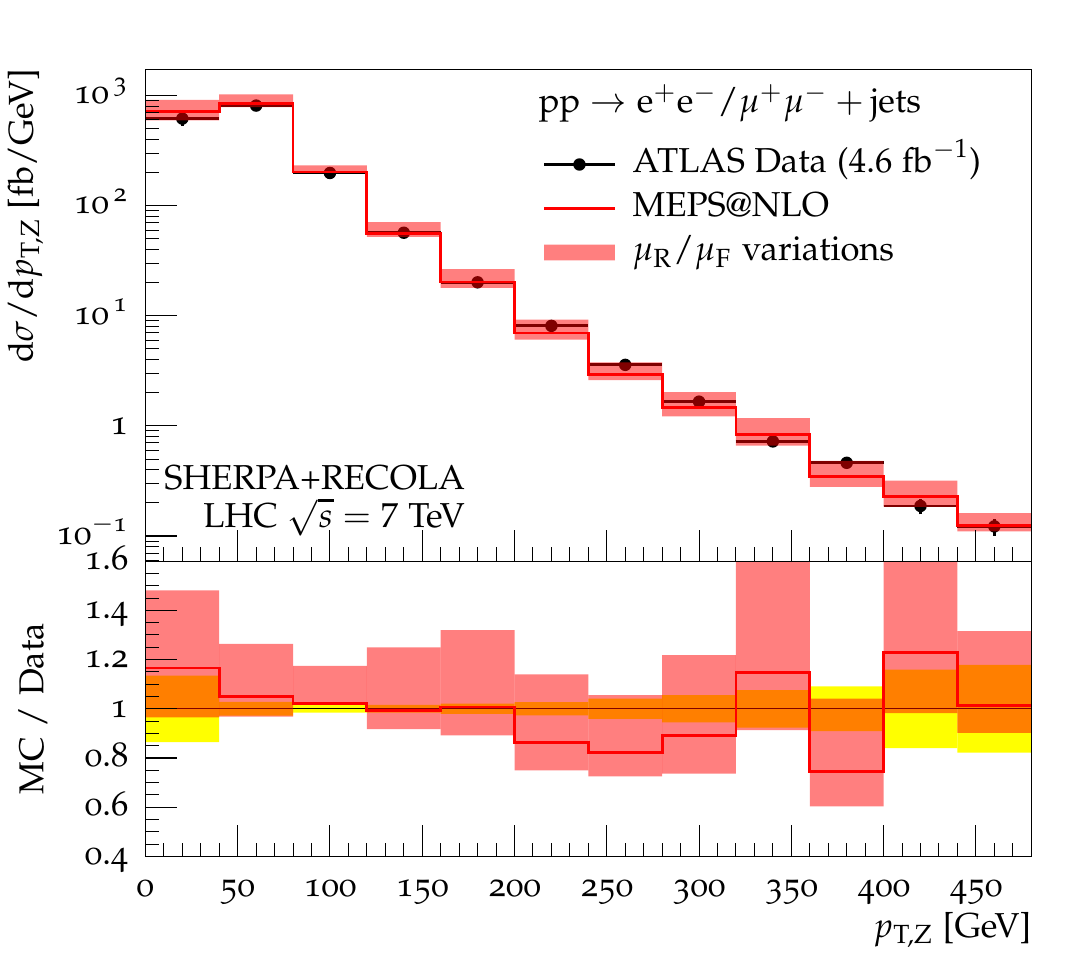}
  \includegraphics[width=0.45\textwidth]{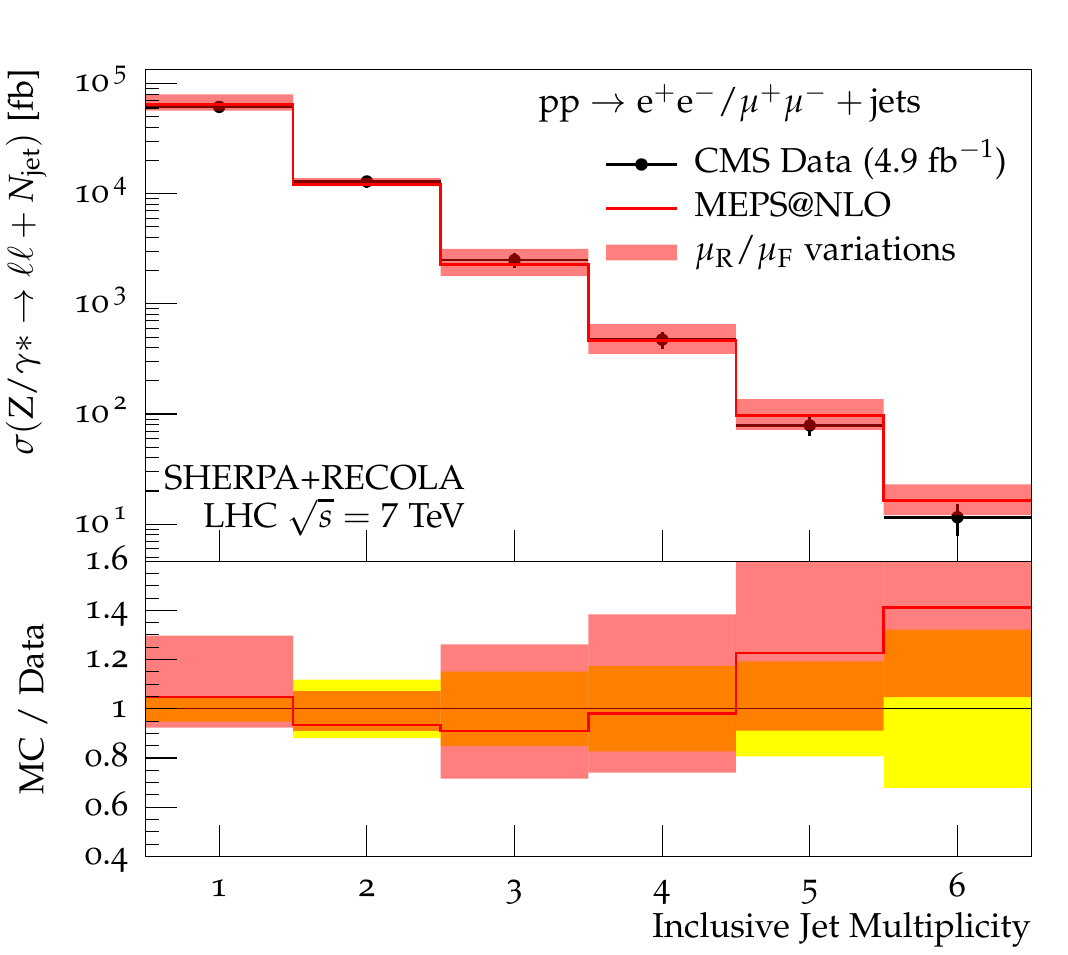}
  \caption{\label{fig:MATCHMERGEMULTIZ} 
    The left-hand figure compares the MEPS@NLO predictions based on \Sherpa\!\!+\Recola for the
    transverse-momentum distribution of the reconstructed ${\rm Z}$ boson compared to $\sqrt{s}=7\TeV$
    LHC data from ATLAS~\cite{Aad:2013ysa}. The right-hand figure shows the inclusive jet
    multiplicity in comparison to $\sqrt{s}=7\TeV$ CMS data~\cite{Khachatryan:2014zya}.  The
    theoretical-uncertainty band is obtained through 7-point scale
    variations of $\mu_{\rm R}$ and $\mu_{\rm F}$.
    The yellow bands in the lower panels indicate the measurements combined statistical
    and systematic uncertainty.}
\efig

\bfig
  \center
  \includegraphics[width=0.45\textwidth]{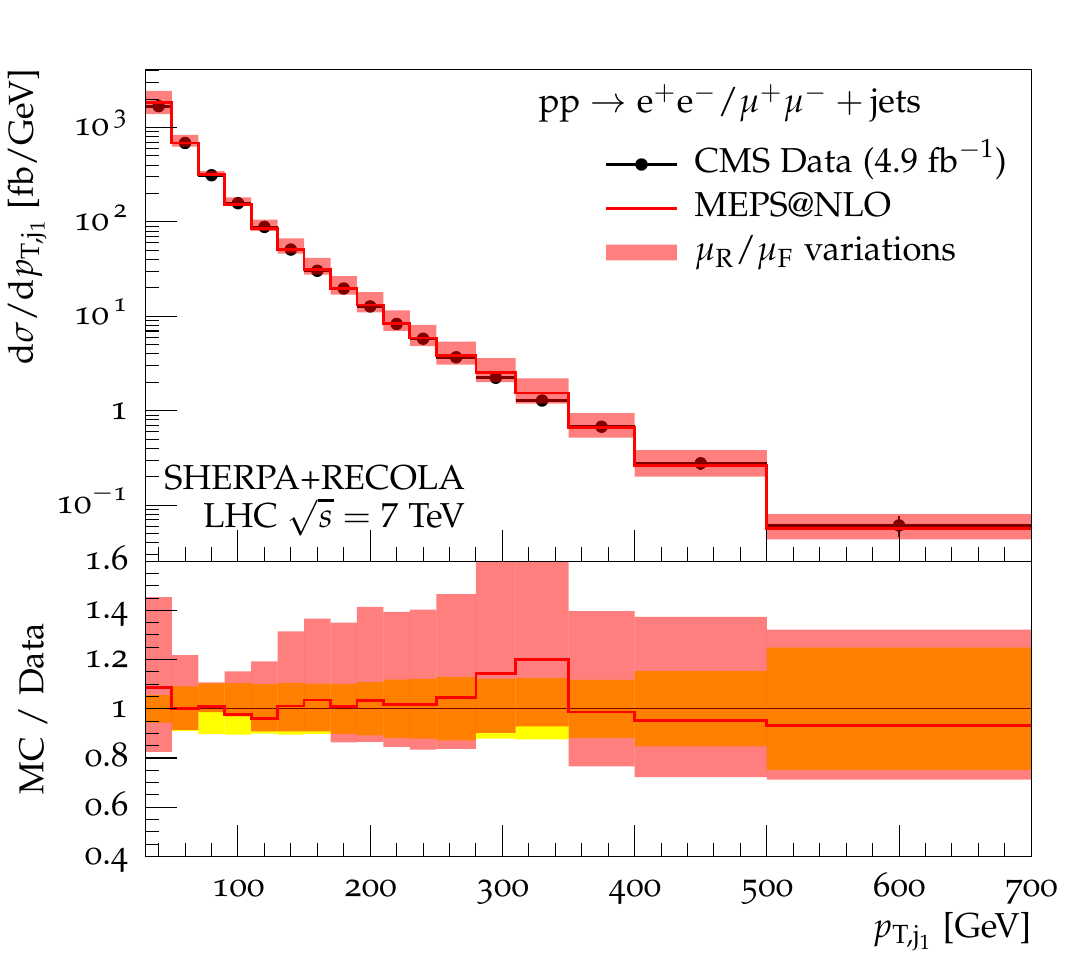}
  \includegraphics[width=0.45\textwidth]{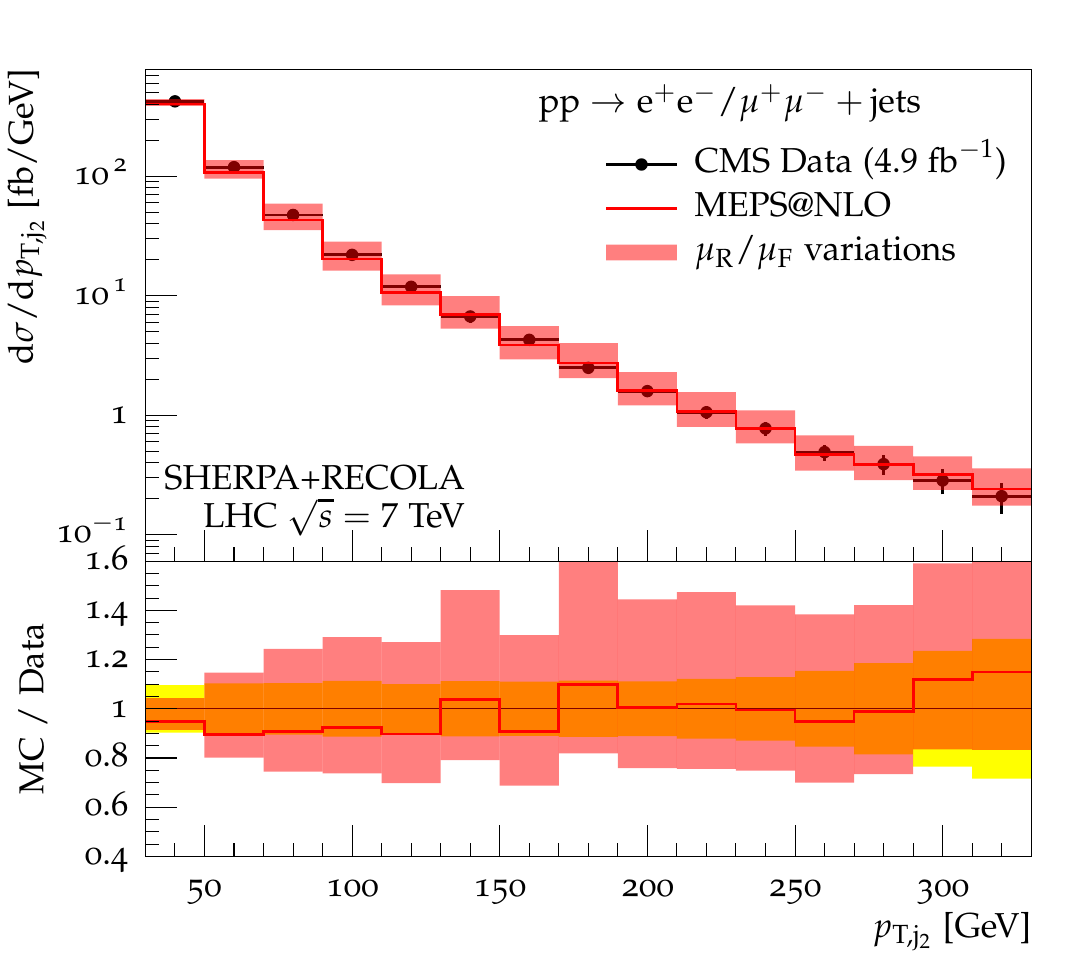}
  \caption{\label{fig:MATCHMERGEJETPT} Transverse-momentum distribution of the first
    (left panel) and second (right panel) hardest jet for Drell--Yan production at the LHC with
    $\sqrt{s}=7\TeV$.  The MEPS@NLO predictions from \Sherpa\!\!+\Recola are compared
    to CMS data from~\citere{Khachatryan:2014zya} (uncertainty bands as in Fig.~\ref{fig:MATCHMERGEMULTIZ}).}
\efig

The results for the jet-associated Drell--Yan process based on \Sherpa\!\!+\Recola presented here
confirm related comparisons of simulations based on NLO QCD matrix elements matched with QCD parton
showers with data, see \eg \citeres{Khachatryan:2016crw,Aaboud:2017hbk}. 
This nicely illustrates the usability of the \Recola loop-amplitude generator in these highly non-trivial
multi-scale  calculations. To this end, the same set-up has been explicitly compared, but
using the \OpenLoops generator for the one-loop amplitudes. For all distributions considered, embracing
many more than presented here, full agreement between \Sherpa\!\!+\Recola and \Sherpa\!\!+\OpenLoops
has been observed. Besides, in this ``real-life'' application no significant run-time differences
have been observed. However, when using \Recola instead of \OpenLoops the allocated memory increases by
about $50\%$ for the process set-up considered. 

\section{NLO EW validation and combined predictions}
\label{sec:processes}

To illustrate the capabilities of the combination of \Sherpa with \Recola, three processes have been computed
at both NLO QCD and EW accuracy. These comprise off-shell vector-boson production in association with jets,
the production of two off-shell ${\rm Z}$ bosons, and the on-shell production of a top-quark pair in association
with a Higgs boson. These three channels represent phenomenologically very important LHC processes. Furthermore,
they are highly non-trivial, and each process features different technical challenges regarding their evaluation
to NLO QCD and EW accuracy, thus demonstrating the generality of our implementation. Since up to now there is no
public code that allows the computation of arbitrary processes at NLO EW accuracy, the number of processes that
have been checked is smaller than for QCD.

For each process, a short introduction is given followed by the
description of the calculational set-up and the actual NLO EW validation.
Finally, the cross sections and differential distributions for combined NLO QCD and EW accuracy are presented.
The cross sections including NLO QCD or EW corrections read
\begin{equation}
 \sigma^{\mathrm{NLO}}_{\mathrm{QCD}} = \sigma^{\mathrm{LO}} + \delta \sigma^{\mathrm{NLO}}_{\mathrm{QCD}} \qquad \text{and} \qquad
 \sigma^{\mathrm{NLO}}_{\mathrm{EW}} = \sigma^{\mathrm{LO}} + \delta \sigma^{\mathrm{NLO}}_{\mathrm{EW}},
\end{equation}
respectively.
The additive combination of the two types of corrections is straight-forward,
\begin{equation}
\label{additionNLO}
 \sigma^{\mathrm{NLO}}_{\mathrm{QCD+EW}} = \sigma^{\mathrm{LO}} + \delta \sigma^{\mathrm{NLO}}_{\mathrm{QCD}} + \delta \sigma^{\mathrm{NLO}}_{\mathrm{EW}},
\end{equation}
while a multiplicative combination can be defined as
\begin{equation}
\label{productNLO}
 \sigma^{\mathrm{NLO}}_{\mathrm{QCD}\times\mathrm{EW}} = \sigma^{\mathrm{NLO}}_{\mathrm{QCD}} \left( 1 + \frac{\delta \sigma^{\mathrm{NLO}}_{\mathrm{EW}}}{\sigma^{\mathrm{LO}}} \right)
 = \sigma^{\mathrm{NLO}}_{\mathrm{EW}} \left( 1 + \frac{\delta \sigma^{\mathrm{NLO}}_{\mathrm{QCD}}}{\sigma^{\mathrm{LO}}} \right) .
\end{equation}
The difference between these two ways of combining NLO QCD and EW
corrections provides an estimate of the missing higher orders
resulting from mixed QCD--EW contributions.
The NLO QCD$\times$EW combination can be understood as an improved prediction when the typical scales of the QCD and EW corrections are well separated.

\subsection{Vector-boson production in association with jets}

As a first process, the production of a vector boson plus jets is considered.
This process has already been  computed for the LHC running at a centre-of-mass energy of $13\TeV$ for both on- and 
off-shell vector bosons at NLO QCD and EW~\cite{Denner:2014ina,Kallweit:2014xda,Kallweit:2015dum},
and therefore allows a comparison to existing results.
From the technical point of view, the process is a mixture of QCD and EW contributions.
In particular, for vector-boson production in association with more than one jet at NLO EW, 
interferences appear between QCD and EW production channels. This makes 
vector-boson-plus-jets production a good testing ground of the interface, although it is also an important process
 in its own right. 

\paragraph{Input parameters:}

The input parameters for this process class are taken from Ref.~\cite{Kallweit:2015fta} 
to allow a tuned comparison. For completeness, these parameters as well
as the analysis cuts are detailed here.
The renormalisation and factorisation scale for off-shell W-boson production is $\hat{H}'_{{\rm T},{\rm W}}/2$, defined via
\begin{equation}
\hat{H}'_{{\rm T},{\rm W}}=\sum_{i=\Pq,\Pg}p_{{\rm T}, i}+p_{{\rm T},\gamma}+\sqrt{p_{{\rm T},\ell\nu}^2+m_{\ell\nu}^2}.
\end{equation}
The scales used in off-shell ${\rm Z}$-boson production, $\hat{H}'_{{\rm T},{\rm Z}}/2$,
are similarly given by
\begin{equation}
\hat{H}'_{{\rm T},{\rm Z}}=\sum_{i=\Pq,\Pg}p_{{\rm T},i}+p_{{\rm T},\gamma}+\sqrt{p_{{\rm T},\ell\ell}^2+m_{\ell \ell}^2}.
\end{equation}
For on-shell weak-boson production, a slightly different scale,
 $\hat{H}_{{\rm T}}/2$, is used, defined via
\begin{equation}
 \hat{H}_{\rm T}=\sum_{i=\Pq,\Pg} p_{{\rm T}, i}+p_{{\rm T}, \gamma}+\sqrt{p_{{\rm T}, { V}}^2+M_{ V}^2},
\end{equation}  
where ${ V}$=${\rm W}, {\rm Z}$ for ${\rm W}$+jets and ${\rm Z}$+jets production, respectively. 
No scale variations have been considered for this validation, although a variation of a factor of $1/2$ 
and $2$ on the central scale was considered in the original article.

For on-shell vector-boson production, the bosons are decayed to leptons in a
factorised approach, thereby preserving the spin correlations \cite{Hoche:2014kca}.
For ${\rm W}$-boson production, the leptonic decay channels  ${\rm W}^\pm\to {\rm e}^{\pm(}\overline{\nu}^{)}_e$ and 
${\rm W}^\pm\to \mu^{\pm(}\overline{\nu}^{)}_\mu$ are considered. Similarly, for ${\rm Z}$-boson
production, the allowed leptonic decay channels are ${\rm Z}\to{\rm e}^+{\rm e}^-$ and 
${\rm Z}\to\mu^+\mu^-$. 
For all processes in this section, the masses 
for the ${\rm Z}$ boson, ${\rm W^\pm}$ bosons, ${\rm H}$ boson and 
top quark read
\begin{equation}
 M_{\rm Z}=91.1876 \GeV,\qquad M_{\rm W}=80.385\GeV,\qquad M_{\rm
   H}=126\GeV,\qquad m_{\rm t}=173.2\GeV.
\end{equation}
The Fermi constant is taken to be $G_\mu=1.1667\times 10^{-5}\GeV^{-2}$, and the $G_\mu$ scheme is 
used to consistently define the EW parameters.
The NNPDF-2.3QED NLO PDF set with a variable flavour-number scheme, QED corrections and
$\alpha_s(M_{\rm Z})=0.118$~\cite{Ball:2013hta,Carrazza:2013bra,Carrazza:2013wua} 
has been used for both LO and NLO calculations.

For on-shell vector-boson production, the widths of the external bosons are set to zero in general.
An exception to this rule is in the QCD--EW interference term
introduced in the EW real-subtracted 
contribution which includes electroweakly produced jets leading to a poorly 
converging phase-space integration. Because this enters only as an interference
term, it does not give rise to a true resonance and maintaining a width of zero is
theoretically acceptable. Following Ref.~\cite{Kallweit:2014xda}, a small, 
artificial width for the ${\rm W}$ and ${\rm Z}$ bosons is introduced in this 
case in order to control the phase-space integration.
In this publication we use $0.3\GeV$.

For the off-shell vector-boson production processes, physical values of the vector-boson widths are
used, as well as for other unstable particles such as the Higgs boson and the top quark,
\begin{equation}
 \Gamma_{\rm Z}=2.4955\GeV, \qquad \Gamma_{\rm W}=2.0897\GeV, \qquad
 \Gamma_{\rm H}=4.07\MeV, \qquad \Gamma_{\rm t}=1.339\GeV.
\end{equation}
Furthermore, the complex-mass scheme is employed for the unstable particles in this case, and
a unit CKM matrix is assumed.

Photons within a rapidity--azimuthal-angle distance of $R_{\gamma \Pq/\ell}=0.1$ from a quark or lepton are recombined with a 
simple cone-like algorithm with the closest charged particle. Jets are defined with the anti-$k_\mathrm{T}$ algorithm using $R=0.4$ and
\begin{equation}
\label{eq:Vj_incl_cuts}
 p_{{\rm T}, {\rm j}}>30 \GeV,\qquad |\eta_{\rm j}|<4.5\,.
\end{equation}
Any jet with more than $50\%$ of its energy originating from a photonic contribution is removed from
the jet list.

For the cross section validation for on-shell ${\rm W+j}$ production, two phase-space regions, 
 other than the inclusive cross section subject to the cuts
 \refeq{eq:Vj_incl_cuts}, are considered, defined by the additional cuts
\begin{equation}
 p_{{\rm T}, {\rm W}}>1\TeV, \qquad p_{{\rm T}, {\rm j}}>1\TeV .
\end{equation}

Distributions have been analysed with \Rivet following 
Ref.~\cite{Kallweit:2015dum} and using the cuts shown in 
\refta{TAB:Vjetscuts}.

\begin{table}
   \begin{center}
   \begin{tabular}{|c|c|c|c|c|}
   \hline \rule{0ex}{2.4ex}%
    cut variable & $\ell^{\pm(}\overline{\nu}^{)}$ & $\ell^+\ell^-$  \\
   \hline \rule{0ex}{2.4ex}%
    $p_{\rm{T},\ell^\pm}/\GeV>$ & $25$  & $25$ \\
    $E_{{\rm T}}^{\rm miss}/\GeV>$ & $25$ & --\\
    $m_{\rm T}^{\rm W}/\GeV>$ & $40$ & --\\
    $|\eta_{\ell^\pm}|<$ & $2.5$ & $2.5$ \\
    $\Delta R_{\ell^\pm {\rm j}}>$ & $0.5$ & $0.5$ \\
    $\Delta R_{\ell^+\ell^-}>$ & -- & $0.2$ \\
\rule[-1.2ex]{0ex}{1ex}%
    $m_{\ell^+\ell^-}/\GeV\in$ & -- & $[66,116]$ \\
   \hline
   \end{tabular}
   \end{center}
   \caption{\label{TAB:Vjetscuts} Event selections used for the \Rivet analyses for the
   ${V+jets}$ distributions, where  
   $m_{\rm T}^{\rm W}=\sqrt{2p_{{\rm T},{\ell}}p_{{\rm T},\nu}(1-\cos\Delta\Phi_{{\ell}\nu})}$.}
\end{table}

\paragraph{NLO EW validation:}
Next, we present the validation of the \Sherpa\!\!+\Recola interface against published cross sections obtained with \Sherpa\!\!+\OpenLoops~\cite{Kallweit:2014xda,Kallweit:2015fta,Kallweit:2015dum}.
First, on-shell ${\rm W+j}$ production at a centre-of-mass energy of $13\TeV$ at the LHC is considered.

\begin{table}
\begin{center}
\begin{tabular}{|c|c|c|c|c|}
\hline \rule[-1.4ex]{0ex}{3.9ex}%
$\sigma^\mathrm{NLO}_{\mathrm{QCD+EW}}$ & \Sherpa\!\!+\OpenLoops [pb] & \Sherpa\!\!+\Recola [pb] & $\delta [\%$]\\
\hline inclusive & $15621$ & $15592(30)$ & $0.19$\\
$p_{{\rm T}, {\rm W}}>1\TeV$ & $0.040$ & $0.0400(2)$ & $0$\\
\rule[-1.2ex]{0ex}{1ex}%
$p_{{\rm T}, {\rm j}}>1\TeV$ & $0.195$ & $0.194(1)$ & $0.51$\\
\hline
\end{tabular}
\end{center}
\caption{\label{TABLE:WjValidation} Cross sections at $13\TeV$
 for $\Pp\Pp\to{\rm W}^\pm {\rm j}$ at the LHC at NLO QCD and EW with 
 \Sherpa\!\!+\Recola, compared to published results with \Sherpa\!\!+\OpenLoops\cite{Kallweit:2014xda,Kallweit:2015dum,Kallweit:2015fta}.
}
\end{table}

Table~\ref{TABLE:WjValidation} shows the relative difference between 
total cross-section calculations, with different phase-space cuts, 
for the \Sherpa\!\!+\Recola interface against the published numbers 
from \Sherpa\!\!+\OpenLoops. The errors quoted on the
\Sherpa\!\!+\Recola cross sections are statistical in origin.
While for the \Sherpa\!\!+\OpenLoops results scale uncertainties were 
presented, the comparably negligible statistical uncertainties were
not listed.
However, for the
validation of the codes, it is necessary to demonstrate close
statistical agreement with the published numbers for the central
scale choice.  Taking the statistical errors of \Sherpa\!\!+\Recola as benchmark, good agreement between the two calculations for the 
NLO QCD and EW total cross sections is observed in all cases.

Besides the integrated cross sections, Ref.~\cite{Kallweit:2014xda} also 
provides distributions, which can be used for a more qualitative validation 
over a large phase space. Figure~\ref{FIG:WjNLOEW} displays the $p_{\rm T}$ 
distribution of the hardest jet in on-shell $\PW^++\Pj$ production. The left-hand 
side of Fig.~\ref{FIG:WjNLOEW} shows the inclusive prediction, and
the right-hand side the effect of a phase-space cut 
$\Delta\Phi({\rm j},{\rm j})<3\pi/4$. This effect is
very large, in line with the findings in Ref.~\cite{Kallweit:2014xda}. The 
Sudakov behaviour in the large-$p_{\rm T}$ region is clearly recovered once this 
$\Delta\Phi({\rm j},{\rm j})$ cut is included, because it removes 
the contributions from dijet-like structures with a soft W boson emitted. 
At NLO EW, these types of contributions can be better viewed
as a real EW correction to dijet production.

\bfig
  \includegraphics[width=0.5\textwidth]{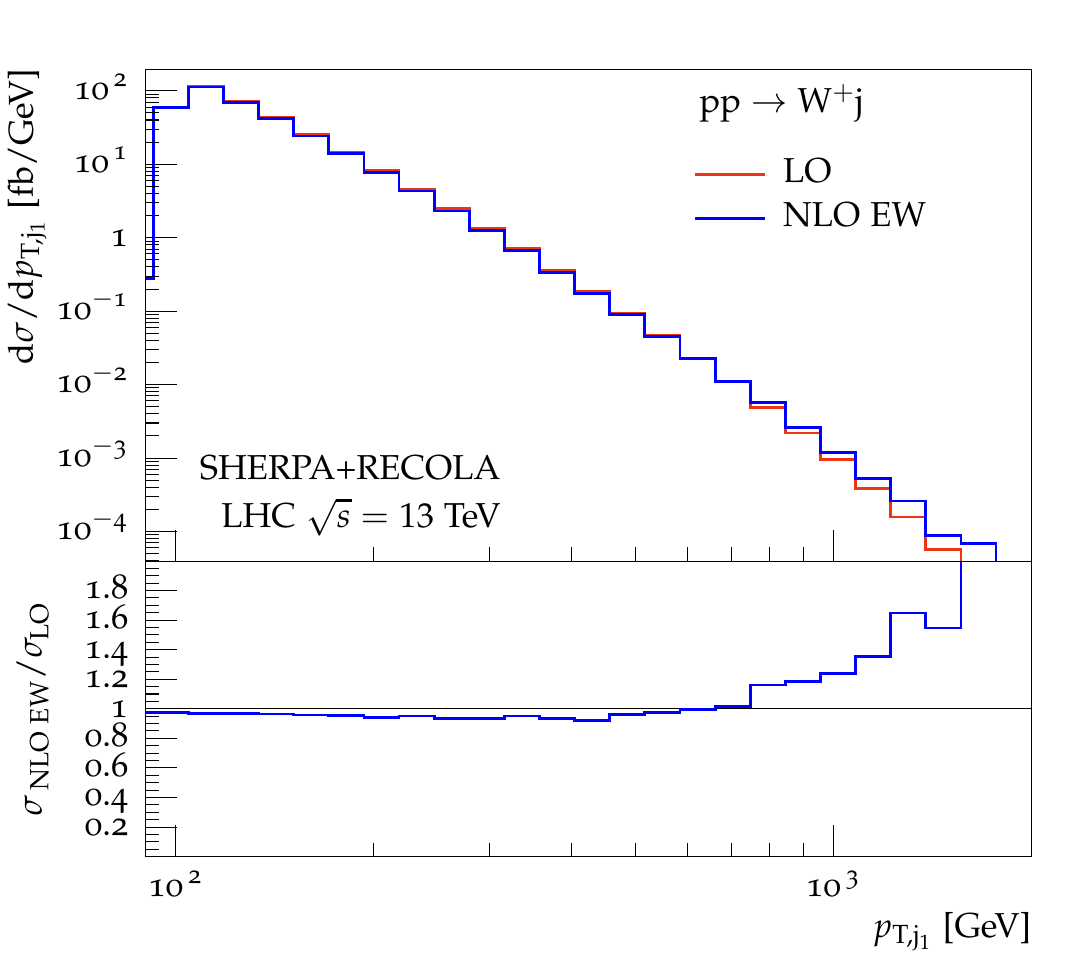}
  \includegraphics[width=0.5\textwidth]{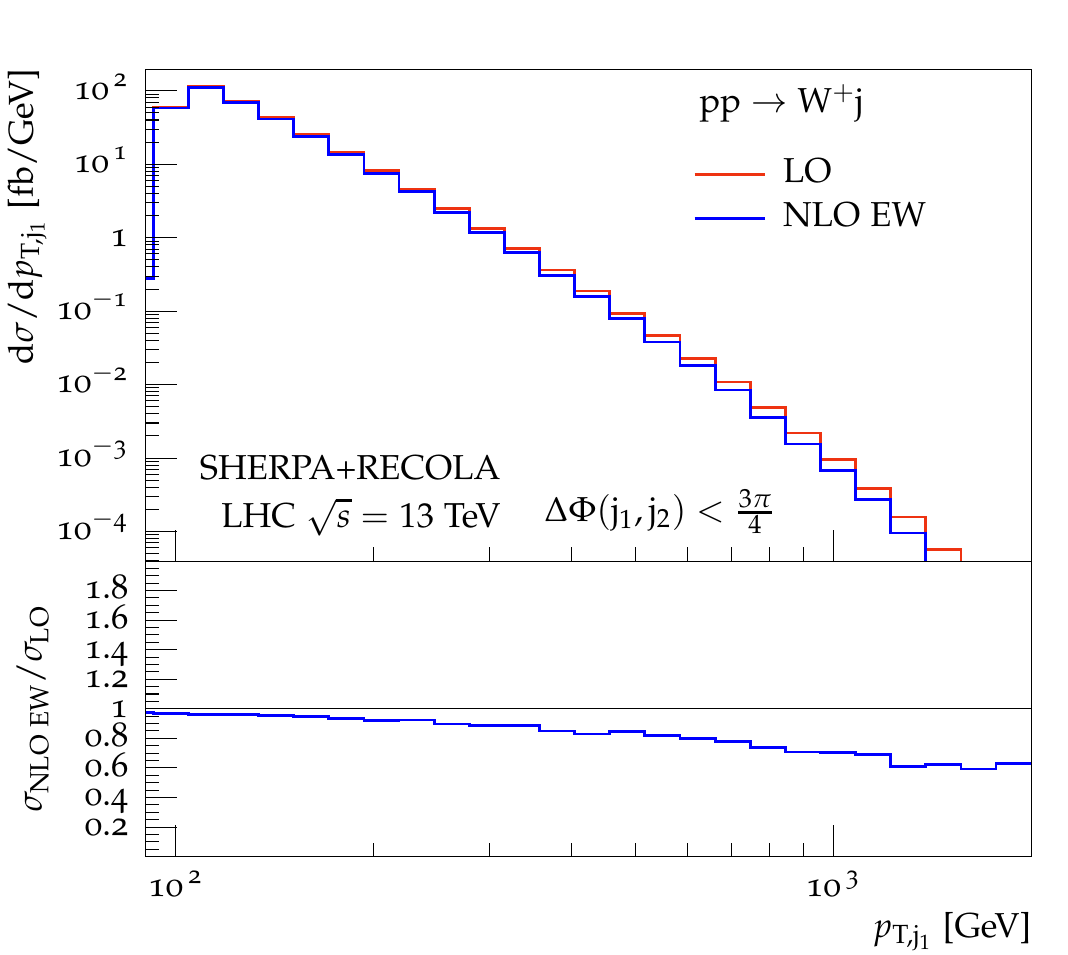}
  \caption{\label{FIG:WjNLOEW} Differential distributions at $13\TeV$ for 
   $\Pp\Pp\to{\rm W}^+{\rm j}$ at the LHC, both at LO and NLO EW. 
   The left-hand figure shows the
   leading-jet $p_{\rm T}$ with minimal cuts and the right-hand plot
   the leading-jet $p_{\rm T}$ with an additional cut 
   $\Delta\Phi({\rm j},{\rm j})<3\pi/4$. The lower panels display the ratio of the
   NLO EW calculation to the corresponding LO result.}
\efig

\begin{figure}
 \includegraphics[width=0.5\textwidth]{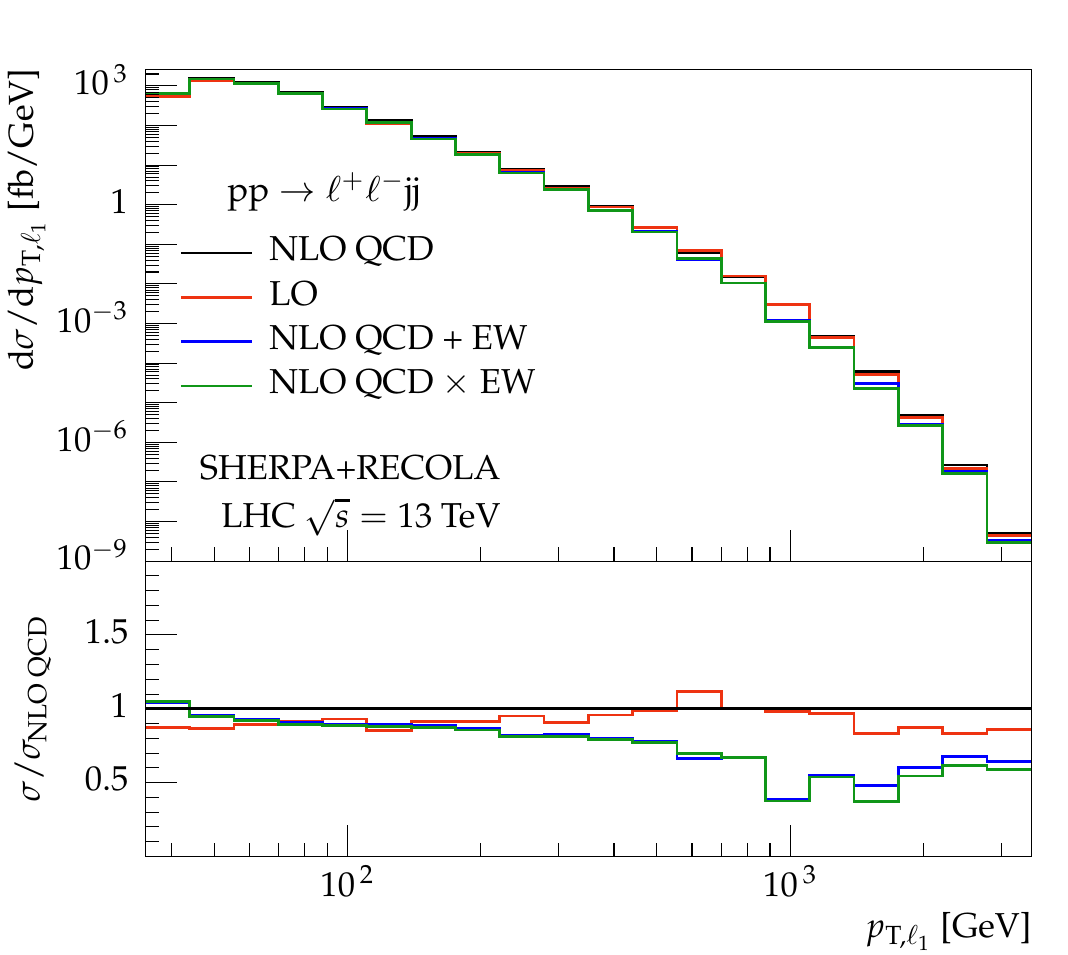}
 \includegraphics[width=0.5\textwidth]{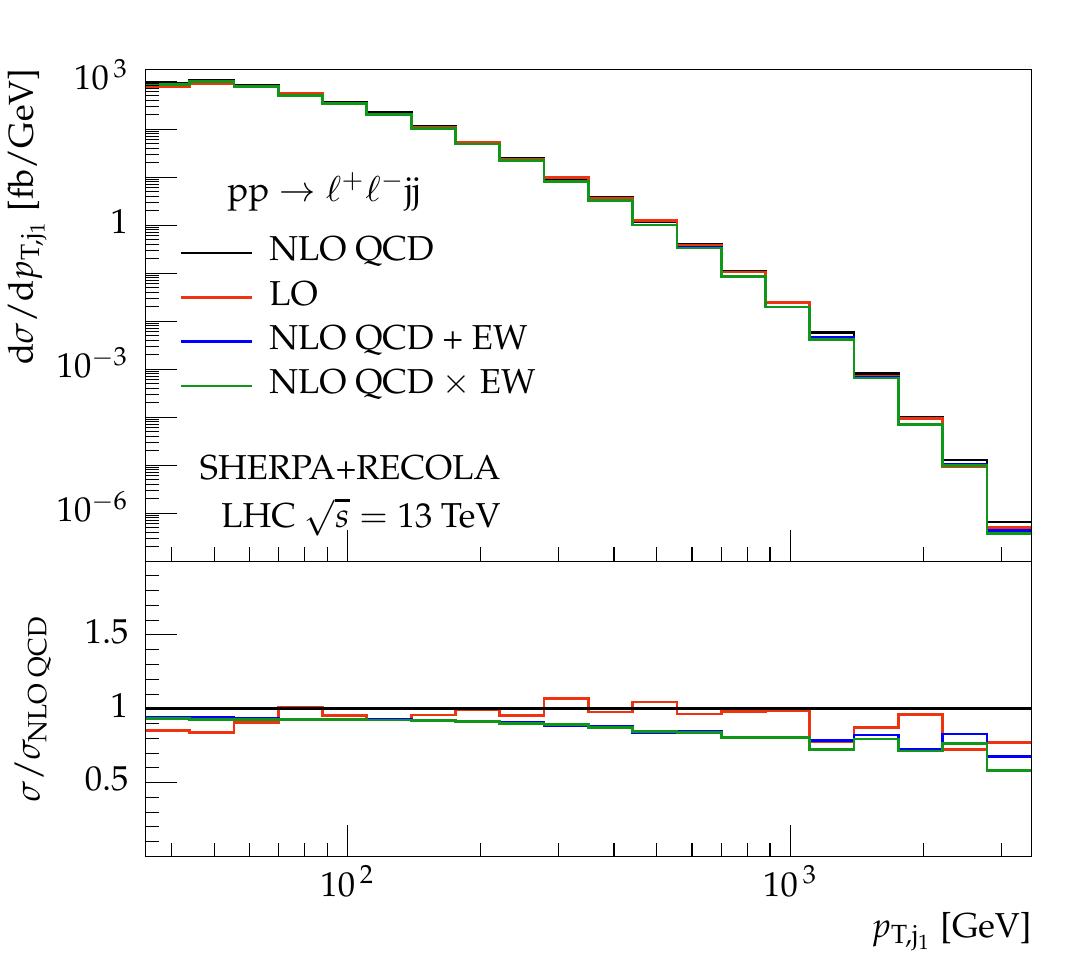}
 \caption{\label{FIG:pTslljj} Differential distributions at $13\TeV$ for 
 $\Pp\Pp\to\ell^+\ell^-{\rm j}{\rm j}$ at the LHC. The left-hand plot
 shows the $p_{\rm T}$ of the hardest lepton and the right-hand plot
 the $p_{\rm T}$ of the hardest jet. The LO and NLO QCD distributions are
 plotted along with both the additive (NLO QCD + EW) and multiplicative
 (NLO QCD $\times$ EW) prescriptions for combining the NLO corrections.
 The ratio of the distributions with respect to NLO QCD is presented in the
 lower panels.}
\end{figure}

Secondly,  ${\rm Z}$+jets production is considered. Distributions have been 
published for off-shell ${\rm Z}$+jets processes in 
Ref.~\cite{Kallweit:2015dum}. Although this is not such a rigorous 
validation as the total cross section
or phase-space point comparisons for other processes, it allows a 
qualitative assessment of the agreement between the calculations
across a large phase space. 
From the wide range of 
distributions presented in Ref.~\cite{Kallweit:2015dum}, we select here the distributions in the transverse momenta of
the leading lepton, $p_{{\rm T}, {\ell}_1}$, and leading jet,  $p_{{\rm T}, {\rm j}_1}$, (according to $p_{{\rm T}}$ ordering) 
(Fig.~\ref{FIG:pTslljj}). The $\Pp\Pp\to\ell^+\ell^-{\rm jj}$ process 
provides a particularly
good test of the \Sherpa\!\!+\Recola interface, because it 
includes interference terms between EW and QCD produced jets
contributing at NLO EW. This makes the process a lot more
challenging than the $\ell^+\ell^-{\rm j}$ or $\ell\nu {\rm j}$ final state.
 For both plots in Fig.~\ref{FIG:pTslljj}, the behaviour across the full
$p_{\rm T}$ range is in agreement with the observations in Ref.~\cite{Kallweit:2015dum}.

\paragraph{Combined predictions:}
Table~\ref{TABLE:VjetsPredictions} presents combined predictions for NLO QCD and
EW corrections to the integrated cross sections for
 ${\rm Z}/\ell^+\ell^-+{\rm jets}$ processes. These cross sections include all 
of the cuts from the analyses, and correspond to the 
distributions presented in this section. The on-shell calculation of 
$\PZ+{\rm jets}$ includes the branching ratios to $\Pe^+\Pe^-/\mu^+\mu^-$.
The large NLO corrections are dominated, in all cases, by the NLO QCD 
contribution.

\begin{table}
\begin{center}
\begin{tabular}{|c|c|c|c|c|c|}
\hline \rule[-1.4ex]{0ex}{3.8ex}%
 inclusive & $\sigma^\mathrm{LO}$~[pb] & $\sigma^\mathrm{NLO}_{\mathrm{QCD}+\mathrm{EW}}$~[pb] & $\delta^\mathrm{NLO}_{\mathrm{QCD}+\mathrm{EW}}$~[\%] & $\sigma^\mathrm{NLO}_{\mathrm{QCD}\times\mathrm{EW}}$~[pb] & $\delta^\mathrm{NLO}_{\mathrm{QCD}\times\mathrm{EW}}$~[\%]\\
\hline \rule{0ex}{2.4ex}%
$\Pp \Pp \to \PZ {\rm j}$ & $1.97 \cdot 10^{2}$ & $2.68 \cdot 10^{2}$ & $36.2$ & $2.72 \cdot 10^{2}$ & $38.3$ \\
$\Pp \Pp \to \PZ {\rm jj}$ & $6.66 \cdot 10^{1}$ & $7.70 \cdot 10^{1}$ & $15.6$ & $7.68 \cdot 10^{1}$ & $15.3$ \\
$\Pp \Pp \to \ell^+\ell^-{\rm j}$ & $1.93 \cdot 10^{2}$ & $2.34 \cdot 10^{2}$ & $21.6$ & $2.31 \cdot 10^{2}$ & $19.7$ \\
 \rule[-1.4ex]{0ex}{1.8ex}%
$\Pp \Pp \to \ell^+\ell^- {\rm jj}$ & $5.53 \cdot 10^{1}$ & $6.93 \cdot 10^{1}$ & $0.04$ & $6.88 \cdot 10^{1}$ & $0.03$ \\
\hline
\end{tabular}
\end{center}
\caption{\label{TABLE:VjetsPredictions}
Integrated cross sections for ${\rm p} {\rm p} \to {\rm Z+jets} $ for a centre-of-mass energy of $\sqrt{s} = 13\TeV$ calculated with \Sherpa\!\!+\Recola.
The cross sections at LO as well as for the additive and multiplicative combinations of NLO QCD and EW corrections are given.
The cross sections are expressed in pb while the relative corrections are given in percent.
}
\end{table}

\begin{figure}
 \includegraphics[width=0.5\textwidth]{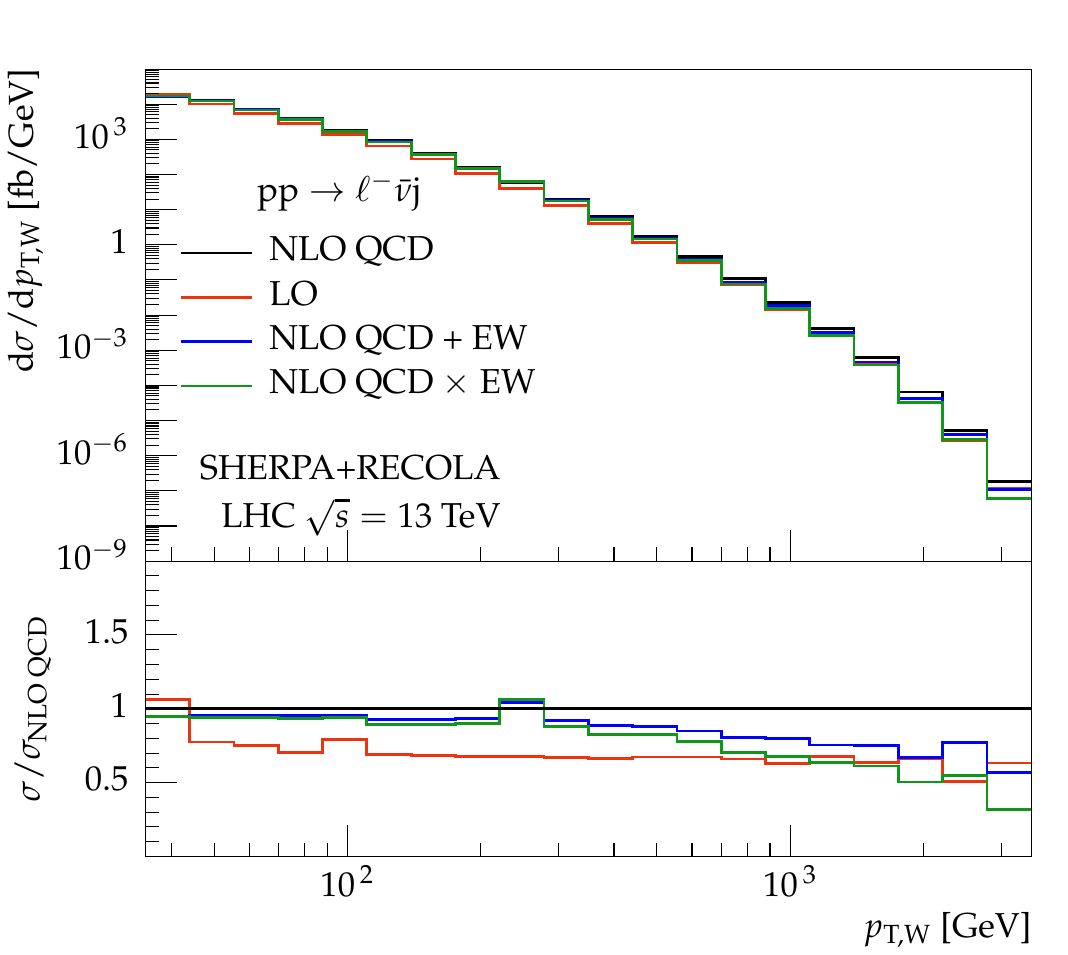}
 \includegraphics[width=0.5\textwidth]{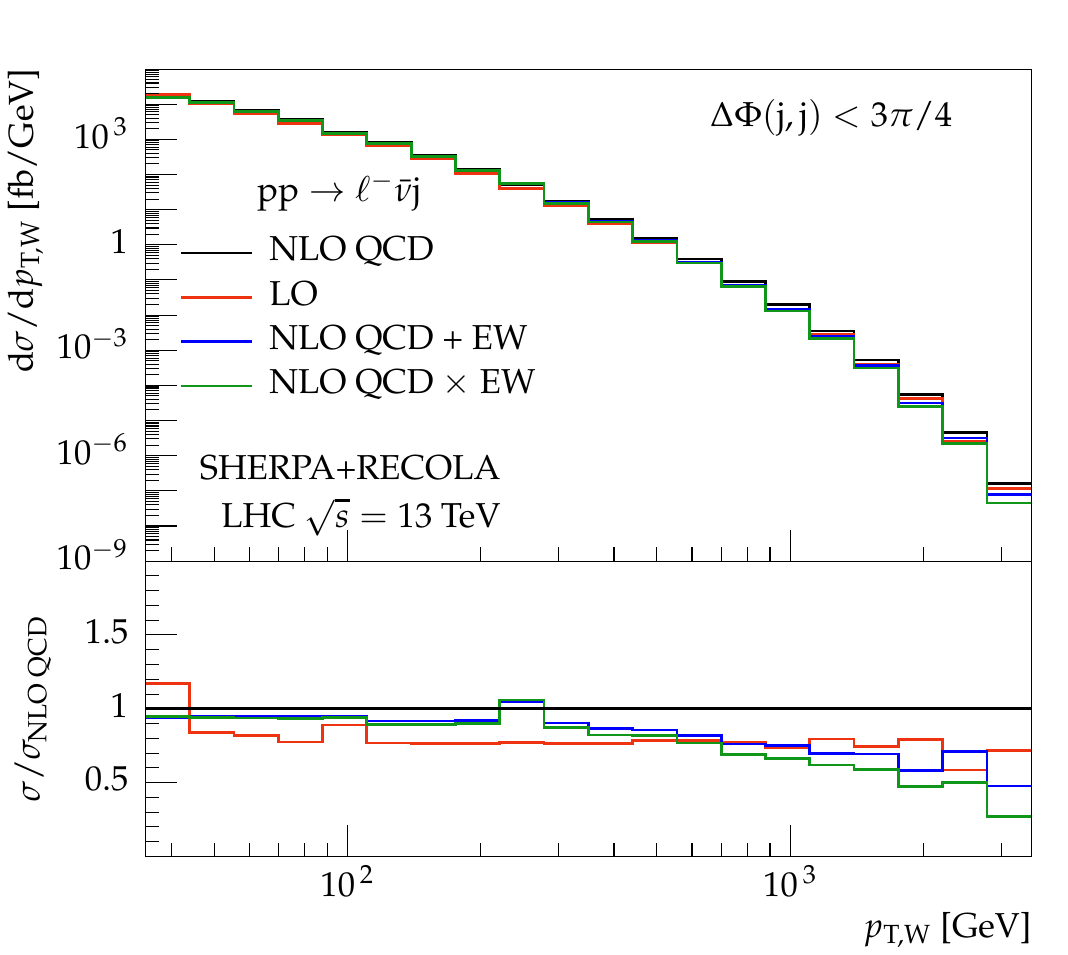}
 \caption{\label{FIG:offshellWpt} Differential distributions at $13\TeV$
 for off-shell ${\rm W}+{\rm j}$ production at the LHC. The left-hand
  plot shows the $p_{\rm T}$ of the reconstructed ${\rm W}$ boson with
  minimal cuts and the right-hand plot the result with a cut of 
  $\Delta\Phi({\rm j},{\rm j})<3\pi/4$.}
\end{figure}
\begin{figure}
 \includegraphics[width=0.5\textwidth]{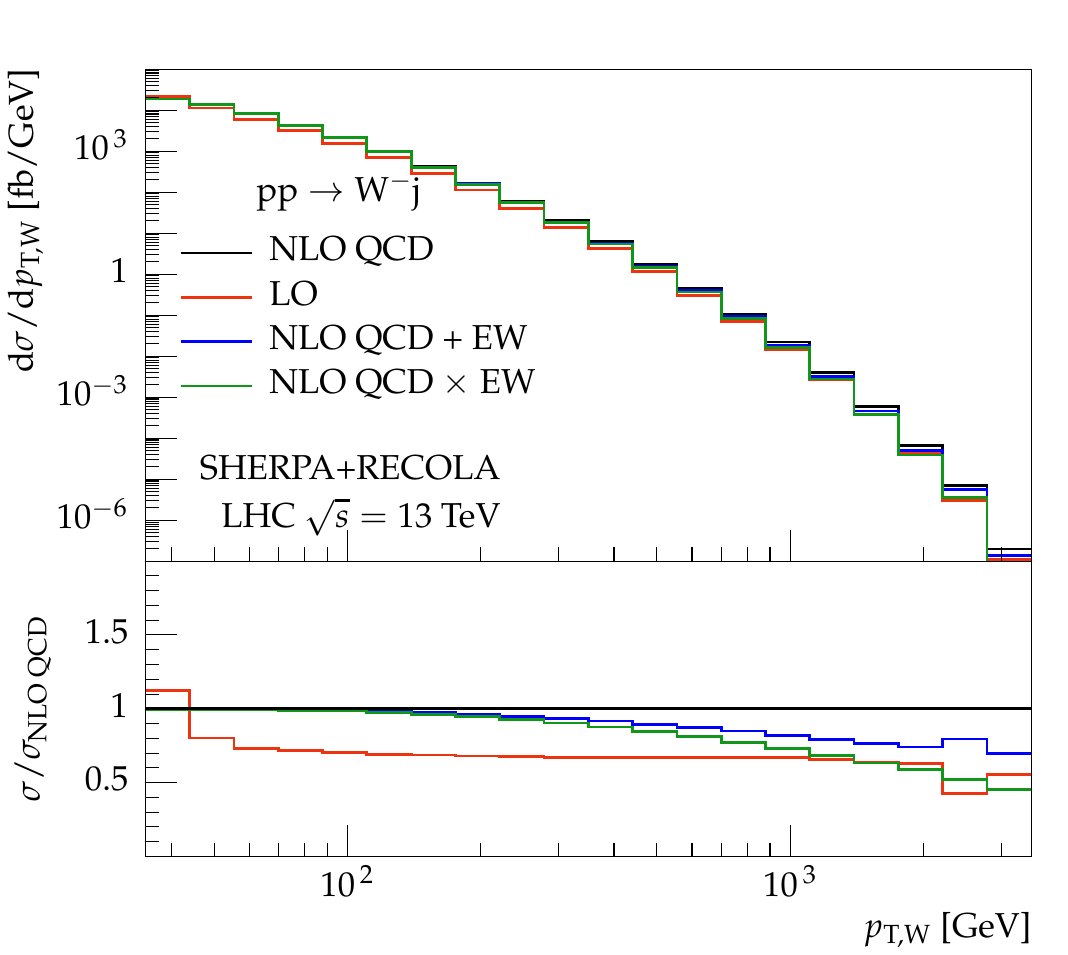}
 \includegraphics[width=0.5\textwidth]{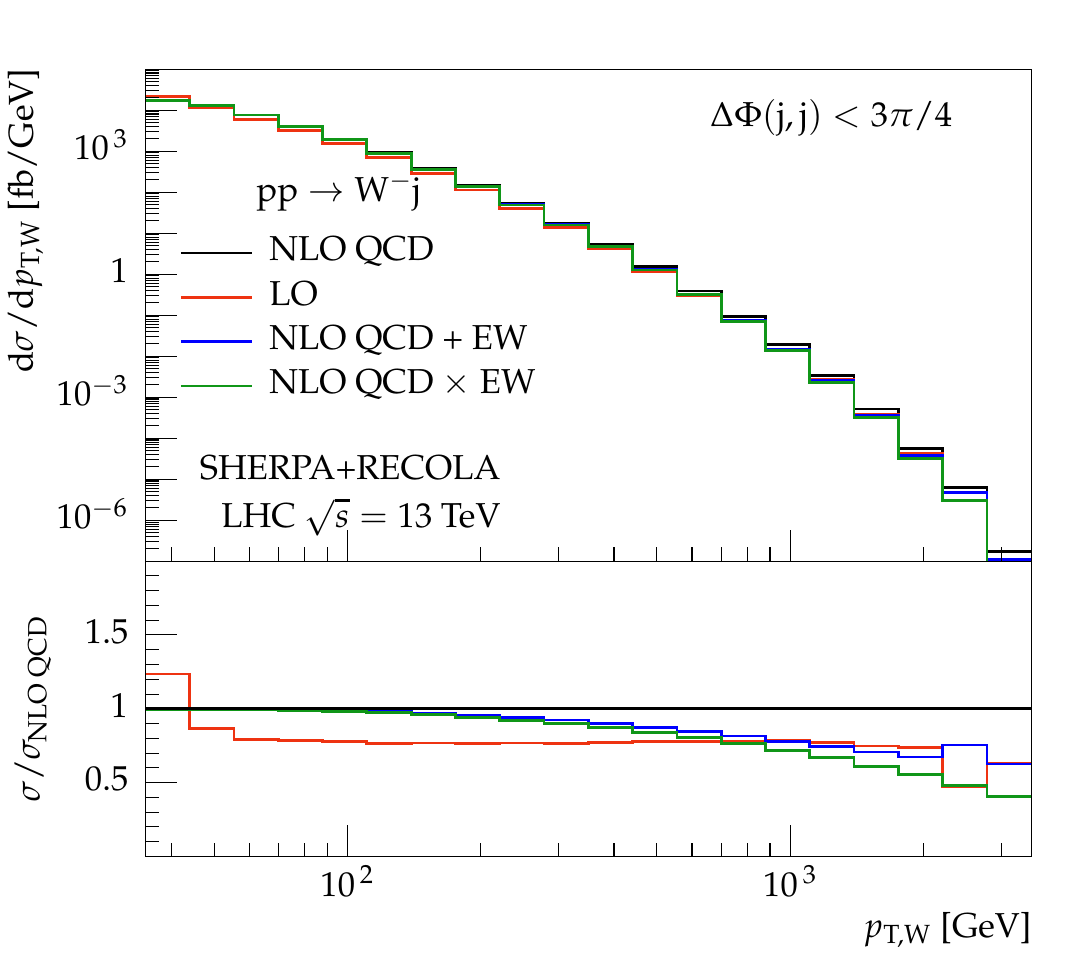}
 \caption{\label{FIG:onshellWpt} Differential distributions at $13\TeV$
 for on-shell ${\rm W}+{\rm j}$ production at the LHC. The left-hand
  plot shows the $p_{\rm T}$ of the ${\rm W}$ boson with
  minimal cuts and the right-hand plot the result with a cut of 
  $\Delta\Phi({\rm j},{\rm j})<3\pi/4$, as in 
  Fig.~\ref{FIG:offshellWpt}.}
\end{figure}

Beginning again with ${\rm W+j}$ production, distributions
at a centre-of-mass energy of $13\TeV$ at the LHC are presented for both on-shell and off-shell ${\rm W+j}$ production.
Figure~\ref{FIG:offshellWpt} shows the transverse
momentum $p_{\rm T,\PW}$ of the reconstructed W~boson for both additive
and multiplicative combination of NLO QCD and EW effects. The prescription used to
combine the NLO corrections clearly
has a large effect in the high-$p_{\rm T}$ region of the plot, which
is indicative of large higher-order corrections. The right-hand plot shows the
effect of imposing a cut $\Delta\Phi({\rm j},{\rm j})<3\pi/4$ on the two jets in
the NLO case. This removes the contribution from the production of two
hard jets and a soft ${\rm W}$~boson. At NLO EW, such cuts are useful in order
to differentiate processes which should more correctly be
considered as an EW real correction to dijet production.

Similarly, Fig.~\ref{FIG:onshellWpt} shows the distributions for on-shell 
${\rm W}+\Pj$ production.
The decays to leptons are treated in a factorised approach, 
and the NLO QCD and NLO EW corrections are applied only to the on-shell ${\rm W+j}$
final state. The NLO QCD and NLO EW corrections show very similar behaviour,
whether on-shell or off-shell ${\rm W}$-boson production is considered.

\begin{figure}
 \includegraphics[width=0.5\textwidth]{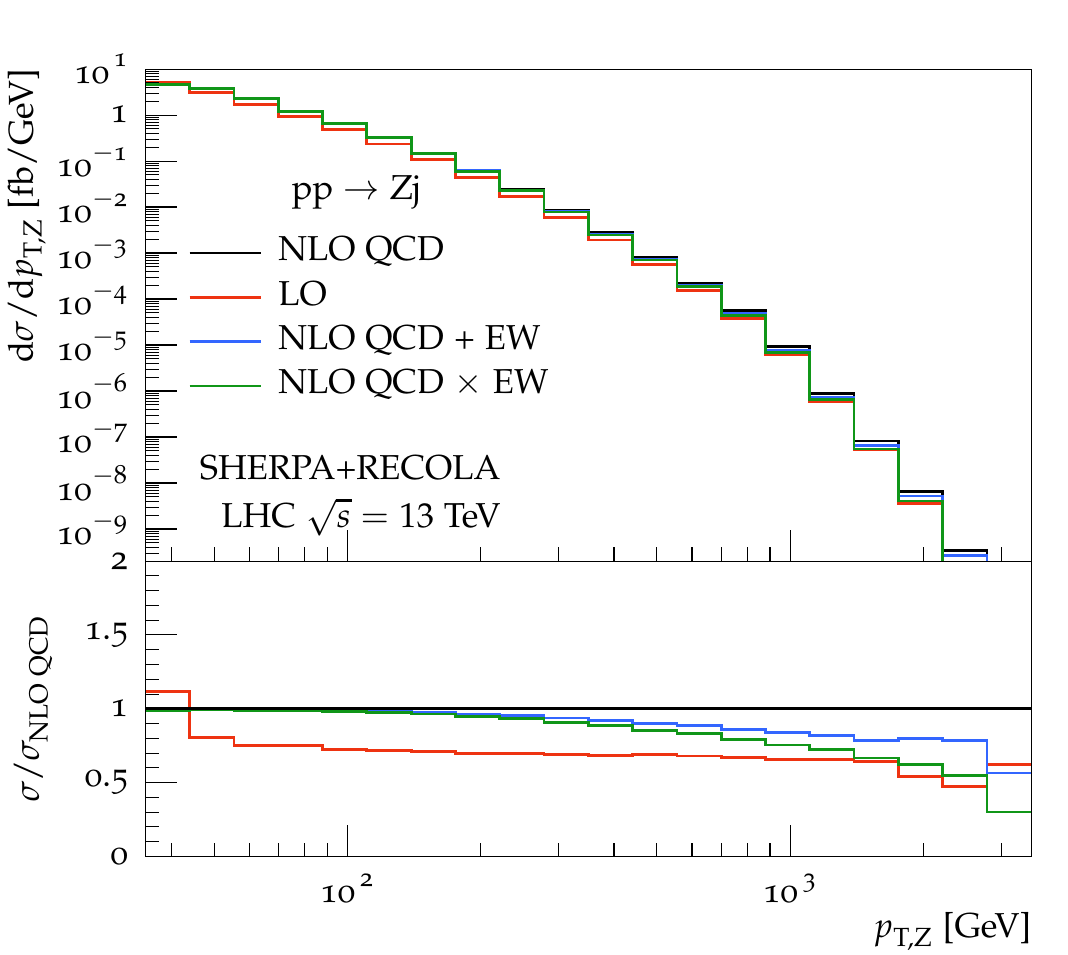}
 \includegraphics[width=0.5\textwidth]{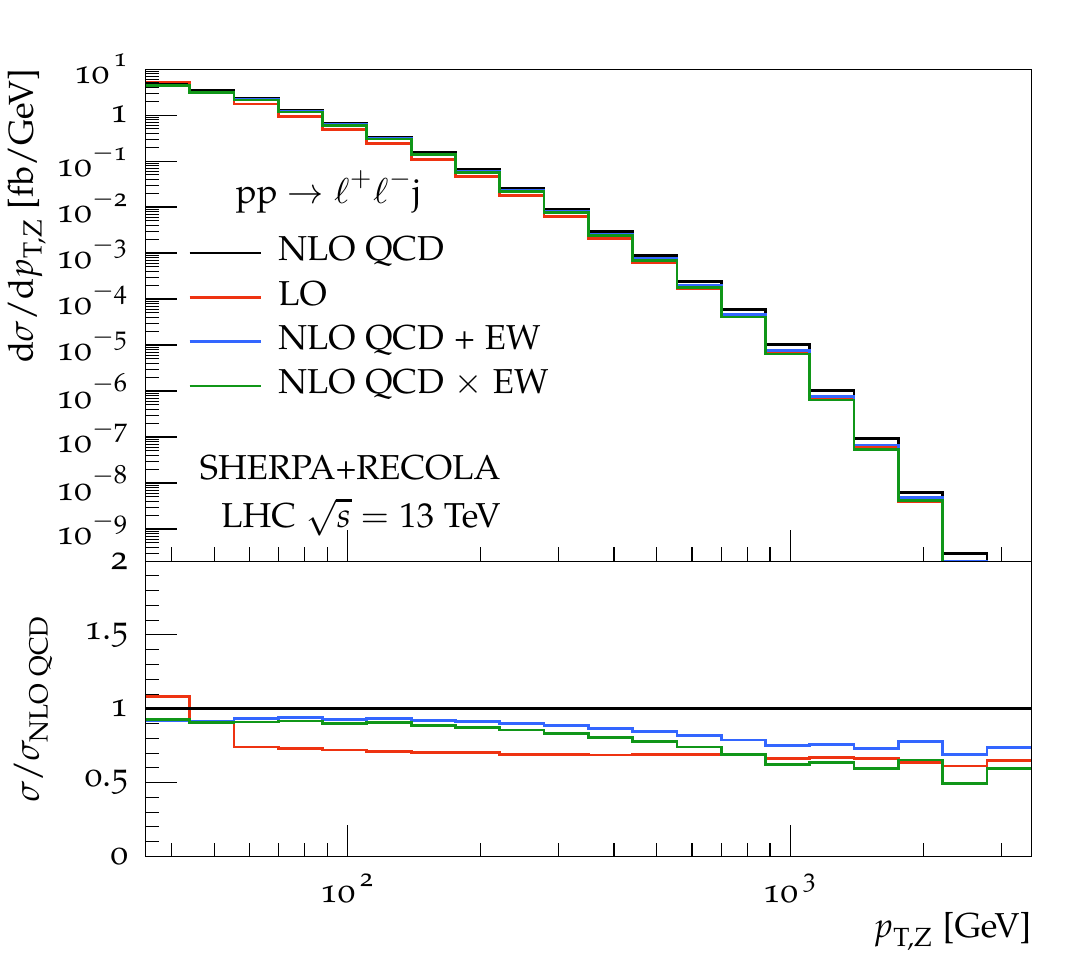}
 \caption{\label{FIG:ZptNLOEWZj} Differential distributions at $13\TeV$
 for ${\rm Z}+{\rm j}$ production at the LHC. The left-hand
  plot shows the $p_{\rm T}$ of the ${\rm Z}$ boson for on-shell 
  ${\rm Z}$
  production and the right-hand plot the same observable for the
  reconstructed ${\rm Z}$ boson for off-shell 
  ${\rm Z}$ production. The lower panels display the ratio of the distributions
  to the NLO QCD result.}
\end{figure}

\begin{figure}
 \includegraphics[width=0.5\textwidth]{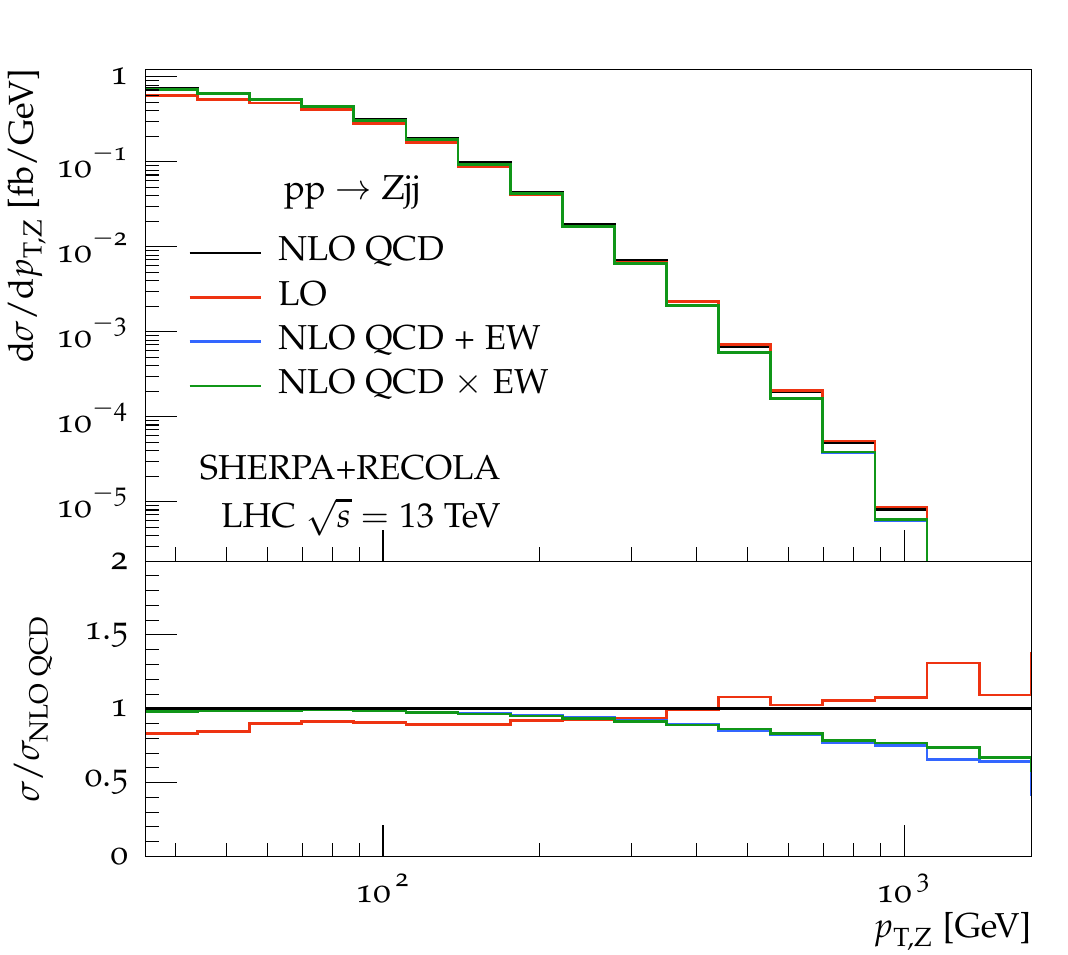}
 \includegraphics[width=0.5\textwidth]{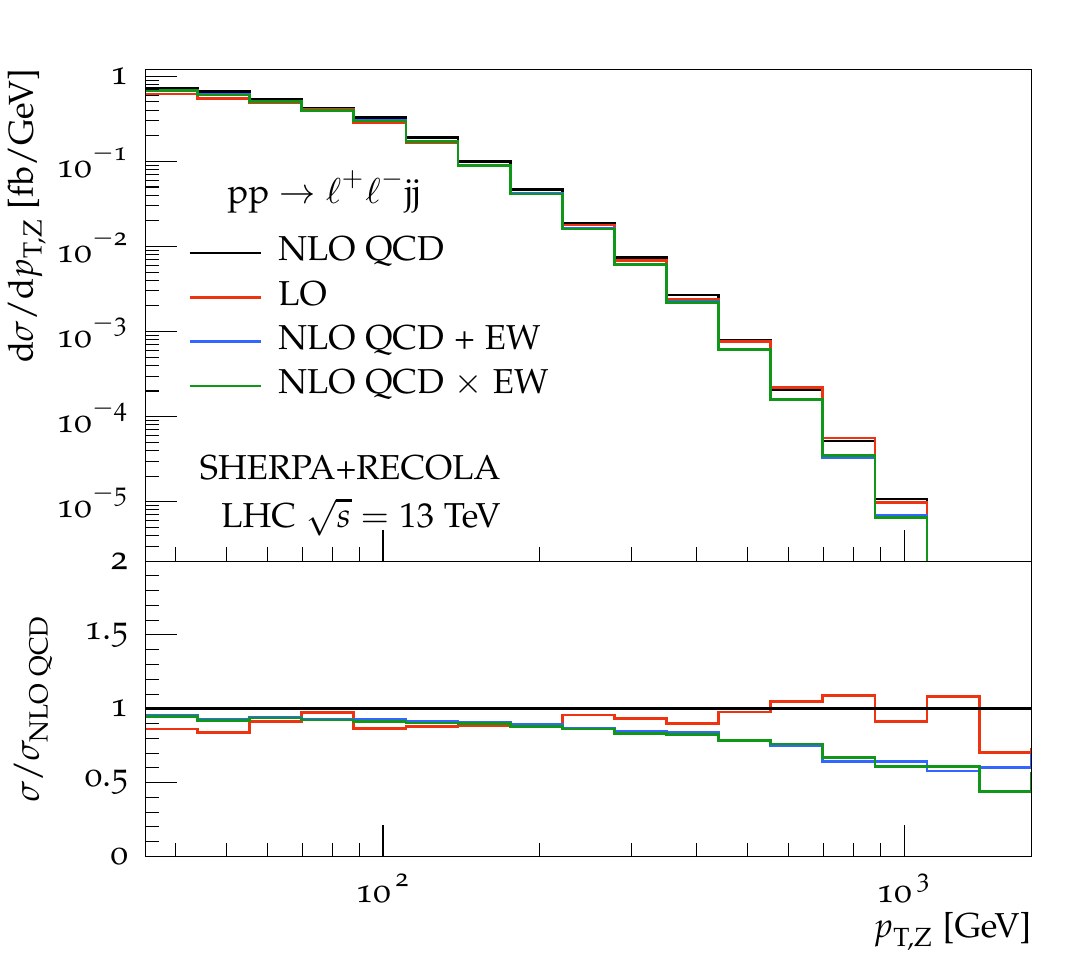}
 \caption{\label{FIG:ZptNLOEWZjj} Differential distributions at $13\TeV$
 for ${\rm Z}+{\rm jj}$ production at the LHC. The left-hand
  plot shows the $p_{\rm T}$ of the reconstructed ${\rm Z}$ boson for on-shell 
  ${\rm Z}$
  production and the right-hand plot the same observable for off-shell 
  ${\rm Z}$ production. The lower panels display the ratio of the distributions
  to the NLO QCD result.}
\end{figure}

Results are also presented for ${\rm Z}$+jets production in
Figs.~\ref{FIG:ZptNLOEWZj} and \ref{FIG:ZptNLOEWZjj}, but this
time considering both the $1$-jet and $2$-jet channels as well as
the on-shell vs.\ off-shell effects.
The off-shell process naturally
includes the effects from $\gamma^*\to\ell^+\ell^-$ interference.
 As was mentioned in the validation
section for this process, the ${\rm Z}+2\,$jets final state introduces 
interference terms between NLO QCD and NLO EW, which are taken
into account automatically.
These plots display the (reconstructed) $p_{\rm T}$ of the 
${\rm Z}$ boson. The distribution for $\ell^+\ell^-{\rm jj}$ shown in Fig.~\ref{FIG:ZptNLOEWZjj}, 
was also presented in Ref.~\cite{Kallweit:2015dum}, and can therefore be 
additionally viewed as a further validation of the \Sherpa\!\!+\Recola
interface.

Figure~\ref{FIG:ZptNLOEWZj} presents the ${\rm Z}$-boson $p_{\rm T}$ in ${\rm Z+j}$ production at a 
$13\TeV$ LHC. Here, there is a larger impact from the NLO EW corrections
in the low-$p_{\rm T}$ region for $\ell^+\ell^-{\rm j}$ production than for on-shell ${\rm Z+j}$
production. This effect is smaller  for the $\ell^+\ell^-{\rm jj}$
final state shown in 
Fig.~\ref{FIG:ZptNLOEWZjj}, where little difference is observed 
between on-shell and off-shell ${\rm Z}$-boson
production across the entire phase space. In all cases, the Sudakov behaviour in the large-$p_{\rm T}$ region
is clearly observed. Also, the NLO QCD + EW and NLO QCD $\times$ EW curves 
for ${\rm Z+jj}$ in Fig.~\ref{FIG:ZptNLOEWZjj}
show a very good agreement with each other, indicating that the higher-order corrections
in this case are a lot smaller than for the distributions of ${\rm Z+j}$ production, where a
large difference is observed in the high-$p_{\rm T}$ Sudakov region. 
It is reassuring that these differences are largely removed
for both the on-shell and off-shell ${\rm Z}$+jets processes once higher jet
multiplicities are included. This implies that a merged NLO QCD and EW
sample would give an accurate picture of both NLO QCD and NLO EW 
corrections to this process.

\subsection{${\rm Z}$-boson pair production}
\label{sec:zzvalidation}

As second process we consider the production of two off-shell Z bosons with subsequent decays into pairs of different-flavour charged leptons.
At leading order (LO), this is a purely EW process implying that interferences in different orders of the strong and EW coupling first occur at NNLO.
As a consequence, the NLO QCD corrections at ${\cal{O}}(\alphas$) and the NLO EW corrections at ${\cal{O}}(\alpha)$ may be computed independently.
The NLO QCD corrections are known \cite{Ohnemus:1990za,Mele:1990bq,Dixon:1999di,Campbell:1999ah}, and the complete NLO EW computations have recently been published \cite{Biedermann:2016yvs,Biedermann:2016lvg}.

\paragraph{Input parameters:} We use the same set-up as for the QCD
validation of Z-boson pair production in Section \ref{sec:NLOQCDfixedorder}.
In addition, real photons from QED bremsstrahlung and charged leptons are recombined to dressed leptons if their separation in the rapidity--azimuthal-angle plane fulfils $\Delta R_{\ell\gamma} < 0.2$, following the prescription of \citere{Biedermann:2016lvg}. In contrast to Ref.~\cite{Biedermann:2016lvg}, we refrain from including photon-induced contributions. 

\paragraph{NLO EW validation:}
In \refta{tab:ZZEWValidation}, a comparison of the NLO EW total cross section is shown once obtained from \Sherpa\!\!+\Recola and once by combining \Recola with an independent private multi-channel Monte Carlo integrator, \ie with the results from Ref.~\cite{Biedermann:2016lvg}.
Perfect agreement is found between the two calculations within the statistical uncertainty at (sub-)permille level.
\begin{table}
\begin{center}
\begin{tabular}{|l|c|c|c|}
\hline \rule[-1.4ex]{0ex}{3.8ex}%
$\Pp\Pp\to\mu^+\mu^-\Pe^+\Pe^-$ & \Sherpa\!\!+\Recola [fb] & private MC+\Recola [fb] & std. dev. [$\sigma$] \\
\hline\rule{0ex}{2.4ex}%
$\sigma^{\rm LO}$ & $11.498(1)$ & $11.4964(1)$ & $1.6$ \\
 \rule[-1.4ex]{0ex}{1.8ex}%
$\sigma^{\rm NLO}_{\rm EW}$ & $10.890(1)$ & $10.8888(2)$ & $1.2$ \\
\hline
\end{tabular}
\end{center}
\caption{\label{tab:ZZEWValidation} Total cross sections calculated for ${\rm pp}\to\mu^+\mu^-{\rm e}^+{\rm e}^-$ at LO and NLO EW with the \Sherpa\!\!+\Recola interface, compared against the benchmark numbers from Ref.~\cite{Biedermann:2016lvg}. 
The difference is expressed in standard deviations.}
\end{table}

\bfig
  \center
  \includegraphics[width=0.45\textwidth]{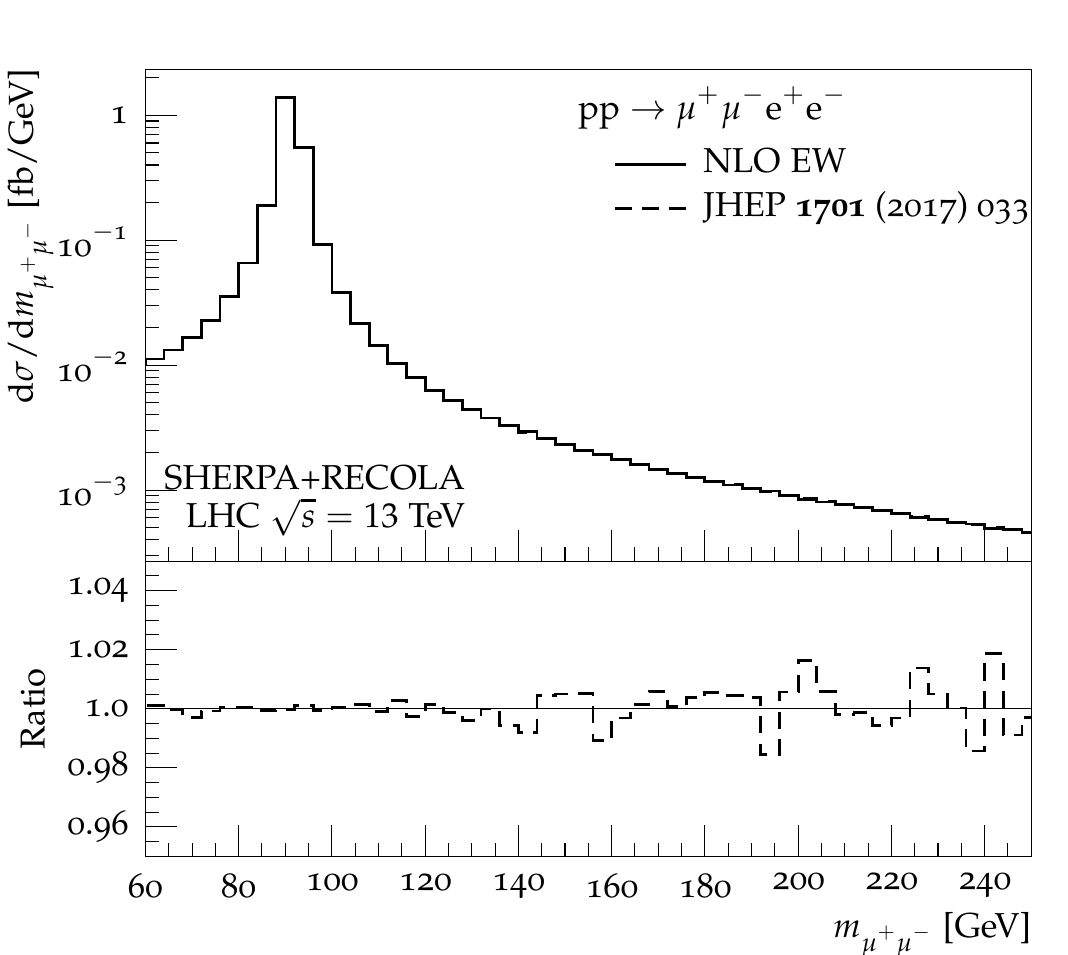}
  \includegraphics[width=0.45\textwidth]{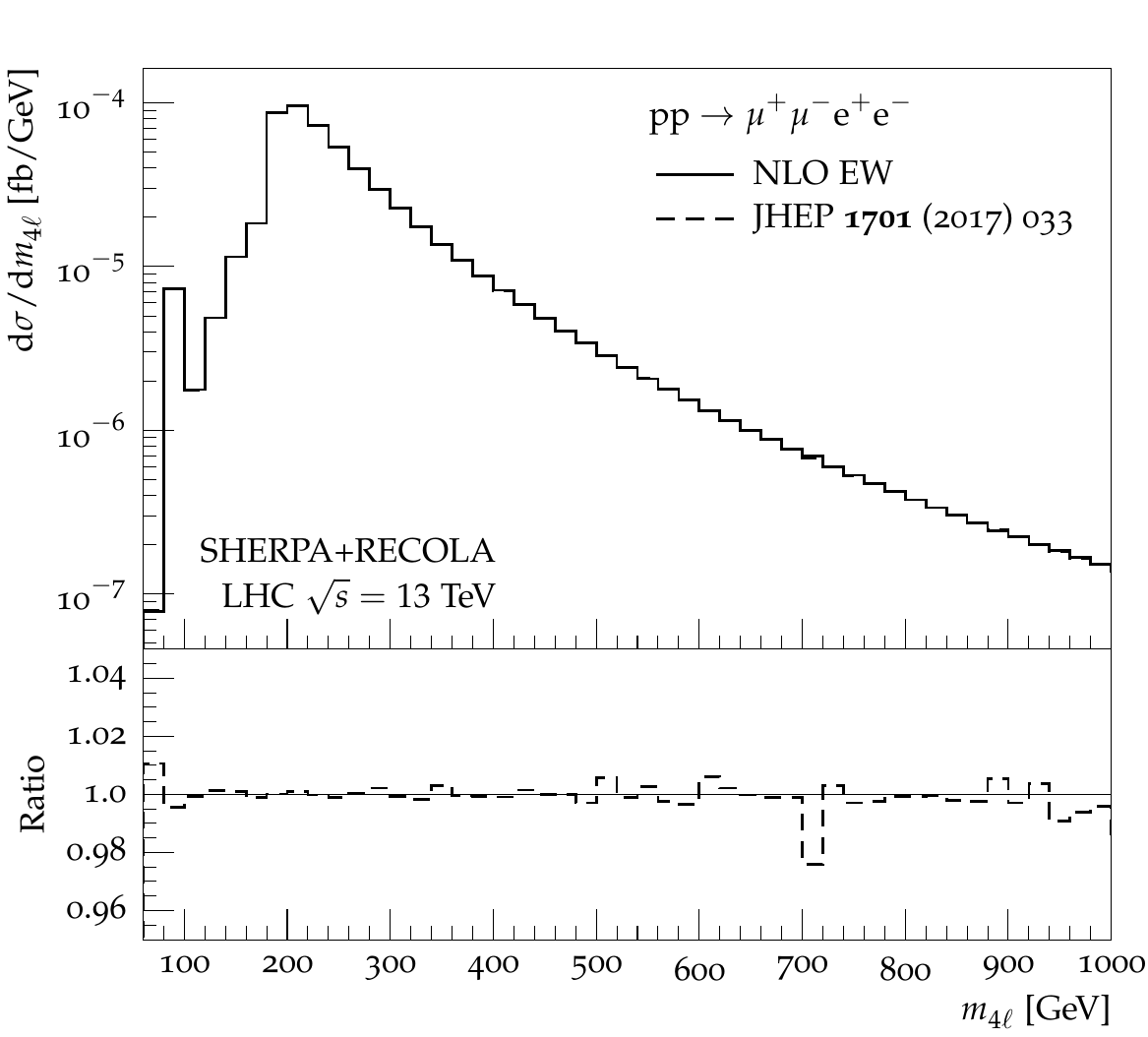}
  \caption{\label{FIG:SHERPAVSINHOUSEMC} Comparison of the di-muon and four-lepton invariant mass in the process ${\rm pp}\to\mu^+\mu^-{\rm e}^+{\rm e}^-$ at NLO EW with the \Sherpa\!\!+\Recola interface compared against the benchmark numbers from Ref.~\cite{Biedermann:2016lvg}. The upper panel shows the absolute prediction, while the lower panel indicates the ratio between both computations.}
\efig
In Fig.~\ref{FIG:SHERPAVSINHOUSEMC}, a comparison of the two independent calculations at the level of NLO EW differential distributions via the di-muon mass $m_{\mu^+\mu^-}$ and the four-lepton invariant mass $m_{4\ell}$ is presented.
Like in the corresponding QCD validation in Section~\ref{sec:NLOQCDfixedorder}, the difference in the absolute prediction in the upper panel is almost invisible. The ratio of individual histogram bins in the lower panel shows statistical percent-level fluctuations which illustrate again the excellent agreement.
This is a highly non-trivial check, since the benchmark calculation of \citere{Biedermann:2016lvg} has been cross-checked internally by two independent calculations both at the level of the employed matrix elements and at the level of the phase-space integration.
Furthermore, the results from \citere{Biedermann:2016lvg} were generated in mass regularisation and slightly different conventions of the dipole subtraction terms.

\paragraph{Combined predictions:}
The combined predictions for the total cross section including the QCD corrections from Section~\ref{sec:ZZQCDvalidation} and the EW corrections from this section are stated in Tab.~\ref{TABLE:ZZPredictions}. 
\begin{table}
\begin{center}
\begin{tabular}{|c|c|c|c|c|c|}
\hline \rule[-1.4ex]{0ex}{3.8ex}%
 ${\rm pp} \to {\mu^+\mu^-\Pe^+\Pe^-}$ & $\sigma^\mathrm{LO}$ &
$\sigma^\mathrm{NLO}_{\mathrm{QCD}+\mathrm{EW}}$ & $\delta^\mathrm{NLO}_{\mathrm{QCD}+\mathrm{EW}}$ &
$\sigma^\mathrm{NLO}_{\mathrm{QCD}\times\mathrm{EW}}$ &
$\delta^\mathrm{NLO}_{\mathrm{QCD}\times\mathrm{EW}}$\\
\hline \rule{0ex}{2.4ex}%
inclusive & $11.498(1)\,$fb & $15.18(1)\,$fb & $32.0(1)\,$\% & $14.96(1)\,$fb & $30.1(1)\,$\% \\
\hline
\end{tabular}
\end{center}
\caption{\label{TABLE:ZZPredictions}
Integrated cross sections for ${\rm p} {\rm p} \to {\mu^+\mu^-\Pe^+\Pe^-}$ for a centre-of-mass energy of
$\sqrt{s} = 13\TeV$, calculated with \Sherpa\!\!+\Recola for the set-up of Ref.~\cite{Biedermann:2016lvg}.
The cross sections at LO as well as for the additive and multiplicative combinations of NLO QCD and EW
corrections are given.
The cross sections are expressed in fb while the relative corrections are given in percent.
The integration errors of the last digits are provided in parentheses.}
\end{table}
In Fig.~\ref{FIG:ZZNLOEW},
different NLO predictions are presented for distributions in the
di-muon and four-lepton invariant mass. In the four-lepton
invariant-mass distribution, we observe the typical pattern of large
negative EW corrections of around $-20\%$ in the high-energy regime at
around $1\TeV$. The radiative tail below the pair-production threshold
at $m_{4\ell}=2M_\PZ$ with corrections around $+30\%$ is due to the
fact that resonant contributions are shifted to lower values by real photon radiation. Since the LO cross section is falling off steeply in this region, the photonic corrections become large. A similar radiative tail is observed also in the di-muon prediction that amounts to positive corrections of up to $+60\%$.
\bfig
  \center
  \includegraphics[width=0.45\textwidth]{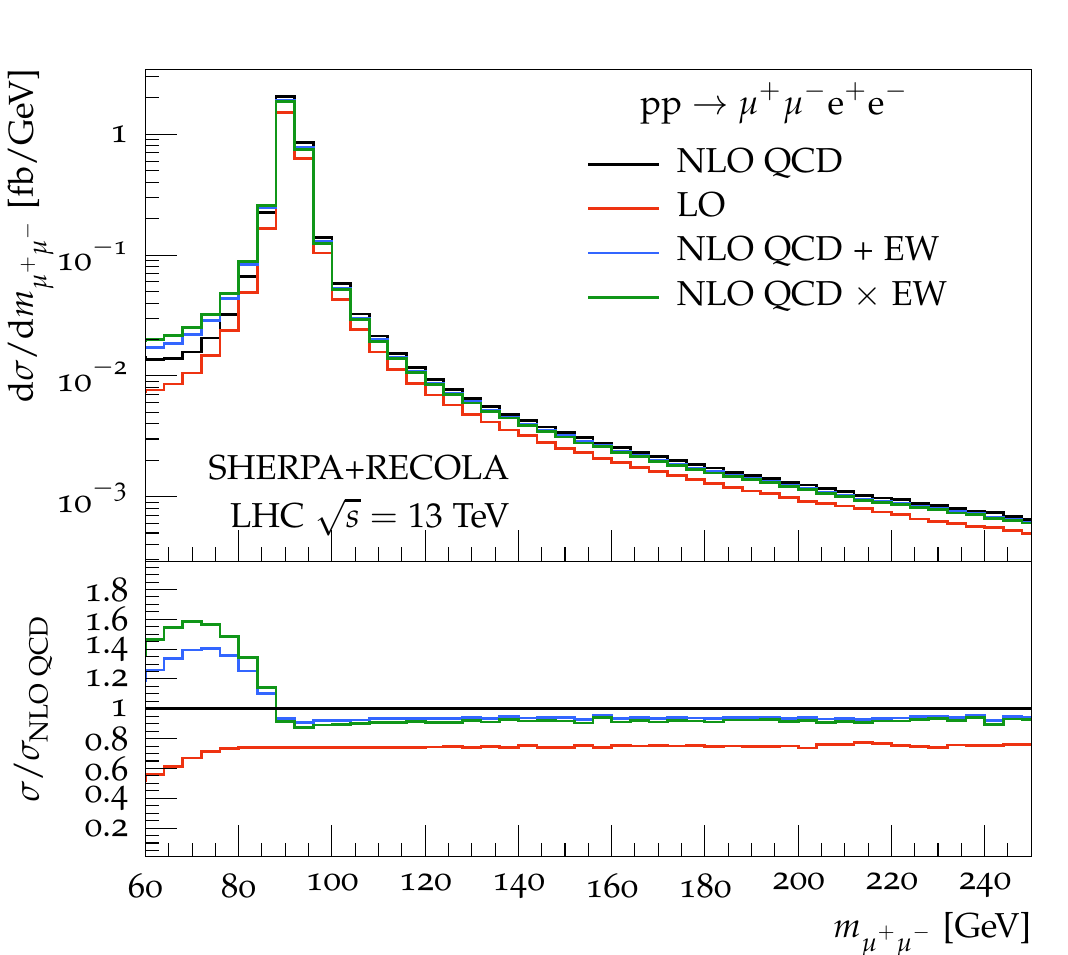}
  \includegraphics[width=0.45\textwidth]{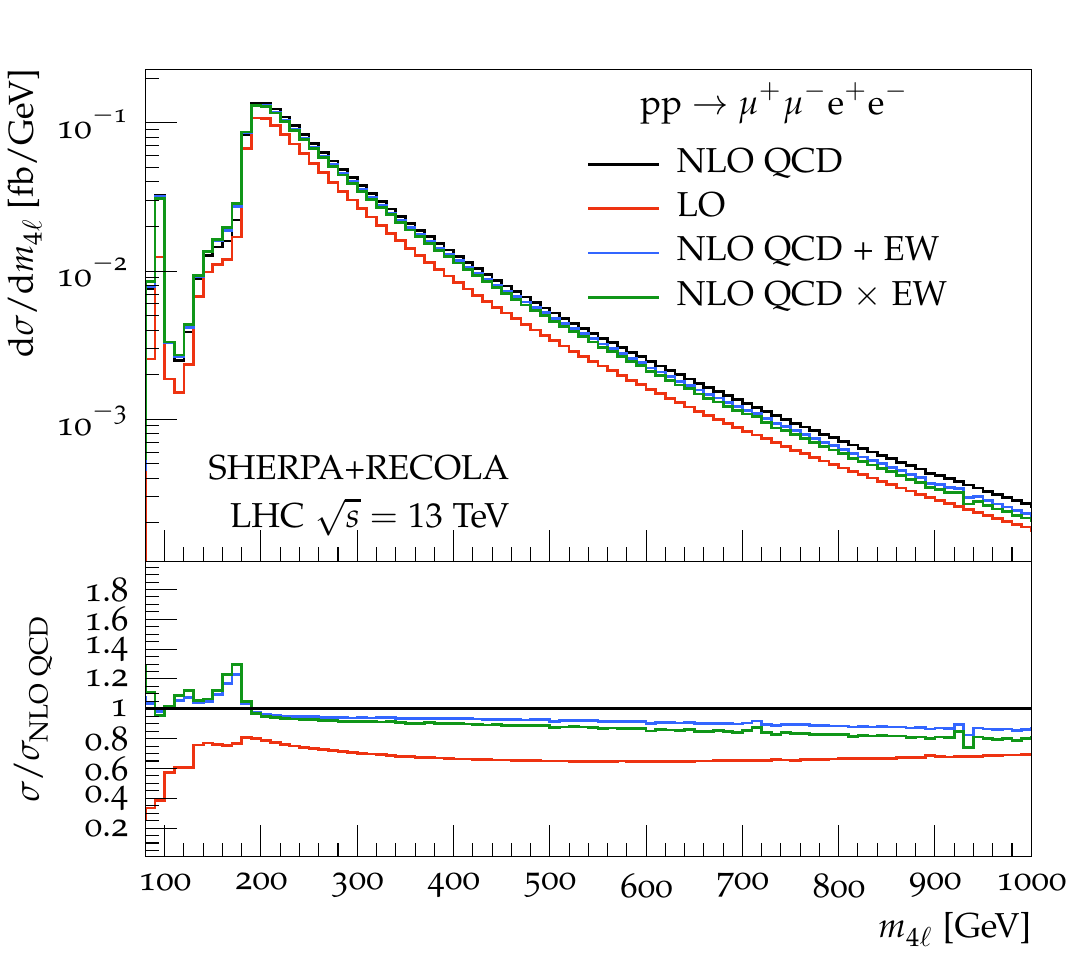}
  \caption{\label{FIG:ZZNLOEW}
 Distributions of the di-muon and four-lepton invariant mass in the
 process ${\rm pp}\to\mu^+\mu^-{\rm e}^+{\rm e}^-$. Results are given for LO, NLO QCD, NLO QCD$+$EW and NLO QCD$\times$EW.  }
\efig
While the QCD corrections are positive over the whole range of both
the di-muon and the four-lepton invariant-mass distribution at the
order of $+30\%$, the EW corrections exhibit a non-trivial sign change
at the Z-boson resonance $m_{\mu^+\mu^-}=M_\PZ$ and  at the pair-production threshold $m_{4\ell}=2M_\PZ$, respectively. Further discussion of this issue can be found in Ref.~\cite{Biedermann:2016lvg}.

\subsection{Higgs production in association with a top-quark pair}

As last example, we consider the on-shell production of a pair of top quarks in association with a Higgs boson.
The LO cross section is of order $\order{\alphas^2\alpha}$. Hence the NLO QCD and EW corrections contribute
at order $\order{\alphas^3\alpha}$ and $\order{\alphas^2\alpha^2}$, respectively.
At NLO EW, this computation features QCD--EW interferences. In addition to computing the EW corrections to
the QCD-mediated production of the top-quark pairs, one must
also consider the QCD corrections to the interference of the QCD and electroweakly produced top-quark pairs.
Moreover, the final state consists exclusively of massive particles which is not the case for
any of the previously presented processes. This process constitutes thus a non-trivial validation of
the implementation. Concerning on-shell top quarks, the process has already been computed at NLO
QCD \cite{Beenakker:2001rj,Beenakker:2002nc,Reina:2001sf,Dawson:2002tg,Dawson:2003zu} and at NLO
EW \cite{Frixione:2014qaa,Yu:2014cka,Frixione:2015zaa}.  It has also
been matched to a parton
shower \cite{Frederix:2011zi,Garzelli:2011vp,Hartanto:2015uka}.
On the other hand, for off-shell top quarks, the NLO QCD \cite{Denner:2015yca} and EW \cite{Denner:2016wet} corrections have been computed only recently.

\paragraph{Input parameters:} We use the set-up of Ref.~\cite{Badger:2016bpw} which has been described in Section~\ref{sec:tthQCD}.
Concerning the electromagnetic coupling $\alpha$, the $\alpha(M_{\rm
  Z})$ scheme is employed.
Note that contributions originating from initial-state photons are neglected in order to match one of the set-ups of Ref.~\cite{Badger:2016bpw}.

\paragraph{NLO EW validation:}
As for the QCD validation, we compare five NLO EW cross sections that
have been computed by \madgraph and \Sherpa\!\!+\OpenLoops in Ref.~\cite{Badger:2016bpw}.
The obtained cross sections are reported in \refta{table:ttHNLOEW}.
\begin{table}
\begin{center}
\begin{tabular}{|c|c|c|c|c|}
\hline \rule{0ex}{2.4ex}%
& \multicolumn{2}{c|}{\Sherpa\!\!+\Recola} & \madgraphbis & \Sherpa\!\!+\OpenLoops \\
\cline{2-5} \rule[-1.3ex]{0ex}{3.8ex}%
 & $\sigma^\mathrm{NLO}_{\rm EW}$~[fb] & $\delta_{\rm EW} [\%]$ & {$\delta_{\rm EW} [\%]$} & $\delta_{\rm EW} [\%]$ \\
\hline \rule{0ex}{2.4ex}%
inclusive & $356.7(2)$ & $-1.2(2)$ & $-1.4$ & $-1.4$ \\
$p_{{\rm T}, {\rm t/\bar t/H}}>200\GeV$ & $12.244(3)$ & $-8.5(1)$ & $-8.5$ & $-8.4$ \\
$p_{{\rm T}, {\rm t/\bar t/H}}>400\GeV$ & $0.3435(3)$ & $-14.1(2)$ & $-13.9$ & $-14.0$ \\
$p_{{\rm T}, {\rm H}}>500\GeV$ & $1.7798(9)$ & $-11.7(1)$ & $-11.6$ & $-11.7$ \\
\rule[-1.2ex]{0ex}{1ex}%
$\left| y_{\rm t} \right| > 2.5$ & $5.035(3)$ & $0.3(2)$ & $0.5$ & $0.5$ \\
\hline
\end{tabular}
\end{center}
\caption{\label{table:ttHNLOEW}
Integrated cross sections for ${\rm p} {\rm p} \to {\rm t} {\rm \bar
  t} {\rm H}$ at NLO EW for a centre-of-mass energy of $\sqrt{s} = 13\TeV$ calculated with \Sherpa\!\!+\Recola, \madgraph, and \Sherpa\!\!+\OpenLoops for the set-up of Ref.~\cite{Badger:2016bpw}.
The cross sections are expressed in fb, while the relative corrections are given in percent.
The integration errors of the last digits are provided in parentheses for the \Sherpa\!\!+\Recola predictions.}
\end{table}
Generally good agreement has been found with the results presented in Ref.~\cite{Badger:2016bpw}.
Nonetheless, as no statistical errors are stated in the aforementioned reference, an exact comparison has not been possible.

\paragraph{Combined predictions:}
Next, combined NLO QCD and EW predictions for the inclusive
cross section as well as for a few distributions are presented.
No event selection is applied to the final state, meaning that we consider the fully-inclusive production process.
The total cross sections at LO and NLO for an additive and
multiplicative combination of NLO QCD and EW corrections as defined in
Eqs.~(\ref{additionNLO}) and (\ref{productNLO}), respectively, are listed in \refta{TABLE:ttHPredictions}.
\begin{table}
\begin{center}
\begin{tabular}{|c|c|c|c|c|c|}
\hline \rule[-1.4ex]{0ex}{3.8ex}%
 ${\rm pp} \to {\rm t \bar t H}$ & $\sigma^\mathrm{LO}$~[fb] & $\sigma^\mathrm{NLO}_{\mathrm{QCD}+\mathrm{EW}}$~[fb] & $\delta^\mathrm{NLO}_{\mathrm{QCD}+\mathrm{EW}}$~[\%] & $\sigma^\mathrm{NLO}_{\mathrm{QCD}\times\mathrm{EW}}$~[fb] & $\delta^\mathrm{NLO}_{\mathrm{QCD}\times\mathrm{EW}}$~[\%]\\
\hline \rule{0ex}{2.4ex}%
inclusive & $361.2(4)$ & $459.8(4)$ & $27.3(2)$ & $458.6(4) $ & $27.0(2)$ \\
\hline
\end{tabular}
\end{center}
\caption{\label{TABLE:ttHPredictions}
Integrated cross sections for ${\rm p} {\rm p} \to {\rm t} {\rm \bar t} {\rm H}$ for a centre-of-mass energy of $\sqrt{s} = 13\TeV$ calculated with \Sherpa\!\!+\Recola for the set-up of Ref.~\cite{Badger:2016bpw}.
The cross sections at LO as well as for the additive and multiplicative combinations of NLO QCD and EW corrections are given.
The cross sections are expressed in fb, while the relative corrections are given in percent.
The integration errors of the last digits are provided in parentheses.}
\end{table}
As the EW corrections are moderate, there are no big differences between the two combinations.
This seems to indicate that the missing higher orders of mixed QCD--EW
type are small in this case.

In Fig.~\ref{FIG:HFIGURES}, the transverse momentum as well as the rapidity distribution of the Higgs boson are displayed.
\bfig
  \center
  \includegraphics[width=0.45\textwidth]{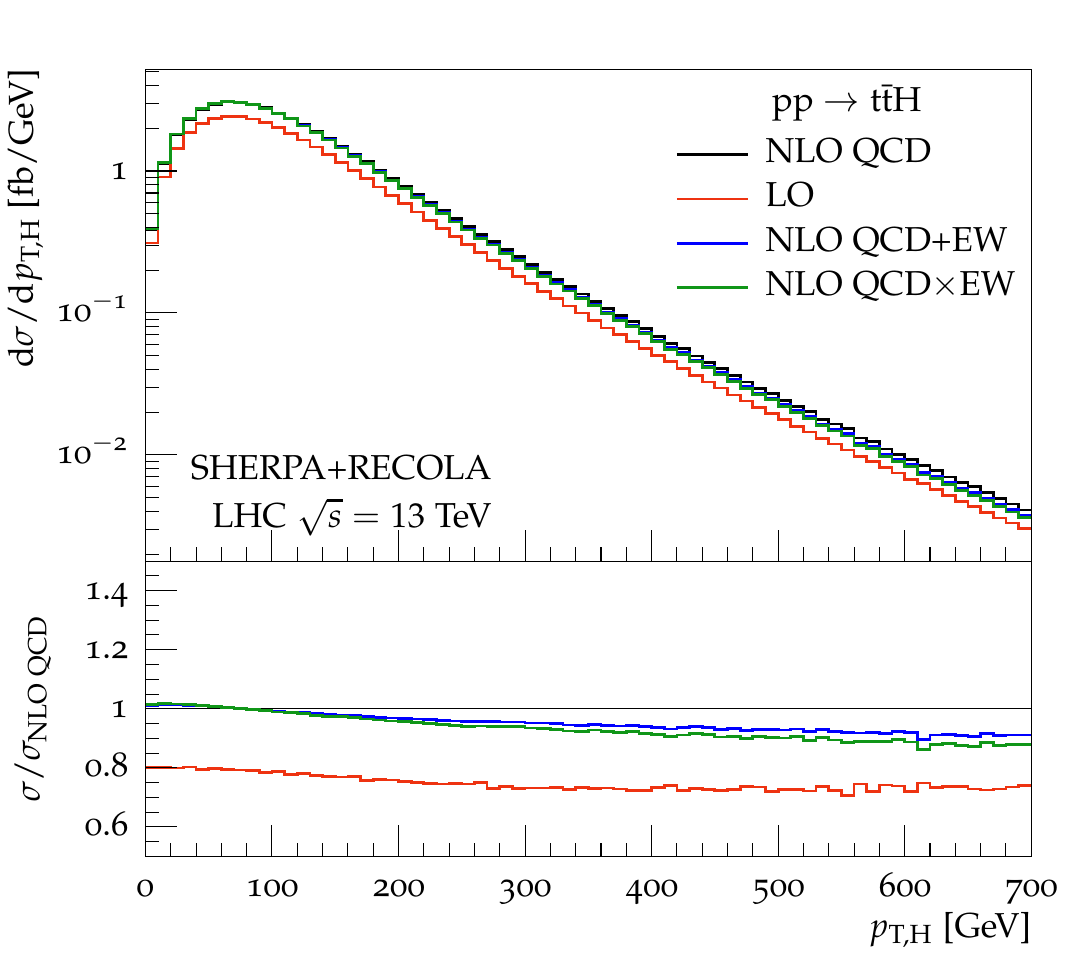}
  \includegraphics[width=0.45\textwidth]{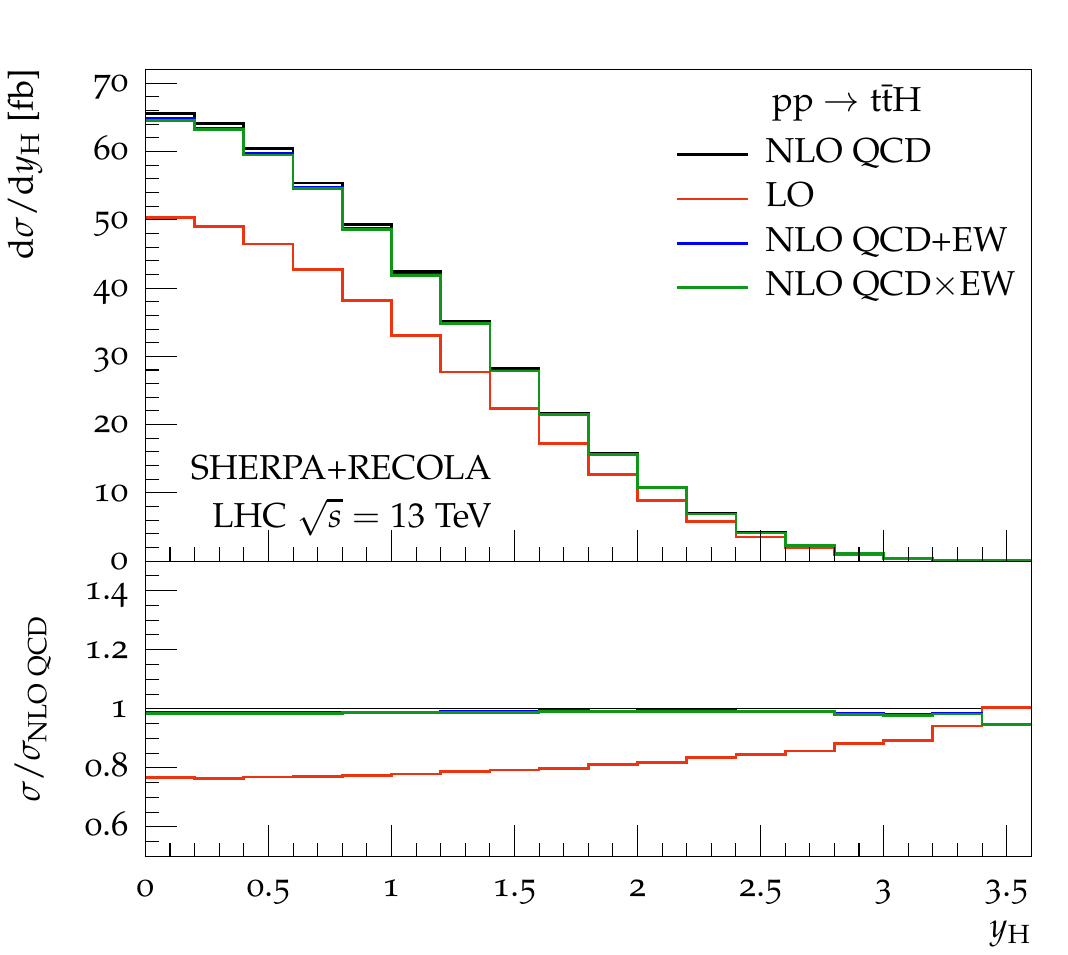}
  \caption{\label{FIG:HFIGURES} 
  Differential distributions at a centre-of-mass energy $\sqrt{s} = 13\TeV$ for ${\rm pp} \to {\rm t \bar t H}$ at the LHC:
  transverse-momentum distribution (left) and rapidity distribution of the Higgs boson (right).
  The lower panels show the ratios $\sigma^\mathrm{LO}/\sigma^\mathrm{NLO}_{\mathrm{QCD}}$, $\sigma^\mathrm{NLO}_{\mathrm{QCD}+\mathrm{EW}}/\sigma^\mathrm{NLO}_{\mathrm{QCD}}$ and $\sigma^{\mathrm{NLO}}_{\mathrm{QCD}\times\mathrm{EW}}/\sigma^{\mathrm{NLO}}_{\mathrm{QCD}}$.}
\efig
The effects of the NLO QCD corrections are dominant over the whole
transverse-momentum range and are typically of the order of $25\%$. The EW corrections vary from about $1.2\%$ at zero transverse momentum to $-8.9\%$ at $700\GeV$.
This behaviour is characteristic for Sudakov logarithms that grow large when all invariants involved in the process become large.
In Fig.~\ref{FIG:TTFIGURES}, the distribution of the transverse momentum as well as the rapidity of the top quark are shown.
\bfig
  \center
  \includegraphics[width=0.45\textwidth]{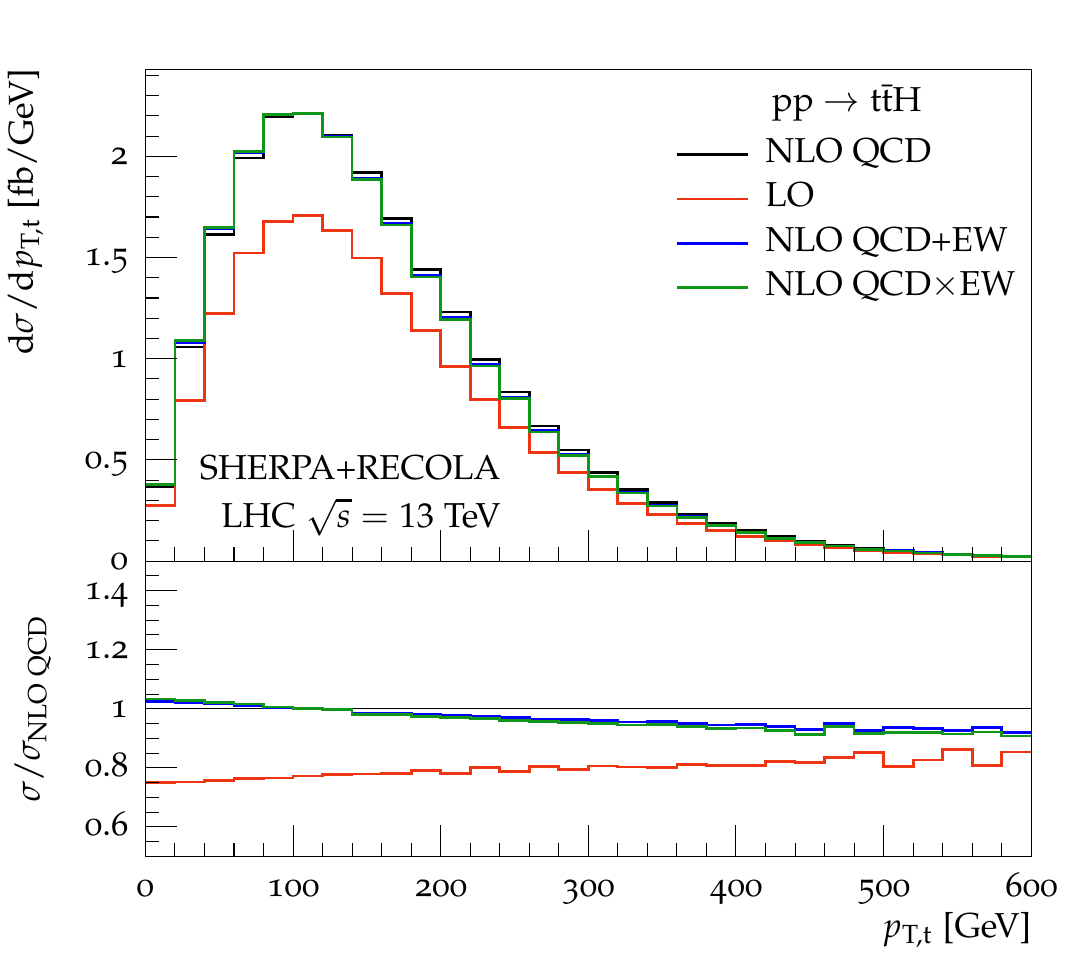}
  \includegraphics[width=0.45\textwidth]{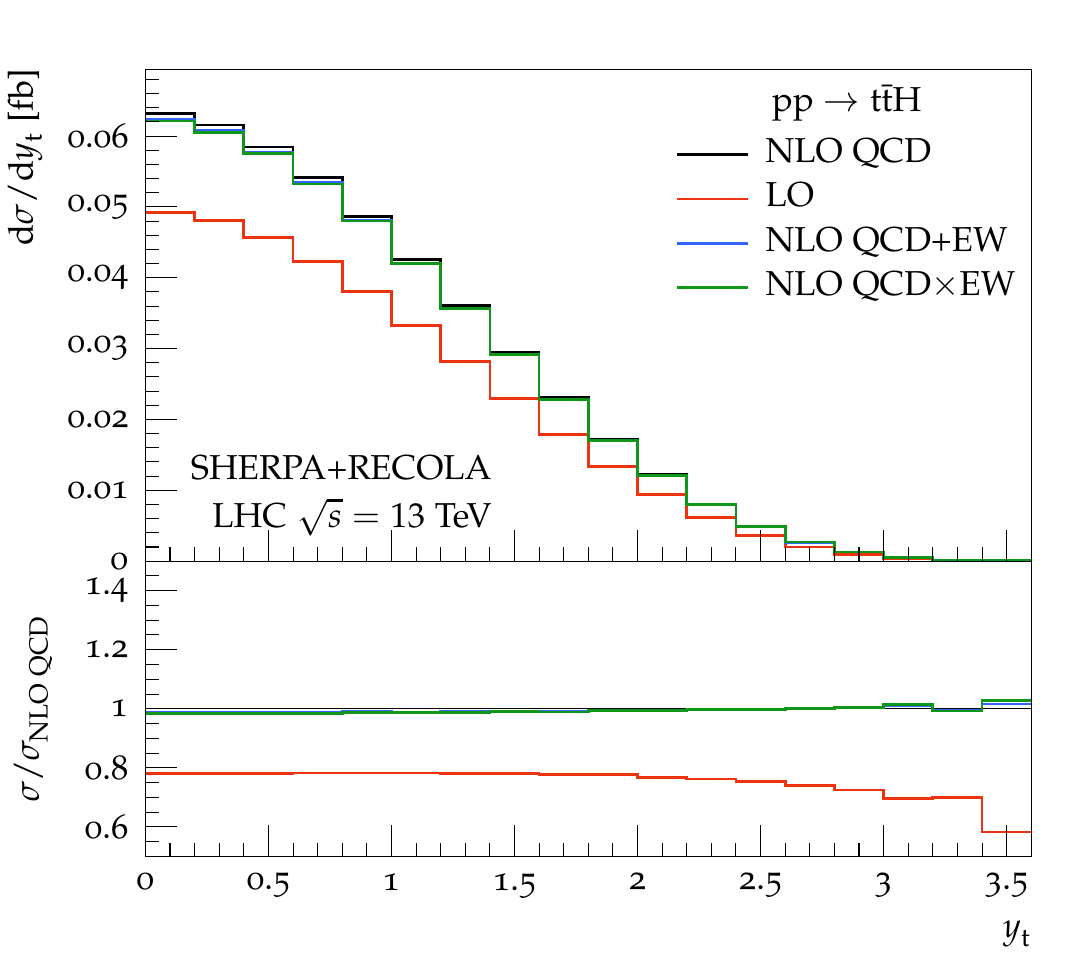}
  \caption{\label{FIG:TTFIGURES}
  Differential distributions at a centre-of-mass energy $\sqrt{s} = 13\TeV$ for ${\rm pp} \to {\rm t \bar t H}$ at the LHC:
  transverse-momentum distribution (left) and rapidity distribution of the top quark (right).
  The lower panels show the ratios $\sigma^\mathrm{LO}/\sigma^\mathrm{NLO}_{\mathrm{QCD}}$, $\sigma^\mathrm{NLO}_{\mathrm{QCD}+\mathrm{EW}}/\sigma^\mathrm{NLO}_{\mathrm{QCD}}$ and $\sigma^\mathrm{NLO}_{\mathrm{QCD}\times\mathrm{EW}}/\sigma^\mathrm{NLO}_{\mathrm{QCD}}$.}
\efig
Again, the QCD corrections are large over the whole transverse momentum range
and amount to at $35\%$
at low transverse momentum to go down to $22\%$ at $p_{{\rm T},\Pt}=600\GeV$.
The relative EW corrections also decrease from about $2.6\%$ to reach $-6.7\%$ at $600\GeV$.

\section{Conclusion}
\label{sec:conclusion}

After the very successful Run~I, culminating in the discovery of the Higgs boson, the LHC is now operating in the
Run~II phase. This phase at $\sqrt{s}=13\TeV$ might ultimately lead to the discovery of New Physics or confirm, to an
even higher degree of accuracy, the theory of the Standard Model of
particle physics. In either case, precise (including at least NLO
QCD and electroweak corrections) and appropriate (with event selections reproducing the experimental set-ups) predictions for
a plethora of Standard Model processes are needed. Such state-of-the-art predictions are, on the one hand, required
for precision measurements in the Standard Model. On the other hand, in New Physics searches, they are compulsory to
obtain realistic estimates of the Standard Model expectations in order to discriminate possible New Physics
contributions. To provide such predictions, public Monte Carlo programs are the ideal tools as they can be used
by both the experimental collaborations and the theory community.

In this publication, the combination of the one-loop matrix-element generator \Recola with the multipurpose
Monte Carlo program \Sherpa has been presented. In particular, a short presentation of both codes as well as the
main features of the interface have been given. The \Sherpa\!\!+\Recola framework is designed for Standard Model
predictions and offers the possibility to compute---in principle---any process at NLO QCD and electroweak accuracy.
A large fraction of this article is devoted to the validation of the implementation of \Recola in \Sherpa.
This entails comparisons of squared matrix elements for individual phase-space points, fixed-order cross sections,
and differential distributions, as well as the merging/matching of NLO QCD matrix elements with \Sherpa's
QCD parton shower. These comparisons are performed against public and private codes as well as results presented in the
literature. 

Following this validation, predictions have been presented at NLO QCD and electroweak accuracy for three specific
processes:  both on- and off-shell vector-boson production in association with jets, off-shell ${\rm Z}$-boson pair production, 
and on-shell production of a
top-quark pair in association with a Higgs boson. In addition to their distinguished physical relevance, these
processes constitute a good testing ground for this fully automatised implementation. They feature both massive and
massless final states, as well as strongly and electroweakly interacting final-state objects. In addition to fixed-order
computations, all other functionalities of \Sherpa (the QCD parton shower, hadronisation etc.) can be used along
with \Recola. To demonstrate this, NLO QCD matrix elements for Drell--Yan production in association with multiple
QCD jets have been merged and matched to the parton shower. For illustrative purposes, some resulting predictions have
been compared to actual LHC data. 

The \Sherpa\!\!+\Recola combination is readily publicly available for NLO QCD predictions. The required methods
to perform NLO electroweak calculations on the \Sherpa side will be made public soon. This ultimately makes it an ideal
tool for both experimentalists and theorists to obtain NLO QCD and electroweak accurate predictions for Standard Model
processes. It opens the possibility to perform systematic studies on the impact of electroweak corrections for a
multitude of LHC production processes.

\section*{Acknowledgements}

We thank Jean-Nicolas Lang and Sandro Uccirati for developing and supporting the code \Recola.
This work has benefited from useful interactions with them. 
We thank our colleagues from the \Sherpa collaboration for fruitful discussions and
technical support. In particular, we are grateful to Silvan Kuttimalai and Marek Sch{\"o}nherr
for their assistance.
Moreover, we would also like to thank Jonas Lindert for his help.

BB, AD and MP acknowledge financial support from  BMBF under contract 05H15WWCA1.
SB, SS and JT acknowledge financial support from the EU research network MCnetITN funded by the
Research Executive Agency (REA) of the European Union under Grant Agreement PITN-GA-2012-315877
and from BMBF under contract 05H15MGCAA.

\appendix

\section{Installation procedures}
\label{sec:installation}

\Recola-1.2 and subsequent versions are compatible with the interface to \Sherpa described above.
\Recola in association with \collier can be downloaded from\\
\url{http://recola.hepforge.org}.\\
Once downloaded, the following command lines have to be issued:
\begin{verbatim}
tar -zxvf recola-collier-1.2.tar.gz
cd recola-collier-1.2 
cd build 
cmake .. 
make
\end{verbatim}
\Recola is then installed in association with \collier.
Various compilation options can be found in the respective manuals \cite{Actis:2016mpe,Denner:2016kdg}. We note that the \Recola library has to be compiled dynamically (this is the default setting) to be used with \Sherpa.

The first version of \Sherpa compatible with \Recola is \Sherpa-v2.2.3.
It can be downloaded from \url{http://sherpa.hepforge.org}.\\
The installation commands read
\begin{verbatim}
tar -zxvf SHERPA-MC-2.2.3.tar.gz
cd SHERPA-MC-2.2.3
autoreconf -i
./configure --enable-recola=/PATH_TO_RECOLA/recola-collier-1.2/recola-1.2 \
[other Sherpa configure options]
make
make install
\end{verbatim}
Extra configuration options can be found in the manual of \Sherpa available on the website.
After this installation procedure, NLO computations can be readily performed.

\section{Specific run-card commands}
\label{sec:commands}

As mentioned in Section~\ref{sec:interface}, some commands allow the user to deviate from the default settings.
This appendix is thus devoted to the description of these commands.

\subsection*{On-shell masses for W and Z boson}

By default, the input masses for the W and Z bosons are the pole masses.
Nonetheless, it is possible to set the on-shell masses instead by
including the line 
\begin{verbatim}
RECOLA_ONSHELLZW=1
\end{verbatim}
in the input run card.

\subsection*{Flavour scheme and quark masses}

In order to allow all possible combinations of masses and flavour schemes, a few flags for the run card exist.
The first one is 
\begin{verbatim}
RECOLA_FIXED_FLAVS.
\end{verbatim}
The values 4, 5, and 6 correspond to the corresponding fixed-flavour schemes.
Setting the flag to 14, 15, 16 allows for a dynamical scheme up to the number of flavours (4, 5 or 6).
Finally, the value $-2$ sets a fixed-flavour scheme according to the masses set in the run card.

It is also possible to set explicitly the masses used in the renormalisation [see Eq.~\eqref{eq:gsct}] of the strong coupling using 
\begin{verbatim}
 RECOLA_AS_REN_MASS_C/B/T, 
\end{verbatim}
corresponding to the charm, bottom, and top-quark mass, respectively.
On the other hand, the quark masses used for the hard matrix element are read out only from the run card.
This means that the user has to take care of the consistency of
her/his computation between the matrix element computed and the PDF
set used.

\subsection*{UV and IR scales}

The UV and IR scales are both set, by default, to a fixed value of 
$100\GeV$. These technical parameters do not impact physical results. However, the choice of the IR scale
does change the individual contributions of the virtual corrections and
the integrated subtraction terms. In the
\Sherpa\!\!+\Recola interface, it is possible to set these scales 
in the run card, \eg to directly compare the 
virtual and real subtraction contributions separately to
independent code. These UV and IR scales are set with the keywords 
\begin{verbatim}
UV_SCALE
\end{verbatim}
and 
\begin{verbatim}
IR_SCALE, 
\end{verbatim}
respectively.

\bibliographystyle{utphys.bst}
\bibliography{sherpa_recola}
\end{document}